\documentstyle[12pt,epsf]{article}

\input epsf

\newcommand{\be}{\begin{eqnarray}}

\newcommand{\ee}{\end{eqnarray}}

\newcommand{\nn}{\nonumber\\ }
\def\labe{\label}

\def\simge{\mathrel{%
   \rlap{\raise 0.511ex \hbox{$>$}}{\lower 0.511ex \hbox{$\sim$}}}}
\def\simle{\mathrel{
   \rlap{\raise 0.511ex \hbox{$<$}}{\lower 0.511ex \hbox{$\sim$}}}}
\def\bigs{\mathrel{
   \rlap{\raise 0.531ex \hbox{$>$}}{\lower 0.531ex \hbox{$<$}}}}
\def\buildchar#1#2#3{{\null\!
    \mathop#1\limits^{#2}_{#3}
    \!\null}}

\def\grad{\nabla}                               
\def\del{\partial}                              

\def\frac#1#2{{#1 \over #2}}

\def\half{\ifinner {\scriptstyle {1 \over 2}}
   \else {1 \over 2} \fi}

\def\simge{\mathrel{%
   \rlap{\raise 0.511ex \hbox{$>$}}{\lower 0.511ex \hbox{$\sim$}}}}
\def\simle{\mathrel{
   \rlap{\raise 0.511ex \hbox{$<$}}{\lower 0.511ex \hbox{$\sim$}}}}
\def\bigs{\mathrel{
   \rlap{\raise 0.531ex \hbox{$>$}}{\lower 0.531ex \hbox{$<$}}}}

\def\buildchar#1#2#3{{\null\!                   
   \mathop#1\limits^{#2}_{#3}                   
   \!\null}}                                    

\def\slashchar#1{\setbox0=\hbox{$#1$}           
   \dimen0=\wd0                                 
   \setbox1=\hbox{/} \dimen1=\wd1               
   \ifdim\dimen0>\dimen1                        
      \rlap{\hbox to \dimen0{\hfil/\hfil}}      
      #1                                        
   \else                                        
      \rlap{\hbox to \dimen1{\hfil$#1$\hfil}}   
      /                                         
   \fi}                                         %

\def\subrightarrow#1{
  \setbox0=\hbox{
    $\displaystyle\mathop{}
    \limits_{#1}$}
  \dimen0=\wd0
  \advance \dimen0 by .5em
  \mathrel{
    \mathop{\hbox to \dimen0{\rightarrowfill}}
       \limits_{#1}}}                           

\begin{document}
\begin{titlepage}
\begin{flushright}
Saclay-T00/166 \\BNL-NT-00/24\\hep-ph/0011241
\end{flushright}
\vspace*{1.2cm}
\begin{center}
{\Large{\bf Nonlinear Gluon Evolution in the Color Glass Condensate: I}}
\vskip0.3cm
 Edmond Iancu,\footnote{E-mail: eiancu@cea.fr}\\
{\small\it Service de Physique Th\'eorique\footnote{Laboratoire 
de la Direction des
Sciences de la Mati\`ere du Commissariat \`a l'Energie
Atomique}, CEA Saclay, 91191 Gif-sur-Yvette, France}
        
Andrei Leonidov\footnote{E-mail: leonidov@lpi.ru }\\
         {\small\it P. N. Lebedev Physical Institute,
          Moscow, Russia} 

Larry McLerran\footnote{E-mail: mclerran@quark.phy.bnl.gov }\\
       {\small\it Physics Department, Brookhaven National Laboratory,
                 Upton, NY 11979, USA  }

\end{center}

\date{\today}
\vskip 1.2cm

\parindent=20pt

\begin{abstract}
We consider a nonlinear evolution equation recently proposed
to describe the small-$x$ hadronic physics in the regime of very 
high gluon density.  This is a functional Fokker-Planck equation
in terms of a classical random color source, which represents
the color charge density of the partons with large $x$. 
In the saturation regime of interest, the coefficients of this
equation must be known to all orders in the source strength.
In this first paper of a series of two, we carefully derive
the evolution equation, via a matching between classical 
and quantum correlations, and set up the framework for the
exact background source calculation of its coefficients.
We address and clarify many of the subtleties and
ambiguities which have plagued past attempts at an explicit 
construction of this equation.  
We also introduce the physical interpretation of the saturation
regime at small $x$ as a Color Glass Condensate.
In the second paper we shall evaluate the expressions
derived here, and compare them to known results in various limits.

\end{abstract}
\end{titlepage}

\section{Introduction}

There has been much activity in the last few years in an attempt to
understand the physics of nuclear and hadronic processes in the
regime where Bjorken's $x$ becomes very small 
\cite{MV94}--\cite{Elena00} (see also \cite{AM1} for a pedagogical
presentation and more references). 
The remarkable feature about this
regime is that the gluon density in the hadron wavefunction becomes 
so high that perturbation theory breaks down even for a small
coupling constant, by strong non-linear effects. 

The important new phenomenon which is expected in 
these conditions is {\it parton saturation} \cite{GLR,MQ,FS}, which is
the fact that the density of partons per unit phase space 
$(dN/d^2p_\perp\,dy\,d^2x_\perp)$
cannot grow forever as $x$ becomes arbitrarilly small
[here, $y\equiv \ln(1/x)$ is the parton rapidity].
If this growth were to occur, then the cross section for deep
inelastic scattering at fixed $Q^2$ would grow unacceptably large and
eventually violate unitarity bounds. One rather expects the cross
section to approach some asymptotic value at high energy, that is, 
to saturate.  It is also intuitively
obvious that this should happen, since if the phase space density
becomes too large, repulsive interactions are generated among the
gluons, and eventually it will be
energetically unfavorable to increase the density anyfurther. 

As $x$ becomes smaller and smaller, the gluon density increases faster,
and is the driving force towards saturation. This is the reason why, 
in what follows, we shall concentrate exclusively on the gluon dynamics.
Saturation is expected at a gluon phase-space density of order 
$1/\alpha_s\/$ \cite{GLR,JKMW97,KM98}, a value typical of condensates.
This estimate, together with the fact that gluons are massless bosons, 
leads naturally to the expectation that the saturated gluons
form a new form of matter, which is a Bose condensate.
Since gluons carry color which is a 
gauge-dependent quantity, any gauge-invariant formulation
will neccessarily involve an average over all colors, to restore
the invariance. This averaging procedure bears a formal resemblance to 
the averaging over background fields done for spin glasses \cite{PS79}.
The new matter will be therefore
called the {\it Color Glass Condensate} (CGC). This matter is
universal in that it should describe the high gluon density part of any
hadron wavefunction at sufficiently small $x$. 

The CGC picture holds
in a frame in which the hadron propagates at the speed of light and,
by Lorentz contraction, appears 
as an infinitesimally thin two-dimensional sheet located at
the light-cone. The formalism supporting this picture is in
terms of a {\it classical effective theory} valid for some 
range of $x$, or longitudinal momentum\footnote{See Sect. 2.1
for the definition of light-cone coordinates and momenta.} $p^+$.
In this theory, originally proposed by McLerran
and Venugopalan to describe the gluon distribution in large nuclei
\cite{MV94}, the valence quarks of the hadron (and, more generally,
the ``fast'' partons, i.e., the partons which carry a large fraction of the
hadron longitudinal momentum) are treated as sources for a classical
color field which represents the small-$x$, or ``soft'',
gluons in the hadron wavefunction. 
The classical approximation is appropriate since the
saturated gluons have large occupation numbers $\sim 1/\alpha_s$,
and can thus be described
by {classical} color fields with large amplitudes $A^i\sim 1/g$.

When going to smaller values of $x$, or lower longitudinal 
momenta $p^+$, 
one must integrate out the degrees of freedom
associated with this change of scale \cite{JKMW97}.
This is done in perturbation theory for the quantum gluons
(to leading logarithmic accuracy), 
but to {\it all} orders in the strong background
fields $A^i$ which represent the condensate.
This entails a change in the effective theory, which is governed
by a functional, {\it non-linear}, evolution equation originally
derived by Jalilian-Marian, Kovner, Leonidov and Weigert (JKLW)
\cite{JKLW97,JKLW99a}. 
This is a functional Fokker-Planck equation for the statistical
weight function associated with the random variable $\rho_a(x)$
($\equiv$ the effective color charge density of the fast gluons).

One of the consequences of the JKLW renormalization group equation
it that it predicts the evolution of the gluon distribution function.
In the low density, or weak-field limit, this equation
linearizes and reduces to the BFKL equation \cite{BFKL}, 
as shown in Ref. \cite{JKLW97}. But in the saturation
regime, where $A^i\sim 1/g$,  the JKLW equation is fully
non-linear, and will be discussed in great detail 
in this and the accompanying paper
(to be referred to as ``Paper I'' and ``Paper II'' in what follows).
In particular, in Paper II,
we will show that this leads to the Balitsky-Kovchegov
equation in the large-$N_c$ limit (with $N_c$ the number of colors).
There is currently a discrepancy between the computations
of Refs.  \cite{JKW99,JKLW99b,KM00,KMW00} 
and those by Balitsky   \cite{B} and Kovchegov \cite{K}.
The additional terms argued in \cite{KMW00} do actually not show up
in our computation \cite{ILM00II}.

In the previous arguments, we have implicitly assumed the QCD
coupling constant $\alpha_s$ to be small, which can be justified as 
follows: Associated with saturation, there is a characteristic
transverse momentum scale $Q_s$ (a function of $x$),
which is proportional to the density of gluons per unit area:
\be
      Q_s^2\, \propto \,{ 1 \over {\pi R^2}}\,{{dN} \over {dy}}\,,
\ee
where $R$ is the radius of the hadron\footnote{This
can certainly be defined for a nucleus, but even for a proton it is
a meaningful quantity since,
 as we shall see, the typical transverse momentum at
small $x$ becomes large, so that one can properly resolve the
transverse extent of the hadron.}. $Q_s$ is the typical
momentum scale of the saturated gluons, and increases with $1/x$
 (see, e.g., Refs. \cite{JKMW97,KM98,AM1}). 
Thus, for sufficiently small $x$,  $Q_s^2 \gg \Lambda^2_{QCD}$,
and the coupling ``constant'' $\alpha_s(Q^2_s)$ is weak indeed.
(In any explicit calculation, the  condition
$Q_s^2 \gg \Lambda^2_{QCD}$ must be
verified a posteriori, for consistency.)

To summarize, the system under consideration
is a weakly coupled high density system described
by classical gluon fields. This is still non-perturbative,
in the sense that the associated non-linear effects
cannot be expanded in perturbation theory. But the
non-perturbative aspects can now be studied in the simpler
setting of a classical theory, via a combination of 
analytical and numerical methods
(see, e.g., Refs. \cite{MV94,KMW,JKMW97,KM98,KV00,AM1}).
In particular, in this context, it has been demonstrated
\cite{JKMW97,KM98} that saturation occurs indeed,
via non-linear dynamics.

The purpose of this paper and its sequel is twofold:
first, to carefully rederive the JKLW evolution equation, 
by matching classical and quantum correlations;
second, to render this equation precise and 
fully explicit, 
by computing its coefficients to all orders in the color field
which describes the Color Glass Condensate. 
This requires an exact background field
calculation which will be prepared by the analysis in Paper I,
and presented in detail in Paper II.

Many of the results we shall present (especially in Paper I)
are known in one form or another, and can be found in the 
literature. We shall nevertheless attempt to be comprehensive 
and pedagogical in our present treatment, in order to clarify 
several subtle points 
which have been often overlooked in the previous literature.
To be specific, let us spell out here the new features of our
analysis, to be presented below in this paper:

\begin{itemize}

\item
The effective theory for the soft fields is stochastic Yang-Mills
theory with a random color source localized near the light-cone
($x^-\simeq 0$) [see Sect. 2.1]. In previous work,
this source has been generally taken as a delta function at
$x^-=0$. Here, we find that in order to do a
sensible computation, it is necessary at intermediate steps to
spread out the source in $x^-$, that is, to take its longitudinal
structure into account.

\item
For our `spread out' source, we describe in
detail the structure of the classical solution for the background field.
We do this by relating the solution in the light cone gauge
($A^+=0$) to the solution in the covariant gauge ($\partial_\mu A^\mu=0$). 
We thus obtain an explicit expression for the background field
in terms of the source in the {\it covariant} gauge [see Sect. 2.3]. 
This indicates that the source which is generated by quantum 
evolution must be evaluated in the covariant gauge as well.

\item
The quantum evolution of the effective theory is obtained by
integrating out quantum fluctuations in a rapidity interval
$\Delta y=\ln(1/b)$ (with $b\ll 1$) to leading order in
$\alpha_s\ln(1/b)$, but to all orders in the strong background
fields [cf. Sect. 3.2.1]. 
The quantum correlations induced in this 
way are then transferred to the stochastic classical theory, 
via a renormalization group equation of the Fokker-Planck
type [cf. Sect. 3.2.2].
To prove the equivalence between the classical and
quantum theories, we explicitly compute the induced field correlators
$\langle A^iA^j\dots\rangle$ in both theories and show that they
are the same, to the order of interest [cf. Sect. 3.2].

\item
The original JKLW renormalization group equation is written in terms
of the color source in the {\it light-cone} gauge \cite{JKLW99a}.
When reexpressing this equation in terms of the 
source in the {\it covariant} gauge, we find that we must subtract 
out a term related to classical vacuum polarization, 
so that the classical and quantum theory properly match [see Sect. 3.3].

\item
The non-linear evolution equation involves functional
differentiations with respect to the source strength.
With sources spread out and with explicit results from the
covariant gauge, we have an explicit and transparent definition 
of these functional differentiations [see Sect. 3.2.2].

\item
The background fields are independent of the light-cone time $x^+$,
but inhomogeneous, and even singular, in $x^-$. To cope with that, 
we find that it is more convenient to integrate the
quantum fluctuations in layers of $p^-$ (the light-cone energy),
rather than of $p^+$ [cf. Sect. 3.4]. Accordingly, we have no
ambiguity associated with possible singularities at $p^- = 0$.

\item
The gauge-invariant action describing the coupling of the
quantum gluons to the classical color source is non-local
in time \cite{JKLW97}. Thus, the proper formulation of the
quantum theory is along a Schwinger-Keldysh 
contour in the complex time plane [cf. Sect. 4].
However, in the approximations of interest, and given the 
specific nature of the background, 
the contour structure turns out not to be essential,
so one can restrict oneself to the previous
formulations in real time \cite{JKLW97,JKLW99a}.

\item
We carefully fix the gauge in the quantum calculations. 
In the light-cone gauge, the gluon propagator has singularities
at $p^+ = 0$ associated with the residual
gauge freedom under $x^-$--independent gauge transformations. We resolve
these ambiguities by using a retarded $i\epsilon$ prescription
[cf. eq.~(\ref{LCPROP}) and Sect. 6.3], which is chosen for consistency
with the boundary conditions imposed on the classical background field.
We could have equally well used an advanced prescription. However,
other gauge conditions, such as
Leibbrandt-Mandelstam or principal value prescriptions,
are for a variety of technical reasons shown to be
unacceptable \cite{ILM00II}.

\item
The choice of a gauge prescription has the interesting
consequence to affect the spatial distribution of the source in
$x^-$, and thus to influence the way we visualize the generation of
the source via quantum evolution. With our retarded prescription,
the source has support only at positive $x^-$ [see Sect. 5.1].

\item
We verify explicitly that the obtention of the
BFKL equation as the weak-field limit of the general 
renormalization group equation
is independent of the gauge-fixing prescription [cf. Sect. 5].

\end{itemize}

Once these various points are properly understood, the
background field calculation in Paper II becomes unambiguous
and well defined.
As a result of this calculation, the coefficients in the
JKLW equation will be obtained explicitly in terms of the color
source in the covariant gauge \cite{ILM00II}.
With such explicit expressions in hand,
we hope that the functional evolution equation can be solved,
via analytic or numerical methods, or both,
at least in particular limits. In fact, as we shall see in Paper II,
in the large $N_c$ limit our results become equivalent to the equation
proposed by Balitsky \cite{B} and Kovchegov \cite{K},
for which an analytic solution has been recently given \cite{W}.
 
{ The ultimate goal of this and related analyses is to 
derive the properties of hadronic physics at arbitrarily high energies
from first principles, that is, from QCD.
We conjecture that there will be universal behaviour for all hadrons in the
high energy limit. The work we present here is a step in proving this, 
and in developing a formalism which allows for explicit calculations
in the saturation regime.

The present paper is organized as follows: 
In Section 2, we discuss the solution to the classical
Yang-Mills equations  with a color source
which define the Mclerran-Venugopalan model.
In Section 3, we derive the non-linear evolution equation
by matching classical and quantum correlations.
In Section 4, we formulate the quantized version of the 
Mclerran-Venugopalan model (on a complex-time contour), and
derive Feynman rules for the coefficients of the
renormalization group equation. In Section 5, we use the
weak field, or BFKL, limit of the general evolution equations
to test properties like the longitudinal distribution of the induced
source, the equivalence between the renormalization group analysis
in $p^+$ and in $p^-$, and the sensitivity to the $i\epsilon$
prescription in the LC-gauge propagator.
In Section 6, we explicitly construct the gluon propagator 
in the presence of the background field describing the
Color Glass Condensate. This propagator is the main ingredient
of the quantum calculations to be presented in Paper II
\cite{ILM00II}.

}
\setcounter{equation}{0}
\section{The classical approximation}

In this section, we shall introduce the McLerran-Venugopalan model
\cite{MV94} 
for the gluon distribution at small $x$. This is classical Yang-Mills
theory with a random color source, which is the effective color charge
of the partons having longitudinal momenta larger than
the scale of interest. This simple model, which is motivated by the
separation of scales in the infinite momentum frame (see Sect. 2.1),
offers an explicit scenario for saturation \cite{JKMW97,KM98}
(see Sect. 2.4) which supports the physical picture
of the Color Glass Condensate. The MV model will be further
substantiated by the analysis in Sect. 3, where we shall see
that this model is consistent with the quantum evolution
towards small $x$  \cite{JKMW97,JKLW97,JKLW99a}.

\subsection{The McLerran-Venugopalan model}

We consider a hadron in the infinite momentum frame (IMF),
and in the light-cone (LC) gauge $A^+_a=0$,
where the parton model is conventionally formulated (this gauge 
choice will be further motivated in Sect. 2.4). 
The hadron four-momentum
reads $P^\mu=(P,0,0,P)$ (we neglect the hadron mass), or, in LC vector 
notations, which we shall use systematically from now on\footnote{These
are as defined as follows: for an arbitrary
4-vector $v^\mu$, we write $v^\mu=(v^+,v^-,{\bf v}_\perp)$, with
$v^+\equiv (1/\sqrt 2)(v^0+v^3)$,
$v^-\equiv (1/\sqrt 2)(v^0-v^3)$, and ${\bf v}_\perp
\equiv (v^1,v^2)$. The dot product reads:
$p\cdot x=p^+x^- + p^-x^+-p_\perp\cdot x_\perp\,$.
$p^-$ and $p^+$ are, respectively, the LC energy and
longitudinal momentum; correspondingly, $x^+$ and $x^-$ are
the LC time and longitudinal coordinate.},
$P^\mu\equiv (P^+,P^-,{\bf P}_\perp)=({\sqrt 2} P,0,{\bf 0}_\perp)$.
The {\it fast partons}, that is, the hadron constituents which, like 
the valence quarks, carry a relatively large fraction of the hadron 
longitudinal momentum $P^+$, move as almost free particles
with momenta collinear with $P^\mu$,
and act as sources for {\it soft gluons}, 
i.e., gluons with small longitudinal momenta $q^+\ll P^+$,
which are probed in deep inelastic scattering (DIS).

For what follows, it is useful to have a sharp distinction between
``fast'' and ``soft''  modes in the hadron 
wavefunction, according to their $p^+$-momentum:
to this aim we introduce an intermediate scale $\Lambda^+$
and define fast (soft) modes to have $p^+ > \Lambda^+$
(respectively, $p^+ < \Lambda^+$). 
We choose $\Lambda^+=x_0 P^+$, with $x_0$ not too small; 
in fact, the main constraint on $x_0$ is to be much larger
than the Bjorken parameter $x$ of the external probe that 
the hadron is interacting with. 


For instance, in DIS, this probe
is the virtual photon $\gamma^*$, with
virtuality $Q^2\equiv -q^\mu q_\mu$. The Bjorken $x$ variable
is defined as $x\equiv Q^2/2P\cdot q$ and, by kinematics, it
coincides with the longitudinal momentum fraction $p^+/P^+$
of the struck parton. In what follows, we shall always assume
that $x\ll 1$. Rather than a virtual photon, which
couples to the charged constituents of the hadron
(quarks and antiquarks), it is more convenient to consider a DIS
initiated by the ``current''
$j\equiv -\frac{1}{4}F^{\mu\nu}_aF^a_{\mu\nu}$
which couples directly to gluons \cite{KM98}. Indeed, we are mainly
interested here in the gluon distribution, which is the dominant
component of the hadron wavefunction when $x \ll 1$.

In the MV model, one assumes that the soft gluons can be described
as the classical color fields $A^\mu_a$ 
radiated by the fast partons,
themselves represented by a random color source  $\rho_a$.
As we shall demonstrate in Sect. 3, this assumption is indeed justified (at 
least to lowest order in $\alpha_s$) provided the MV model is used as an 
{\it effective} theory valid for some range of soft momenta. Then, 
the fast degrees of freedom are treated as quantum, but they are 
integrated out (perturbatively) in the construction of the effective 
theory, while
the soft modes are non-perturbative, because of their large occupation
numbers, but can be treated as classical, for this same reason.
Via this construction, to be detailed in Sects. 3 and 4,
the classical source  $\rho_a$ will emerge as a natural way to
describe the effects of the fast modes on the dynamics of the
soft ones. In the original MV model \cite{MV94},
this source has been simply postulated, and its properties have been
inferred via an analysis of the separation of scales in the problem.
Since this separation will play a major role in what 
follows, let us briefly discuss it here:

The fast partons move along the $z$ axis with large $p^+$ momenta.
They can emit, or absorb, soft gluons,
but in a first approximation they do not deviate from their 
light-cone trajectories at $z=t$, or $x^-=0$ 
({\it no recoil}, or {\it eikonal approximation}). 
Thus, they generate a color current only
in the $+$ direction:  $J_a^\mu=\delta^{\mu+}J^+_a$.

As quantum fields, the fast partons are delocalized 
in the $x^-$ direction within a
distance $\lambda^- \sim 1/p^+$. However, when ``seen'' by 
the external probe or by the soft gluons (with momenta $q^+\ll p^+$, 
and therefore a poor longitudinal resolution), they
appear as {\it sharply localized} at the light cone, within
a distance $\Delta x^- \sim 1/\Lambda^+$.
(This can be also understood as an effect of the Lorentz contraction).
In some cases, to be carefully justified later, the associated
current $J^+$ can be represented as a $\delta$ function 
at the light cone: $J^+\propto \delta(x^-)$.
But in general, it turns out that the longitudinal 
extent of the source cannot be neglected: this will be important
both for the classical calculations in this section,
and for the quantum analysis to be performed
in the subsequent sections, and in Paper II.

Consider also the relevant LC-time scales: for on-shell excitations,
$2p^+p^-=p_\perp^2$, which implies that softer gluons have larger
energies $p^-\sim Q^2_\perp/p^+$, and therefore shorther lifetimes
$\Delta x^+\sim 1/p^- \sim p^+ \sim x$. Over such a short time
interval, the ``fast'' partons are only slowly varying (since
they have smaller energies), so their dynamics is essentially
frozen. Thus, the
soft gluons, or the external current in DIS, can probe only
the {\it equal-time} correlators of the fast partons. 
All such correlators can be generated by a 
{\it static}\footnote{We will see below 
that, strictly speaking, this time independence assumption will have
to be relaxed for a general solution of the classical
equations of motion:  The source turns out to be only 
{\it covariantly} time independent.  We will however always
be able to find a classical solution of the equations of motion which
has a truly time-independent source.} current $J^+$ 
({\it quenched approximation}). 

To summarize, the soft color current due to the
fast partons is expected with the following structure:
\be\labe{jclas}
J_a^\mu(x)=\delta^{\mu+}\rho_a(x^-,{\bf x}_\perp),\,\,\,
\,\,\del^-\rho_a\equiv \frac{\del\rho_a}{\del x^+}=0,\,\,\,\,\,
{\rm supp}\,\rho_a=\{x^-| 0 \le x^- \le 1/\Lambda^+ \},\,\,\ee
which will be confirmed by the quantum analysis in
Sect. 3--5 and Paper II.
Note the restriction of the support of $\rho$
to {\it positive} values of $x^-$. We shall see later that
the precise longitudinal structure of the color source
depends upon the gauge-fixing prescription, i.e., the
condition used to completely
fix the residual gauge freedom in the LC gauge.
[Recall that, even after imposing the LC gauge, 
one still has the possibility to perform $x^-$--independent
(or ``residual'') gauge transformations, which preserve
the condition $A^+=0$. In quantum calculations, fixing this
residual gauge freedom amounts to chosing a prescription for
the pole of the gluon propagator at $p^+=0$.]
With the specific prescription that we shall adopt
(cf. Sects. 3.4 and 6.3), the source $\rho$ is located at
positive $x^-$, as anticipated in eq.~(\ref{jclas}).

The current (\ref{jclas}) acts as a source for the Yang-Mills
equations describing the soft gluon dynamics:
\be
[D_{\nu}, F^{\nu \mu}]\, =\, \delta^{\mu +} \rho_a(x^-,{\bf x}_\perp)\,.
\labe{cleq0}
\ee
To have a gauge-invariant formulation, 
the source $\rho_a$ must be treated as a {\it stochastic} variable
with zero expectation value. This is also consistent with the
physical interpretation of $\rho_a$ as the instantaneous color charge 
of the fast partons ``seen'' by the shortlived soft 
gluons, at some arbitrary time. The (spatial) correlators of
the classical variable $\rho_a(\vec x)$,
with ${\vec x}\equiv (x^-,{\bf x}_{\perp})$, are inherited
from the (generally time-dependent) quantum correlations of the
fast gluons. The precise scheme for transforming
quantum into classical correlations will be explained in Sect. 3.
Here, we shall simply summarize the ensuing classical correlations
in some (not yet specified) weight function for $\rho_a$, denoted
as $W_\Lambda[\rho]$. This is gauge-invariant by assumption, and
depends upon the separation scale $\Lambda^+$ since it is
obtained by integrating out quantum fluctuations with longitudinal
momenta $|p^+|> \Lambda^+$ (cf. Sect. 3).

To conclude, in the MV model, the small-$x$ gluon correlation functions
are obtained in two steps: (i) First, one solves the classical 
Yang-Mills equations (\ref{cleq0}) in the light-cone gauge $A^+=0$.
The solution will be some non-linear functional of $\rho$, 
which we denote as ${\cal A}^i(\vec x)[\rho]$.
(Indeed, as we shall see in Sect. 2.3,
one can always construct a solution which
has $A^-=0$ and is time-independent.)
(ii) The correlation functions of interest are 
evaluated on this classical solution, and then averaged over $\rho$,
with the weight function $W_\Lambda[\rho]$ :
\be\labe{clascorr}
\langle A^i_a(x^+,\vec x)A^j_b(x^+,\vec y)
\cdots\rangle_\Lambda\,=\,
\int {\cal D}\rho\,\,W_\Lambda[\rho]\,{\cal A}_a^i({\vec x})
{\cal A}_b^j({\vec y})\cdots\,.\ee
The normalization here is such as
\be\labe{normF}
\int {\cal D}\rho\,\,W_\Lambda[\rho]\,=\,1.\ee
Note that only equal-time correlators of the transverse
fields can be computed in this way; but these are precisely
the correlators which matter for the calculation
of the gluon distribution function, cf. Sect. 2.4.

Note furthermore that the correlations obtained in this way depend
upon the scale $\Lambda^+$. As we shall argue in
Sect. 3 below, the effective theory in eqs.~(\ref{cleq0})--(\ref{clascorr})
is valid only at soft momenta of order $\Lambda^+$
(that is, at momenta $p^+\simle \Lambda^+$, but not too far
below $\Lambda^+$). Indeed, if one goes to the much softer
scale $b\Lambda^+$ with $b\ll 1$, then there are large radiative 
corrections, of order $\alpha_s\ln(1/b)$, which invalidate the
classical approximation at the scale $\Lambda^+$.
In order to compute correlations at the new scale
$b\Lambda^+$, one must first construct the effective theory
valid at this scale, by integrating out the quantum 
degrees of freedom with longitudinal momenta in the strip 
$b\Lambda^+ <|p^+|<\Lambda^+$. Once this is done, the dependence
upon the intermediate scale $\Lambda^+$ goes away, as it should.
This will be explained in detail in Sects. 3 and 4.

Remarkably, the equations (\ref{cleq0})--(\ref{clascorr})
above are those for a glass (here,
a {\it color} glass): There is a source, 
and the source is averaged over. 
This is entirely analogous to what is done for
spin glasses when one averages over background magnetic fields
\cite{PS79}.  To argue that one also has a condensate, 
one has to compute the correlation function (\ref{clascorr}).  
By using a Gaussian approximation for the weight
function, one has found a saturation regime where the 
classical field has a typical strength of order $1/g$
\cite{JKMW97,KM98,AM1} (see also Sect. 2.4 below).  This is
the maximal occupation number for a classical field, since larger
occupation numbers are blocked by repulsive interactions of the gluon
field.  For weak coupling, this occupation number is large, and the gluons
can be thought of as in some condensate.  We are therefore lead to
conclude that the matter which describes the small $x$ part of a hadron
wavefunction is a Color Glass Condensate.

\subsection{The Abelian case, as a warm up}

Before attempting to solve the Yang-Mills equation (\ref{cleq0}), 
it is instructive to consider its Abelian version in some
detail. This reads:
\be
\partial _{\nu} F^{\nu \mu} =  \delta^{\mu +} \rho({\vec x})
\labe{acleq}
\ee
with $F^{\mu \nu} = \partial^\mu A^\nu-\partial^\nu A^\mu$
and ${\vec x}\equiv (x^-,{\bf x}_{\perp})$.

We are interested in LC-gauge ($A^+=0$)
solutions which vanish as $\rho\to 0$.
Since $\rho$ is static (i.e., independent of $x^+$), 
we can look for solutions which are static as well: 
$\partial^- A^{\mu}=0$.
Then, the components of eq.~(\ref{acleq}) with
$\mu=-$ and $\mu=i$ imply that $A^-=0$
(so that $F^{-+}=F^{i-}=0$), and also $F^{ij}=0$.
Thus, only the transverse fields $A^i({\vec x})$ are
non-vanishing, and 
define a ``pure gauge'' in the two-dimensional 
transverse plane\footnote{Of course, this is not a pure gauge
in four dimensions, since $A^+=0\,\ne\,-\partial^+\omega$.}.
We can thus write [throughout this paper, we shall systematically
use calligraphic letters (like ${\cal A}^i$ and ${\cal F}^{i+}$) to denote 
solutions of the classical equations of motion] :
\be
{\cal A}^i({\vec x}) = -\partial^i\omega({\vec x})
\ee
with  $\omega({\vec x})$ satisfying the following equation
(cf. eq.~(\ref{acleq}) with $\mu=+$ and $F^{i+}=-\del^+A^i$):
\be
-\nabla_{\perp}^2(\partial^+ \omega)=\rho
\labe{alambda}
\ee
Eq.~(\ref{alambda}) can be easily solved in momentum space, to give:
\be
{\cal A}^i(p)=i p^i \omega(p) = -{p^i \over p^+} {\rho(p^+,p_{\perp}) 
\over p_{\perp}^2}\,,
\labe{aaimom}
\ee
where, however, one needs a prescription to
handle the pole at $p^+=0$. For instance, with the following,
retarded prescription (with $\varepsilon\to 0_+$):
\be\label{RETARDED}
 {1 \over p^+} \,\equiv\,{1 \over p^+\ +i \varepsilon}\,,\ee
one obtains a solution which vanishes at $x^-\to -\infty$
(thus, the ``retardation'' property refers here
to $x^-$, and not to time) :
 \be
{\cal A}^i (x^-,x_{\perp})=
\int_{-\infty}^{x^-} dy^- \, \partial^i \alpha (y^-,x_{\perp})\,,
\labe{aiab1}
\ee
with $\alpha ({\vec x})$ satisfying
\be
-\nabla_{\perp}^2 \alpha ({\vec x}) = \rho ({\vec x}).
\label{aal}
\ee
This yields $\alpha({\vec p})=\rho({\vec p})/p_{\perp}^2$
(with $\vec p \equiv (p^+,{\bf p}_\perp)$), or, in coordinate space,
\be\labe{alpha}
\alpha ({\vec x})\,=\,-\int \frac{d^2y_\perp}{2\pi}\,
\ln\Bigl(|{\bf x}_\perp - {\bf y}_\perp|\mu\Bigr)\,
\rho (x^-,{\bf y}_\perp),\ee
where $\mu$ is an infrared cutoff which must disappear from
physical quantities.

In general, different prescriptions for the pole in $1/p^+$
correspond to different boundary conditions in $x^-$, which 
nevertheless describe the same physical situation, since they lead 
to the same electric field ${\cal F}^{i+}=-\partial^+
{\cal A}^i =\partial^i\alpha\,$.
Alternatively, solutions corresponding to different 
such prescriptions are related by (residual) gauge transformations.

Note finally that if the external source is localized around $x^-=0$
(cf. eq.~(\ref{jclas})), so is
$\alpha ({\vec x})$, and therefore the associated electric
field ${\cal F}^{i+}({\vec x})=\partial^i \alpha ({\vec x})$.

\subsection{The non-Abelian case}

We now return to the non-Abelian problem, and note first that,
written as it stands, eq.~(\ref{cleq0}) is not really consistent:
The property $[D_\mu, [D_{\nu}, F^{\nu \mu}]]=0$ requires
the color current $J^\mu$ in the Yang-Mills equations
to be {\it covariantly} conserved, $[D_\mu, J^\mu]=0$, which for a
current $J^\mu=\delta^{\mu +}J^+$ amounts to:
\be\labe{COVCON1} 
[D^-, J^+]\,\equiv\,\del^- J^+\,-\,ig[A^-,J^+]\,=\,0\,.\ee
In general, however, this is not satisfied by the static current
(\ref{jclas}) (since $\del^- \rho=0$, but the commutator $[A^-,\rho]$
can be non-zero). Rather, eq.~(\ref{COVCON1}) 
shows that, for a non-zero field $A^-_a$, the current $J^+_a$ 
can be static only up to a color precession. Namely,
if we identify $\rho_a$ in eq.~(\ref{jclas})
with the current $J^+_a$ at some particular time $x^+_0$,
\be\labe{INITJ+}
J^+(x)\,=\,\rho(\vec x)\qquad{\rm at}\quad
x^+=x^+_0\,,\ee
then eq.~(\ref{COVCON1}) can be easily integrated out
to give the current at some later time $x^+>x^+_0\,$:
\be\labe{JTIME}
J^+(x^+, \vec x)\,=\,W(x^+,\vec x)\,\rho(\vec x)
\,W^\dagger(x^+,\vec x).\ee
We have introduced here the temporal Wilson line:
\be\labe{WLINE1}
     W[A^-](x^+,\vec x)\,\equiv\,{\rm T}\, \exp\left\{\,
ig\int_{x^+_0}^{x^+} dz^+ A^-(z^+, \vec x) \right\},
\ee
with T denoting the time ordering of the color matrices
in the exponential (that is, the matrix fields $A^-(x)
\equiv A^-_a(x) T^a$ are ordered from right to left
in increasing sequence of their $x^+$ arguments).

The ensuing equations of motion for the background field:
\be
[D_{\nu}, F^{\nu \mu}](x)\, =\,  \delta^{\mu +}
W(x)\,\rho(\vec x)
\,W^\dagger(x),
\labe{cleq1}
\ee
are generally complicated by the non-locality of the current
in time. As in the Abelian case, however, it is consistent
to look for solutions to eq.~(\ref{cleq1}) which are 
static\footnote{Of course, time-dependent solutions with
non-zero $A^-_a$ can be also constructed, e.g., 
via $x^+$-dependent gauge transformations of the static
solution in eq.~(\ref{Ansatz}). For what follows,
it is simply more convenient to use the residual gauge
freedom in order to choose the solution in the form
(\ref{Ansatz}).} and satisfy $A^-=0$ (in addition to the
LC gauge condition $A^+=0$):
\be\labe{Ansatz}
A^+\,=\,A^-\,=\,0,\qquad A^i\,\equiv\,{\cal A}^i(x^-,x_{\perp})\,.\ee
This Ansatz is preserved by gauge transformations which are
both $x^-$-- and $x^+$--independent, that is, by two-dimensional
gauge transformations in the transverse plane.
With this Ansatz, eq.~(\ref{cleq1}) reduces back to 
eq.~(\ref{cleq0}), which explains the emphasis put on 
this equation in Sect. 2.1. In particular, for $\mu=+$,
\be
[D_i, F^{i+}] \,=\, \rho({\vec x}),
\labe{n+}
\ee
while the $\mu=i$ component, $[D_j, F^{ji}] = 0$,
implies that ${\cal A}^i$ is a pure gauge
in the transverse plane (${\cal F}^{ji} = 0$). We thus write:
\be
{\cal A}^i(x^-,x_{\perp})\,=\,{i \over g}\,
U(x^-,x_{\perp})\,\partial^i U^{\dagger}(x^-,x_{\perp})\,,
\labe{tpg} 
\ee
with the $SU(N)$ group element $U(x^-,x_{\perp})$
implicitly related to $\rho$ by eq.~(\ref{n+}).

To make progress, it is useful to
observe that the fields ${\cal A}^i$ in eq.~(\ref{tpg}) can be gauge
transformed to zero by the gauge transformation with matrix
 $U^{\dagger} ({\vec x})$:
\be\labe{gtr}
A^\mu\longrightarrow {\tilde A}^\mu=U^{\dagger}A^\mu U+
\,{i \over g}\, U^{\dagger} \partial^\mu U.
\ee
This yields indeed ${\tilde A}^i={\tilde A}^- = 0$, together with
\be
{\tilde A}^+ = {i \over g}\, U^{\dagger} \left( \partial^+ U \right ).
\labe{ta+}
\ee
Thus, there exists a gauge where the non-Abelian
field has just one non-trivial component, 
${\tilde A}^\mu=\delta^{\mu +}{\tilde A}^+$.
This gauge will play an important role in what
follows. It will be referred to as the {\it covariant gauge},
since $\partial_\mu {\tilde A}^\mu =0$ for the fields in
eqs.~(\ref{gtr})--(\ref{ta+}).
In this gauge, the general Yang-Mills equations (\ref{cleq1})
reduce to a single {\it linear} equation,
\be\labe{EQTA0}
- \nabla^2_\perp {\tilde A}^+(\vec x)\,=\,{\tilde \rho}(\vec x)\,,
\ee
where
\be\label{COVRHO}
{\tilde \rho}(\vec x)\,\equiv\,U^{\dagger}(\vec x)
\, \rho(\vec x) \,U(\vec x)\ee
is the classical color source in the covariant gauge.
(We preserve the notations without the tilde --- e.g., $\rho$ ---
for quantities in the LC gauge.)

Eq.~(\ref{EQTA0}) is formally the same as eq.~(\ref{aal}) 
for $\alpha$ in QED, and we prefer to rename 
${\tilde A}^+({\vec x})\equiv\alpha({\vec x})$ in what follows.
That is, $\alpha$ satisfies
\be\labe{EQTA}
- \nabla^2_\perp \alpha({\vec x})\,=\,{\tilde \rho}(\vec x)\,,
\ee
and is a linear functional of the {\it covariant gauge}
source ${\tilde \rho}$.
Thus, the classical solution in the covariant gauge is trivial
to obtain. However, physical quantities 
like the gluon distribution function rather involve the 
correlators of the fields $A^i$ in the {\it light-cone}
gauge\footnote{Of course, physical quantities
are gauge invariant, and could be computed in any gauge;
in general gauges, however, they are not simply related to the
Green's functions of the corresponding vector potentials;
see also Sect. 2.4.} (cf. eq.~(\ref{clascorr}) and Sect. 2.4).
Still, the fact that the latter can be expressed as
gauge rotations of the quasi-Abelian field
${\tilde A}^\mu=\delta^{\mu +}\alpha$
will greatly simplify the calculations to follow.
Specifically, eq.~(\ref{ta+}) can be inverted to give (with 
${\tilde A}^+\equiv\alpha$)
\be
U^{\dagger}(x^-,x_{\perp})=
 {\rm P} \exp
 \left \{
ig \int_{x_0^-}^{x^-} dz^-\,\alpha (z^-,x_{\perp})
 \right \},\labe{UTA}
\ee
where the symbol P orders the color matrices $\alpha(\vec x)$ from
right to left, in increasing or  decreasing order of their $x^-$
arguments depending respectively on whether $x^->x_0^-$ or $x^-<x_0^-\,$.
The initial point $x^-_0$ is as yet arbitrary, and will
be specified in a moment. (Different choices for $x^-_0$ lead
to LC-gauge solutions ${\cal A}^i$ which are connected by 
residual, two-dimensional, gauge transformations.)

Together, eqs.~(\ref{tpg}), (\ref{EQTA}) and (\ref{UTA}) provide 
an explicit expression for the  LC-gauge solution ${\cal A}^i$ 
as a non-linear functional of 
${\tilde \rho}\,$, to be denoted as ${\cal A}^i [\tilde \rho]$. 
If one rather tries to express ${\cal A}^i$ in terms of 
$\rho$, then the corresponding functional ${\cal A}^i [\rho]$
 would be more complicated and not explicitly known.
Indeed, $\alpha$ is related to $\rho$ via the following
non-linear equation (cf. eqs.~(\ref{COVRHO})--(\ref{UTA})):
\be
U\Bigl(-\nabla^2_{\perp} \alpha 
\Bigr) U^{\dagger}  \,=\, {\rho}\,,
\labe{nal}
\ee
for which we have no explicit solution.

For what follows, it will be essential
to know the solution ${\cal A}^i$ {explicitly}.
We thus propose
to use the covariant gauge source ${\tilde \rho}$, rather
than the LC-gauge source $\rho$, as the functional variable
that the fields depend upon. This is possible because the measure 
and the weight function in eq.~(\ref{clascorr})
are gauge invariant, so that the final average over the 
classical source can be equally well expressed as a
functional integral over ${\tilde \rho}$.
In other terms, from now on we shall replace eq.~(\ref{clascorr})
with
\be\labe{COVclascorr}
\langle A^i(x^+,\vec x)A^j(x^+,\vec y)
\cdots\rangle_\Lambda\,=\,
\int {\cal D}\tilde\rho\,\,W_\Lambda[\tilde\rho]\,{\cal A}^i_{\vec x}
[\tilde \rho]\,{\cal A}^j_{\vec y}[\tilde \rho]\cdots\,.
\ee

Up to this point, the spatial distribution of the source in $x^-$ 
has been completely arbitrary: the equations above hold for any function
$\rho(x^-)$. For what follows, however, it is useful to recall
that $\rho$ has the localized structure shown in eq.~(\ref{jclas}).
As mentioned after this equation, and it will be demonstrated by the
quantum analysis to follow, the
precise longitudinal structure of $\rho$ is sensitive to the
conditions used to completely fix the gauge, in both 
the classical and quantum calculations. 

To fix the {\it  classical} gauge, i.e., the gauge for the 
classical solution (\ref{tpg}), we shall
adopt retarded boundary conditions in $x^-\,$:
${\cal A}^i(\tilde x)\to 0$ for $x^- \to -\infty\,$. That is,
we choose $x^-_0\to -\infty$ in eq.~(\ref{UTA}).
This amounts to a complete gauge fixing since 
this boundary condition would be violated by any 
$x^-$--independent gauge transformation.
Remarkably, we shall see later that
this classical gauge condition also fixes the
gauge prescription to be used in the {\it quantum} calculations:
indeed, for the above boundary condition 
to be consistent with the quantum evolution, one must adopt a retarded
$i\epsilon$ prescription for the axial pole in the gluon
propagator (see Sect. 6.3).
With these gauge-fixing prescriptions, the ensuing classical
source has support only at positive $x^-$, 
with $0\simle x^-\simle 1/\Lambda^+$ (cf. Sect. 5.1 and Paper II).

In what follows, $\Lambda^+$ will be most often the {\it large}
longitudinal momentum scale in the problem: both the external
current in DIS, and the quantum fluctuations to be considered in Sect. 3,
will have momenta $p^+\ll \Lambda^+$. Because of their poor
longitudinal resolution, such fields are
sensitive only to the gross features of the background fields 
${\cal A}^i$ over large distances $|x^-| \gg 1/\Lambda^+$, where
we can write:
\be\labe{UTAF}
U^{\dagger}(x^-,x_{\perp})\,\equiv\,
 {\rm P} \exp
 \left \{
ig \int_{-\infty}^{x^-} dz^-\,\alpha (z^-,x_{\perp})
 \right \}\approx \,\theta(x^-)\,V^\dagger(x_{\perp}) + \theta(-x^-),
\ee
with 
\be\labe{v}
V^\dagger(x_{\perp})\,\equiv\,{\rm P} \exp
 \left \{
ig \int_{-\infty}^{\infty} dz^-\,\alpha (z^-,x_{\perp})
 \right \}.\ee
Together with eq.~(\ref{tpg}), this gives a background field of the
form:
\be\labe{APM}
{\cal A}^i(x^-,x_\perp)\,\approx\,\theta(x^-)\,
\frac{i}{g}\,V(\del^i V^\dagger)
\,\equiv\,\theta(x^-){\cal A}^i_\infty(x_\perp).\ee
The associated electric field strength is then
effectively a $\delta$--function:
\be\labe{FDELTA}
{\cal F}^{i+}(\vec x) \,\equiv\,
-\partial^+{\cal A}^i\,\approx\,-\delta(x^-)\,
{\cal A}^{i}_\infty(x_\perp).\ee
It is implicitly understood here that the various 
$\delta$-- and $\theta$--functions are regularized over a distance
$\Delta x^-\sim 1/\Lambda^+$.

\subsection{Gluon distribution function and saturation}

According to eq.~(\ref{COVclascorr}), there are two ingredients
needed in order to compute soft gluon correlations in the MV model:
the solution ${\cal A}^i_a[\tilde \rho]$ to the
classical equations of motion and the weight function 
$W_\Lambda[\tilde\rho]$. The classical solution has just
been constructed, and is given by eqs.~(\ref{tpg}), 
(\ref{EQTA}) and (\ref{UTA}) 
(with $x^-_0\to -\infty$). The construction of the weight
function from quantum fluctuations will be the main objective of the
remaining part of this paper together with Paper II.
But before turning to the quantum dynamics, let us briefly
recall the results of some previous calculations in the MV model
 \cite{JKMW97,KM98}, which use a Gaussian weight function 
and show saturation explicitly. This will also give us the 
opportunity to discuss the gauge-invariant definition
of the gluon distribution function.

We first recall the usual definition of the
gluon distribution function, which is written 
in the LC gauge \cite{AM1} :
\be\labe{GDF0}
G(x,Q^2)&\equiv&
\int {d^2k_\perp \over (2 \pi)^2}\,\Theta(Q^2-
k_\perp^2)\int{dk^+ \over 2 \pi}
\,2k^+\,\delta\biggl(x-{k^+\over P^+}\biggr)\nonumber\\
&{}&\qquad\qquad
\Bigl\langle A^i_a(x^+,k^+,{\bf k}_\perp)
A^i_a(x^+,-k^+,-{\bf k}_\perp)\Bigr\rangle,\,\,\,\ee
with the brackets denoting an average over the hadron wavefunction.
Eq.~(\ref{GDF0}) can be understood as follows \cite{AM1}: In light-cone
quantization, and with $\vec k \equiv (k^+,{\bf k}_\perp)$,
\be
\frac{2k^+}{(2 \pi)^3}\, A^i_c(x^+,\vec k)
A^i_c(x^+,-\vec k)\,=\,\sum_\lambda\sum_c 
a^\dagger_{\lambda c}(\vec k)\,a_{\lambda c}(\vec k)\,=\,
\frac{dN}{d^3 k}\,\ee
is the Fock space gluon distribution function, i.e., the
number of gluons per unit of volume in momentum space.
(Here, $a^\dagger_{\lambda c}(\vec k)$ and $a_{\lambda c}(\vec k)$
are, respectively, creation and annihilation operators for gluons
with momentum $\vec k$, color $c$ and transverse polarization
$\lambda$, and we use the normalization in Ref. \cite{AM1}.)
Thus, eq.~(\ref{GDF0}) simply counts all the gluons in the
hadron wavefunction having longitudinal momentum 
$k^+=xP^+$, and transverse momentum up to $Q$.

As emphasized above, eq.~(\ref{GDF0}) is meaningful only in the LC gauge
$A^+_a=0$; indeed, it is only in this gauge that the matrix element
in the r.h.s. has a gauge-invariant meaning, as we discuss now:
Note first that, in this gauge, the electric field $F^{i+}_a$ and
the vector potential $A^i_a$ are linearly related, 
$F^{i+}_a(k)=ik^+A^i_a(k)$,
so that eq.~(\ref{GDF0}) can be also written as a two-point
Green's function of the (gauge-covariant) electric fields.
After performing the integral over $k^+$, one obtains
(with $k^+=xP^+$ from now on) :
\be\labe{GDF}
xG(x,Q^2)\,=\,\frac{1}{\pi}\int {d^2k_\perp \over (2 \pi)^2}\,\Theta(Q^2-
k_\perp^2)\,\Bigl\langle F^{i+}_a(x^+,\vec k)
F^{i+}_a(x^+,-\vec k)\Bigr\rangle.\ee
This does not look gauge invariant as yet: in coordinate space
($\vec x\cdot \vec k = x^-k^+-{\bf x}_\perp\cdot{\bf k}_\perp$),
\be\labe{FF}
F^{i+}_a(x^+,\vec k)F^{i+}_a(x^+,-\vec k)\,=\,
\int d^3x \int d^3y\,\,{\rm e}^{i(\vec x- \vec y)\cdot \vec k}\,
F^{i+}_a(x^+,\vec x)F^{i+}_a(x^+,\vec y)\ee
involves the electric fields at different spatial points
$\vec x$ and $\vec y$. A manifestly gauge invariant operator
can be constructed by inserting Wilson lines:
\be\labe{GIFF}
{\rm Tr}\,\left\{F^{i+}(\vec x)\,U_\gamma(\vec x,\vec y)\,
F^{i+}(\vec y)\,U_\gamma(\vec y,\vec x)\right\},\ee
where (with $\vec A\equiv (A^+,{\bf A}_\perp)$)
\be\labe{UGEN}
U_\gamma(\vec x,\vec y)\,=\,{\rm P}\,{\rm exp}\left\{ig\int_\gamma d\vec z
\cdot \vec A(\vec z)\right\},\ee
and the temporal coordinates $x^+$ are omitted
(they are the same for all the fields). In eq.~(\ref{UGEN}),
$\gamma$ is an arbitrary oriented path going from $\vec y$
to $\vec x$. For any such a path, eq.~(\ref{GIFF}) defines
a gauge-invariant operator. 

In particular, let us choose a path 
made of the following three elements: two ``vertical'' pieces
going along the $x^-$ axis from $(y^-,y_\perp)$ to $(-\infty,y_\perp)$,
and, respectively, from $(-\infty,x_\perp)$ to $(x^-,x_\perp)$,
and an ``horizontal'' piece from $(-\infty,y_\perp)$ to $(-\infty,x_\perp)$.
Along the vertical pieces, $d\vec z \cdot \vec A =
dz^- A^+$, so these pieces do not matter in the LC gauge. 
Along the horizontal piece $d\vec z \cdot \vec A = d{\bf z}_\perp 
\cdot {\bf A}_{\perp}(-\infty, z_\perp)$, and the path $\gamma$
is still arbitrary, since this can be any path 
joining $y_\perp$ to $x_\perp$ in the 
transverse plane at $x^-\to -\infty$. This arbitrariness disappears,
however, for the classical solution constructed in Sect. 2.3: 
this is a two-dimensional pure gauge (cf. eq.~(\ref{tpg})),
so the associated Wilson lines in the transverse plane
are path-independent. Moreover, these Wilson lines become even 
trivial, $U_\gamma(x_\perp, y_\perp)_{ x^-,y^-\to-\infty}=1$,
once we choose the retarded prescription
$x^-_0\to -\infty$ in eq.~(\ref{UTA}).

Thus, within  the classical approximation,
for the particular class of paths mentioned above, 
and with retarded boundary conditions
for the classical solution,
the manifestly gauge-invariant operator product 
in eq.~(\ref{GIFF}) coincides with the simpler operator
which enters the standard definition of the gluon distribution, 
in eqs.~(\ref{GDF})--(\ref{FF}). 
Converserly, the latter can be given a gauge-invariant meaning 
under the conditions mentioned above\footnote{Note also that,
if we let $Q^2\to \infty$ in eq.~(\ref{GDF}), then the corresponding
limit of the gluon distribution is gauge invariant in full generality, 
since the unrestricted integral over $k_\perp$ sets $x_\perp=y_\perp$.}.

In the remaining part of this section, 
we shall be concerned only with this classical approximation,
which is expected to work better at very small $x$, in particular,
in the saturation regime. Then we can replace, in eq.~(\ref{GDF}),
$F^{i+}_a\to {\cal F}^{i+}_a$,
with ${\cal F}^{i+}_a$ the time-independent classical electric field 
constructed in Sect. 2.3 (a functional of $\tilde \rho$):
\be\labe{GCL}
x G_{cl}(x,Q^2)\,=\,\frac{1}{\pi}
\int {d^2k_\perp \over (2 \pi)^2}\,\Theta(Q^2-
k_\perp^2)\,\Bigl\langle\,
|{\cal F}^{i+}_a(\vec k)|^2\Bigr\rangle_\Lambda.\,\ee
Here and from now on, the average is to be understood 
as an average over $\tilde\rho$, in the sense of 
eq.~(\ref{COVclascorr}) where the scale $\Lambda^+$
is now chosen as $\Lambda^+\sim xP^+$
(cf. the discussion after eqs.~(\ref{clascorr})--(\ref{normF})).
With eq.~(\ref{FDELTA}) for ${\cal F}^{i+}$, eq.~(\ref{GCL})
reduces to:
\be\labe{GCL1}
x G_{cl}(x,Q^2)&=&\frac{1}{\pi}
\int {d^2k_\perp \over (2 \pi)^2}\,\Theta(Q^2-
k_\perp^2)\,\Bigl\langle\,
|{\cal F}^{i+}_a(k_\perp)|^2\Bigr\rangle_\Lambda\nn
&=&{R^2}
\int^{Q^2} {d^2k_\perp \over (2 \pi)^2}
\int d^2x_\perp\,{\rm e}^{-ik_\perp\cdot x_\perp}
\Bigl\langle  {\cal A}^{ia}_\infty(0)\,
 {\cal A}^{ia}_\infty(x_\perp)\Bigr\rangle_\Lambda,\ee
where $R$ is the hadron radius, and we have assumed
homogeneity in the transverse plane, for simplicity.

Note that, if there were not for the $x$--dependence
hidden in the weight function for $\tilde\rho$ (that is,
$W_\Lambda[\tilde\rho]$ with $\Lambda^+\sim xP^+$),
the r.h.s. of eq.~(\ref{GCL1}) would be independent of
$x$, and so would be also the quantity $x G_{cl}(x,Q^2)$, 
in agreement with leading-order calculations
in light-cone perturbation theory \cite{AM1}. 
Thus, in the MV model, the actual $x$--dependence of the gluon 
distribution function is encoded in the weight function,
and is a consequence of the quantum evolution (which
makes $W[\tilde\rho]$ a function of $\Lambda^+$; cf. Sect. 3).

To make progress, a model for the weight function
$W_\Lambda[\tilde\rho]$ is required. As argued in Refs.
\cite{JKMW97}, a simple approximation which should
be valid for transverse wavelengths much smaller
than the size of the hadron, is to write 
\be\label{FCLAS}
W_\Lambda[\tilde\rho]\simeq \exp\left\{-
\frac{1}{2}\int d^3 x \,\frac{\tilde\rho_a^2(\vec x)}
{\xi^2_\Lambda(\vec x)}\right\}\,,\ee
where $\xi^2_\Lambda$ is proportional to the total color charge 
density squared of the partons with $p^+>\Lambda^+$.
For instance, in a simple valence quark model,
\be
\xi^2_\Lambda(\vec x)\,\equiv\,g^2\,\frac{n(\vec x)}{2N_c}\,,\ee
where $n(x^-,x_\perp)$ is the quark number density in the hadron, 
normalized such as (for a nucleon)
\be\label{VQ}
\int dx^- \int d^2x_\perp\,n(x^-,x_\perp)\,=\,N_c.\ee

With this approximation for $W_\Lambda[\tilde\rho]$, let us first
compute the gluon distribution function (\ref{GCL}) in the
{\it linear response approximation}, that is, for a source $\rho$
which is weak enough so that one can linearize the results in Sect. 2.3
(this is the case as long as $x$ is not too small). 
One thus obtains 
${\cal F}^{+j}_a \simeq i(k^j/k^2_\perp)\rho_a\,$,
cf. eq.~(\ref{aaimom}), and therefore:
\be\labe{GCLLIN}
x G_{cl}(x,Q^2)\,\simeq\,\frac{1}{\pi}
\int {d^2k_\perp \over (2 \pi)^2}\,
\frac{\Theta(Q^2-k_\perp^2)}{k^2_\perp}\,\Bigl
\langle |\,\rho_a(\vec k)|^2\Bigr\rangle_\Lambda.\,\ee
(Note that there is no difference between
$\rho$ and $\tilde\rho$ in this linear approximation.)
The average over $\rho$ is now easily 
performed by using eqs.~(\ref{FCLAS})--(\ref{VQ}),
which give (with $\alpha_s=g^2/4\pi$ and $C_F=(N^2_c-1)/2N_c$) :
\be
x G_{cl}(x,Q^2)\,\simeq\,\frac{\alpha_s N_c\, C_F}{\pi}\,
\ln{Q^2\over \mu^2}\,,\ee
where $\mu$ is an IR cutoff as in eq.~(\ref{alpha}).
This result is as expected from LC perturbation theory
\cite{AM1}. 

But in the present formalism, one can actually do better: 
with the Gaussian weight function
(\ref{FCLAS}) and the non-linear classical solution
in Sect. 2.3, one has been able to evaluate the 
fully non-linear expectation value in eq.~(\ref{GCL1}), with the
following result \cite{JKMW97,KM98}:
\be\labe{GCLNON}
\Bigl\langle  {\cal A}^{ia}_\infty(0)\,
 {\cal A}^{ia}_\infty(x_\perp)\Bigr\rangle
\,=\,\frac{N_c^2-1}{\pi\alpha_s N_c}\,\frac{1-{\rm e}^{-x_\perp^2
\ln(x_\perp^2 \Lambda_{QCD}^2) Q_s^2/4}}
{x_\perp^2}\,,\ee
where $Q_s\propto \alpha_s\xi_\Lambda$ is the {\it saturation momentum},
and is expected to increase when $\Lambda^+$ (or $x$) decreases.
(The above equation is valid only for $x_\perp \ll 1/\Lambda_{QCD}$.  The
logarithmic dependence upon $\Lambda_{QCD}$ comes from properly regulating
the infrared transverse momenta, and is associated with the over all color
neutrality on scale sizes of order $1/\Lambda_{QCD}$ \cite{LM00}.)

The remarkable feature about eq.~(\ref{GCLNON}) is that it displays
saturation via non-linear effects. Namely,
the LC gauge potential never becomes larger than
\be \labe{SAT}
\sqrt{x_\perp^2}\,{\cal A}^{i}\,\sim\,\frac{1}
{\sqrt{\alpha_s N_c}}\,.\ee
As anticipated in the Introduction, gluon saturation
requires fields as strong as $A^i\sim 1/g$, which supports
the physical picture of a condensate. Together with eq.~(\ref{cleq0}), 
this implies that $\rho\sim 1/g$ at saturation.

This interpretation can be made sharper by going to momentum space.
As obvious from eq.~(\ref{GCL1}),
\be\label{WIGG}
N(k_\perp)\,\equiv\,\frac{d^2(x G_{cl})}{d^2k_\perp\,d^2b_\perp}\,\equiv\,
\int d^2x_\perp\,{\rm e}^{-ik_\perp\cdot x_\perp}
\Bigl\langle  {\cal A}^{ia}_\infty(0)\,
 {\cal A}^{ia}_\infty(x_\perp)\Bigr\rangle,\ee
is the gluon density per unit of $x$ and per unit
of transverse phase-space. By using this and eq.~(\ref{GCLNON}),
one obtains \cite{JKMW97}
\be N(k_\perp) \,\propto\, \alpha_s (Q_s^2/k_\perp^2) \quad{\rm for}
\quad k_\perp^2\gg Q_s^2,\ee
which is the normal perturbative behavior, but
\be N(k_\perp) \,\propto\, {1\over \alpha_s}\,\ln\,
\frac{k_\perp^2}{Q_s^2}\quad{\rm for}\quad k_\perp^2\ll Q_s^2,\ee
which shows a much slower increase, i.e., saturation, at
low momenta (with $k_\perp\gg \Lambda_{QCD}$ though).

Note, however, that the above argument is not rigorous, since
the local Gaussian form for $W$ in eq.~(\ref{FCLAS}) is only valid at
sufficiently large transverse momentum scales so that the effects of
high gluon density are small. It is therefore important to verify
if saturation comes up similarly with a more realistic form for
the weight function, as obtained after including quantum 
evolution towards small $x$. This would also determine the
$x$--dependence of the saturation scale.
We now turn towards this quantum analysis.

\setcounter{equation}{0}
\section{The non-linear evolution equation}

With this section, we begin the study of the quantum dynamics
which allows one to actually derive the MV model as an effective
theory valid for some range of longitudinal momenta.
The central element in this construction is a functional
renormalization group equation (RGE) which describes the flow of
the weight function $W_\Lambda[\tilde\rho]$
with the separation scale $\Lambda^+$.
By means of this equation, the soft correlations induced by the
quantum fields with momenta $p^+>\Lambda^+$ are transferred into
classical correlations of the random source $\rho$.
Thus, this equation governs the quantum evolution of the effective theory.

In Refs. \cite{JKLW97,JKLW99a}, this RGE has been obtained
by studying the correlators
$\langle \delta\rho\, \delta\rho\dots\rangle$ of the
color charge $\delta\rho$ of the quantum fluctuations.
However, physical quantities like the gluon distribution
function are more directly related to {\it field} correlators, 
like $\langle A^i A^j\dots\rangle$ (see, e.g., eq.~(\ref{GDF0})).
Given the different ways that the field correlators develop
out of the charge correlators in the classical and 
quantum settings, it is not a priori obvious that matching
charge correlations should be enough to insure the equivalence
between the two theories.
Our approach here will be therefore different: 
the RGE will be derived by matching directly field correlators,
to the accuracy of interest. This will lead us indeed to the
JKLW evolution equation for $W_\Lambda[\rho]$ when the source
$\rho$ is in the {\it light-cone} gauge.
In addition, we shall also establish the evolution
equation for $W_\Lambda[\tilde\rho]$ where the source 
$\tilde\rho$ is in the {\it covariant} gauge.
According to the discussion in Sect. 2.3, it is this latter
representation which is more suitable for explicit calculations.

\subsection{A quantum extension of the MV model}

To study the quantum evolution of the MV model,
one needs first a quantum extension of this theory.
By definition, such an extension should allow for {\it soft}
($|p^+|<\Lambda^+$) quantum gluons in addition to the
classical color fields ${\cal A}^i$ generated by $\rho$.
Only { soft} fluctuations are permitted since,
by assumption, the fast ($|p^+|>\Lambda^+$) degrees of freedom
have been already integrated out in the construction of the 
effective theory at the scale $\Lambda^+$.

A generic quantum extension of the MV model is defined
by the following generating functional for soft gluon
correlation functions:
\be\labe{PART}
{\cal Z}[j]\,=\,\int {\cal D}\rho\,\,W_\Lambda[\rho]
\,\,Z_{\Lambda}^{-1}[\rho]\int^\Lambda {\cal D}A_a^\mu\,
\delta(A^+_a)\,\,{\rm e}^{\,iS[A,\,\rho]-i\int j\cdot A}\,.\ee
This is written fully in the LC gauge (in particular, $\rho$
is the LC-gauge color source), and involves
two functional integrals: a quantum 
path integral over the soft gluon fields $A^\mu\,$:
\be\labe{PARTQUANT}
Z_{\Lambda}[\rho,j]\,\equiv\,
Z_{\Lambda}^{-1}[\rho]\int^\Lambda {\cal D}A_a^\mu\,
\delta(A^+_a)\,\,{\rm e}^{\,iS[A,\,\rho]-i\int j\cdot A}\,\ee
(with $Z_{\Lambda}[\rho]\equiv Z_{\Lambda}[\rho,j=0]$),
which defines
quantum correlations at fixed $\rho$,  and a classical average 
over $\rho$ with weight function $W_\Lambda[\rho]$.
The upper script ``$\Lambda$'' on the quantum path integral
is to recall the
restriction to soft ($|p^+| < \Lambda^+$) longitudinal momenta. 
Note that the separation between fast and soft degrees of freedom 
according to their longitudinal momenta has a gauge-invariant
meaning (within the LC-gauge) since the residual gauge 
transformations, being independent of  $x^-$,
cannot change the $p^+$ momenta.

All non-trivial information about the quantum dynamics is contained
in the action $S[A,\rho]$. This is chosen such
as to reproduce the classical equations of motion (\ref{cleq1})
in the saddle point approximation $\delta S/\delta A^\mu=0$
(so that, at tree-level, the quantum theory (\ref{PART}) 
reduce to the classical MV model, as it should).
A gauge-invariant action satisfying this property has been
proposed in Refs. \cite{JKLW97,JKLW99a,JKW99}
(see also Ref. \cite{JJV00}), and will be
presented in Sect. 4 below.
As we shall argue there, this original
proposal requires some further refinement (in the form of a
complex-time formulation), but this is not essential for the
general discussion in the present section. In fact, for what
follows, it is enough to know that $S$ has two pieces,
$S[A,\rho]=S_{YM}[A]+S_W[A^-,\rho]$, with $S_{YM}$
the usual Yang-Mills action, and $S_W$ a gauge-invariant
generalization of the eikonal vertex $\int d^4x \,\rho_a A^-_a$.
(See Sect. 4 for more details.)

At tree-level, $A^\mu_a\approx {\cal A}^\mu_a =\delta^{\mu i}
{\cal A}^i_a$, with ${\cal A}^i_a$ the classical field generated by
 $\rho_a$, cf. Sect. 2.3. In general, the total gluon field
in eqs.~(\ref{PART})--(\ref{PARTQUANT})  can be decomposed
into tree-level field plus quantum fluctuations:
\be\labe{fluct0}
 A^\mu_a(x)\,=\,{\cal A}^\mu_a(x) + \delta A^\mu_a(x).\ee
The average field in the system is not just ${\cal A}^\mu$,
but involves also an {\it induced} piece
$\langle \delta A^\mu\rangle$ coming from the
polarization of the quantum gluons by the external source :
\be
\langle A^\mu_a (x)\rangle\,=\,
{\cal A}^\mu_a(x) + \langle \delta A^\mu_a(x)\rangle\,\equiv\,
{\cal A}^\mu_a(x) + \delta {\cal A}^\mu_a(x).\ee
This average field satisfies the following equation
\be\labe{MFE}
\left\langle \frac{\delta S}{\delta A^\mu_a(x)}
\right\rangle\,=\,0,\ee
where the brackets refer to the average over quantum
fluctuations at fixed $\rho$. (Unless otherwise specified, we shall 
always use this convention in what follows.) 

From eq.~(\ref{PART}), the (time-ordered)
two-point function of the soft gluons is obtained as: 
\be\labe{2point} 
<<
{\rm T}\,A^\mu(x)A^\nu(y)>>
\,\,=\,\int {\cal D}\rho\,\,W_\Lambda[\rho]
\left\{\frac{\int^\Lambda {\cal D}A
\,\,A^\mu(x)A^\nu(y)\,\,{\rm e}^{\,iS[A,\,\rho]}}
{\int^\Lambda {\cal D}A\,\,{\rm e}^{\,iS[A,\,\rho]}}
\right\},\ee
where we use double brackets to denote the average
over both quantum fluctuations and the random source.
This double average should be contrasted with
\be
\frac{\int {\cal D}\rho\,\,W_\Lambda[\rho]\,
\int^\Lambda {\cal D}A\,
\,A^\mu(x)A^\nu(y)\,\,{\rm e}^{\,iS[A,\,\rho]}}
{\int {\cal D}\rho\,\,W_\Lambda[\rho]\,
\int^\Lambda {\cal D}A\,\,{\rm e}^{\,iS[A,\,\rho]}}\,,\ee
which would be the correct average
if $\rho$ and $A^\mu$ were two sets of quantum fields which
are coupled by the dynamics (e.g., fast and soft gluon fields). 
In reality, $\rho_a(\vec x)$
is a classical variable which has no dynamics by itself, and
represents only in an average way the effects of the fast partons 
on the dynamics of the soft  gluons.
Then the appropriate averaging is that in eq.~(\ref{2point}),
where one first performs a quantum average at fixed $\rho$,
and then a classical statistical average over $\rho$.
This reflects the physical fact that the changes of $\rho$ 
happen on a time scale which is much larger than the time
scale characterizing the dynamics of the soft gluons.
This is very similar to the average over disorder performed
in the context of amorphous materials (like spin glasses)
\cite{PS79}, and supports the physical picture of the 
saturation regime as a Color Glass Condensate.

Since the intermediate scale $\Lambda$ is arbitrary, 
it must cancel out in any complete calculation of soft correlations. 
That is, the cutoff dependence of the quantum loops must cancel 
against the corresponding dependence of the classical weight function
$W_\Lambda[\rho]$.
This constraint can be formulated as a renormalization group 
equation for $W_\Lambda[\rho]$ \cite{JKLW99a,JKW99}, 
to be constructed in the next subsection.

\subsection{From quantum to classical correlations}

In writing down the quantum effective theory in eq.~(\ref{PART}),
we have assumed that the effects of the fast gluons can be
reproduced --- at least to some accuracy which, as yet, has 
not been specified --- by a classical color source $\rho$
with a peculiar structure (time-independent, and localized
near the light-cone), and some (undetermined) weight function 
$W_\Lambda[\rho]$.
In this subsection, we show how to construct this effective 
theory, step by step, by integrating quantum fluctuations 
in successive layers of $p^+$ (or $p^-$).

As usual with the BFKL evolution \cite{BFKL}, 
the most important quantum corrections at small $x$
are those which are
enhanced by large intervals of rapidity $\Delta \tau=\ln(1/x)
\gg 1$. As we shall see in the explicit calculations in Sect.
5 and Paper II,
the amplitudes with soft ($k^+ < \Lambda^+$) external lines receive 
quantum corrections of order
$ \alpha_s\ln(\Lambda^+/k^+)$ from integration 
over the fluctuations with $p^+$ momenta in the strip
$k^+\ll p^+\ll\Lambda^+$. ($\Lambda^+$ is the upper cutoff
on the longitudinal momenta of the quantum fluctuations,
cf. eq.~(\ref{PART}).) Clearly, this is a large correction as 
long as $\Lambda^+\gg k^+$.
Converserly, if $\Lambda^+$ is small enough (close to 
$k^+$), the quantum effects are relatively small
(ordinary perturbative corrections of 
order $\alpha_s$), and the classical approximation
is justified, at least, to lowest order in $\alpha_s$.

This argument suggests that one can use the classical MV model
to study soft correlations provided all the fast partons 
with momenta larger than the scale $k^+$ of interest have been 
integrated out, and their effects included in the parameters
of the classical theory. In this way, all the potentially
large logarithms are resummed in the structure of
the weight function $W_\Lambda[\rho]$, with $\Lambda^+\sim k^+$.
The resulting classical theory is then
the correct effective theory at the scale $k^+$,
up to corrections of order $\alpha_s$ (with no large logarithms).

To construct the classical theory, we shall study the evolution
of $\rho$ and $W_\Lambda[\rho]$ with decreasing $\Lambda^+$. 
Specifically, we shall consider
a sequence of two effective theories (``Theory I''
and ``Theory II'') defined as in eq.~(\ref{PART}),
but with different separation scales: $\Lambda^+$
in the case of Theory I, and $b\Lambda^+$ for
Theory II, with $b\ll 1$, but such as $\alpha_s\ln(1/b) < 1$. 
(This allows us to treat corrections of order $\alpha_s\ln(1/b)$
 in perturbation theory.)
Theory II differs from Theory I in that the quantum fluctuations  
with longitudinal momenta inside the strip
\be\labe{strip}\,\,
 b\Lambda^+ \,\,<\,\, |p^+|\,\, <\,\,\Lambda^+\,,\ee
have been integrated out, and the associated correlations
have been incorporated at tree-level,
within the new weight function $W_{b\Lambda}$.

The difference $\Delta W
\equiv W_{b\Lambda} - W_\Lambda$ will be obtained by
matching calculations of gluon correlations at the scale 
$k^+\simle b\Lambda^+$ in the two theories.
In Theory II, and to lowest order in $\alpha_s$,
these correlations are found already at tree level, 
i.e., in the classical approximation.
In Theory I, and to the same accuracy,
they involve also the logarithmically enhanced quantum
corrections due to fluctuations with momenta in the strip.
In computing the latter, we shall work to leading order in 
$\alpha_s\ln(1/b)$ [``leading logarithmic accuracy''(LLA)],
but to all orders in the background fields and sources.
Indeed, we are interested here in the regime of saturation,
where ${\cal A}^i \sim 1/g$ and $\rho\sim 1/g$,
so that the associated non-linear effects cannot be expanded 
in perturbation theory.

The final outcome will be a {\it functional},
and {\it non-linear}, differential equation 
for $W_\tau[\rho]$ (with $\tau\equiv \ln(1/b)$) which describes
the flow of the weight function with $\tau$ \cite{JKLW97,JKLW99a}. 
In the weak field limit, this equation can be linearized, and
shown to reduce to the BFKL equation \cite{JKLW97} (see also Sect. 5).

\subsubsection{Quantum corrections in Theory I}

The classical effective theory is expected to reproduce 
{\it equal-time} correlators of the {\it transverse} fields
with {\it soft} ($k^+\simle b\Lambda^+$) longitudinal momenta
(cf. eq.~(\ref{clascorr})).
So, these are the correlations that we shall try to match between
Theory I and Theory II. In turns out that, in order to
establish the evolution equation for $W_\tau[\rho]$,
it is enough to consider the two-point function 
\be\label{2P}
\langle A^i_a(x^+,\vec k) A^i_a(x^+,-\vec k)\rangle,
\ee
where $\vec k \equiv (k^+,{\bf k}_\perp)$ with
$k^+\simle b\Lambda^+$, and 
the brackets denote quantum expectation values at fixed $\rho$.
 This is independent of time
because the background is static.
When computing this function within Theory I, there are
quantum corrections of order $\alpha_s\ln(1/b)$ to be
identified in what follows.
To simplify writing, it is convenient to work
in the coordinate representation, and define (with
$A^i_x\equiv A^i_a(x)$)
\be\labe{2PL0}\,
{\cal G}(\vec x, \vec y) \,\equiv\,
\langle A^i_a(x^+,\vec x)  A^i_a(x^+,\vec y)\rangle\,=\,
\langle ({\cal A}^i+\delta A^i)_x({\cal A}^i+\delta A^i)_y\rangle\,,\ee
where it is understood that the external lines carry soft 
($\simle b\Lambda^+$) longitudinal momenta.

The equal-time correlator (\ref{2P}) or (\ref{2PL0}) 
represents the density of the soft gluons generated by $\rho$ 
(cf. Sect. 2.4). [The vacuum, $\rho$--independent, contributions
to this quantity are to be subtracted away.] It involves a tree-level
piece ${\cal A}^i(\vec x){\cal A}^i(\vec y)$ together
with quantum corrections
--- the induced mean field
$\delta {\cal A}^i_a\equiv \langle \delta A^i_a\rangle$, and
the induced density $\langle\delta A^i_a\delta A^i_a\rangle$
 --- which describe the
polarization of the quantum fluctuations in the presence
of $\rho$. In general, these corrections
receive contributions from all the fluctuations
with momenta $|p^+| \le \Lambda^+$. However, for matching purposes, 
we need only the respective contributions of the
{\it semi-fast} gluons, by which we mean the fluctuations
with $p^+$  momenta inside the strip (\ref{strip}).
Indeed, these are the only quantum effects from Theory I that
should be found at the tree-level of Theory II.
(The remaining quantum effects, 
due to the fluctuations with $|p^+| < b\Lambda^+$, will 
appear as radiative corrections also in Theory II.)

{}From now on, we shall include in
$\delta {\cal A}^i$ or $\langle\delta A^i\delta A^i\rangle$
only the quantum effects generated by the semi-fast gluons.
In fact, to LLA, 
the definition of the ``semi-fast gluons'' --- i.e., the
quantum fluctuations that have to be integrated out in going from
Theory I to Theory II --- can be restricted even further:
These are the nearly on-shell fluctuations with 
longitudinal momenta deeply inside the strip,
$b\Lambda^+\ll |p^+|\ll \Lambda^+$, and energies
$\Lambda^-\ll |p^-|\ll \Lambda^-/b$, where
\be\labe{strips}
 \Lambda^-\,\equiv\,\frac{Q_\perp^2}{2\Lambda^+}\,,\ee
and $Q_\perp$ is a typical transverse momentum.
(To LLA, it makes no difference what is the precise value of $Q_\perp$.)
Indeed, as we shall see later, 
these are the modes which are responsible for the
logarithmic enhancement of the radiative corrections.

For more clarity, we introduce the notation $a^\mu_b(x)$ for
the semi-fast gluons, and preserve the notation 
$\delta A^\mu_a(x)$ for the softer fields, with momenta 
$|k^+|\simle b\Lambda^+$. (The fields
in eq.~(\ref{2PL0}) belong to the latter category.)
That is, we rewrite the total gluon field as follows
(compare to eq.~(\ref{fluct0})) :
\be\labe{fluct1}
A^\mu_a(x)\,=\,{\cal A}^\mu_a(x) + \delta A^\mu_a(x) + a^\mu_a(x).\ee
The relevant quantum effects arise from the interactions
between the soft fields $\delta A^\mu$ and the semi-fast gluons 
$a^\mu$ in the background of the tree-level fields and sources 
${\cal A}^i$ and $\rho$. 
This yields, as we shall see,
$\delta {\cal A}^i\sim \alpha_s\ln(1/b){\cal A}^i$ and 
$\langle\delta A^i\delta A^i\rangle\sim \alpha_s\ln(1/b){\cal A}^i
{\cal A}^i$, so that
\be
{\cal G}(\vec x, \vec y) =
{\cal A}^i(\vec x){\cal A}^i(\vec y) +
{\cal A}^i(\vec x)\delta {\cal A}^i(\vec y) +
\delta{\cal A}^i(\vec x){\cal A}^i(\vec y)+
\langle\delta A^i(x^+,\vec x)\delta A^i(x^+,\vec y)\rangle
,\label{2PL}\ee
where the disconnected piece $\delta{\cal A}^i\delta{\cal A}^i$ 
of $\langle\delta A^i\delta A^i\rangle$ can be discarded,
since of higher order in $\alpha_s$. In general, effects 
non-linear in $\delta {\cal A}^i$ can be ignored since the 
induced fields, in contrast to the tree-level fields, are weak.
Because the background fields ${\cal A}^i$ and $\rho$ are
static, so is the induced mean field 
$\delta{\cal A}^i$ (and any other one-point function), while 
two-point functions like 
$\langle\delta A^i_x\delta A^j_y\rangle$ depend only upon the
time difference $x^+-y^+$.
In particular, $\langle a^\mu\rangle=0$, since the field
$a^\mu$ has no static mode.

In what follows, we shall express the quantum
effects $\delta {\cal A}^i$ and $\langle\delta A^i\delta A^i\rangle$
in terms of correlation functions of the semi-fast gluons.
In doing this, we shall perform simplifications appropriate
to LLA. First, the semi-fast gluons will be treated 
in the Gaussian, or mean-field, approximation, which means
that we shall consider, at most, one-loop diagrams
 (higher loops contribute to higher orders in $\alpha_s$).
Then, we shall perform kinematical approximations
which exploit the hierarchy of scales to retain
only terms of leading order in $\alpha_s\ln(1/b)$.
In the arguments below, we shall anticipate over results
to be fully demonstrated only a posteriori, via the formal
developments in Sect. 4, and the explicit calculations in Sect. 5
and in Paper II (see also Refs. \cite{JKLW97,JKLW99a,JKW99,JKLW99b,KM00}).

The interactions between the various fields are described by the 
action $S[{\cal A}+ \delta A + a]$. To the order of interest, it
is enough to preserve the linear coupling
$\delta A^\mu_a \delta J_\mu^a\,$, where
\be\labe{deltaJ}\,
\delta J_\mu^a (x) &\equiv&-\,\frac{\delta S}{\delta A^\mu_a(x)}
\bigg|_{{\cal A}+a}\\
&\approx&-\,\frac{\delta^2 S}{\delta A^\mu_a(x)\delta A^\nu_b(y)}\bigg|_{
{\cal A}}\,a^\nu_b(y)\,-\,\frac{1}{2}\,
\frac{\delta^3 S}{\delta A^\mu_a(x)\delta A^\nu_b(y)
\delta A^\lambda_c(z)}\bigg|_{
{\cal A}}\,a^\nu_b(y)a^\lambda_c(z)\,\nonumber\ee
is the soft color current generated by the semi-fast quantum
fluctuations. 
It is understood here that only
the soft modes with $k^+\simle b\Lambda^+$ are kept in the
products of fields.
In going from the first to the second line in this
equation, we have expanded in powers of $a^\mu_b$, 
and kept only terms which are linear or quadratic
(Gaussian approximation). Note also that we use compact notations,
where repeated indices (coordinate variables) are 
understood to be summed (integrated) over.

Given the separation of scales in the problem, we expect
the color charge
$\delta \rho_a \equiv\delta J^+_a$ to be the ``large component''
of the current; that is, we expect the dominance of the 
eikonal coupling $\delta A^-_a \delta \rho_a$ (cf. Sect. 2.1).
This will be indeed confirmed by the calculations, which show that
the only current correlators to be retained to LLA are correlators
of $\delta \rho_a$.

Consider first the two-point function
$\langle\delta A^i\delta A^i\rangle$. 
Quite generally, the following relation holds:
\be\label{JJSIG}
\langle\delta A^\mu_a(x) \delta A^\nu_b(y)\rangle\,=\,\int d^4z
\int d^4u\,
G^{\mu\alpha}_{R\,ac}(x,z)\,\langle\delta J_\alpha^{c} (z)
\delta J_\beta^{d} (u)\rangle\,G^{\beta\nu}_{A\,db}(u,y),
\ee
where $G_R$ ($G_A$) is the retarded (advanced) propagator of the
soft fields in the background of the tree-level fields and sources.
To the order of interest, this can be computed in the mean-field
approximation, that is, by inverting the following
differential operator:
\be\labe{invG}
G_{\mu\nu}^{-1}(x,y)[{\cal A},\,\rho]\,\equiv\,
\frac{\delta^2 S [A,\,\rho]}
{\delta A^\mu(x)\delta A^\nu(y)}\bigg|_{{\cal A}}\,\ee
in the subspace of soft fields, in the LC gauge $\delta A^+=0$,
and with appropriate boundary conditions.
Eq.~(\ref{JJSIG}) is simply the statement that the current-current 
correlator acts as a self-energy for the soft field two-point function
(see Sect. 4.2 for a formal proof).
To LLA, this equation can be simplified as follows:

Among the various components of the polarization tensor
$\langle\delta J^\mu \delta J^\nu \rangle$, the logarithmic enhancement
shows up only in the charge-charge correlator:
\be\labe{CHIDEF}\,
\hat\chi_{ab}(x,y)\,\equiv\,\langle\delta \rho_a(x)\,
\delta \rho_b(y)\rangle\,.\ee
Specifically, we shall see that $\hat\chi$ is a quantity
of order O($\alpha_s\ln(1/b)\rho\rho$).
\begin{figure}
\protect\epsfxsize=12.cm{\centerline{\epsfbox{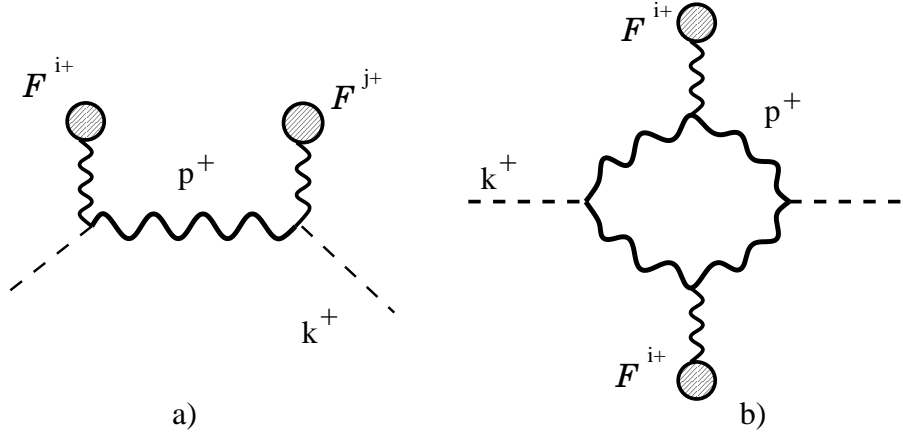}}}
         \caption{Tree (a) and one-loop (b) contributions to 
 $\chi$ to quadratic order in $\rho$. Wavy lines with a blob
denote insertions of the electric background field
${\cal F}^{+i}\,$; the thick internal wavy lines are propagators
of the semi-fast gluons; the external dotted lines
carry soft momenta, and couple to the fields $\delta A^-$.}
\label{CHILIN}
\end{figure} 

{ To gain some more intuition, let us use as an example
the contributions to $\delta \rho_a$ coming from the Yang-Mills
piece of the action, $S_{YM}=\int d^4x(-F_{\mu\nu}^2/4)$ :
\be\label{drhoYM}
\delta \rho_a (x)\big|_{YM}\,=\,
2 gf^{abc}{\cal F}^{+i}_b (\vec x) a^{i}_c (x)
+g f^{abc} (\partial^+ a_{b}^{i}(x) )a_{c}^{i}(x).\ee
(The additional contributions to $\delta \rho_a$, coming from $S_W$,
will be presented in Sect. 4.3.) 
The first term in the r.h.s. is linear in the
quantum fluctuations $a^i$ and in the electric
background field ${\cal F}^{+i}$ (this is
kinematically allowed because the semi-fast longitudinal
momentum can flow from  $a^i$ to ${\cal F}^{+i}$). 
It is only this term that contributes to $\hat\chi$ to
leading order in $\alpha_s$. Indeed, it generates
{\it tree} diagrams like the one depicted in Fig. \ref{CHILIN}.a, 
where the internal 
line represents the propagator $\langle a^ia^j \rangle$ of the semi-fast
gluons, 
and the soft external legs couple to the fields $\delta A^-_a\/$.
Physically, this diagram describes the emission of an on-shell, 
or {\it real}, semi-fast gluon by the classical source, 
which is a radiative correction
to the direct emission of the soft field. 
This diagram will be evaluated in Sect. 5.2, but it is clear
by power counting that it is of O$(\alpha_s\rho\rho)$.
By contrast, {\it one-loop} diagrams which are of the same order 
in $\rho$ are necessarily of higher order in $g$,
and thus negligible for the present purposes.
[An exemple of such a diagram, of O$(\alpha_s^2\rho\rho)$,
is shown in  Fig. \ref{CHILIN}.b; this is generated by the second piece, 
quadratic in $a^i$, in the r.h.s. of eq.~(\ref{drhoYM}).]

}

When computing ${\cal G}(\vec x, \vec y)$, eq.~(\ref{2PL}), to
LLA, the temporal non-locality in $\hat\chi$ can be ignored. To see this,
consider eq.~(\ref{JJSIG}) with  $x^+=y^+$ 
and $\langle\delta J^\mu \delta J^\nu \rangle$ replaced by $\hat\chi$.
Since the external lines $G_R$ and $G_A$ are soft ($k^+\simle b\Lambda^+$),
the typical energy scale which controls the non-localities
 $x^+-z^+$ and $x^+-u^+$ (and therefore also
$z^+-u^+$) is $k^-\equiv Q_\perp^2/2k^+ \simge \Lambda^-/b$
(cf. eq.~(\ref{strips})). This is much larger than the typical energy 
$p^-\ll \Lambda^-/b$ of the semi-fast gluons contributing 
to $\hat\chi$. Therefore, one can neglect the evolution of
$\delta \rho$ during the small time interval $z^+-u^+\simle
b/\Lambda^-$, and replace $\hat\chi(z^+-u^+)$ 
by its equal-time limit, which is independent of time:
\be\labe{chi0} \,\hat\chi(\vec z,\vec u)\,\equiv\,
\hat\chi(z^+=u^+,\vec z,\vec u)\,\equiv\,
\alpha_s\ln{1\over b}\,\,\chi(\vec z,\vec u)\,.\ee
As indicated above, it is only after
taking the equal-time limit inside $\hat\chi$ that the
logarithmic enhancement becomes manifest.
The integrations over $z^+$ and $u^+$ then set
$p^-=0$ in both external propagators, thus finally 
yielding
\be\label{GSIG1} 
\langle\delta A^i(x^+, \vec x) \delta A^j(x^+,
\vec y)\rangle\approx\int d^3\vec z 
\int d^3\vec u\, G^{\,i-}(\vec x,\vec z,p^-=0)
\,\hat\chi(\vec z,\vec u)\,
G^{\,-j}(\vec u,\vec y,p^-=0),\,\,\,\ee
where we have omitted the subscripts $R$ and $A$
on the propagators since the boundary conditions in time
become irrelevant as $p^-=0$.
In particular, since $G^{--}=0$ for $p^-=0$, the
two-point functions involving $\delta A^-_a$
(like $\langle\delta A^- \delta A^- \rangle$) vanish to LLA.

We now consider the induced mean field $\delta {\cal A}^\mu$,
which can be obtained by solving eq.~(\ref{MFE}) to the order of 
interest:
\be\label{MFE1}
0\,=\,\left\langle \frac{\delta S}{\delta A^\mu_a(x)}\bigg|_{
{\cal A}+\delta A+a}
\right\rangle\Bigg|_{\langle a \rangle =0}\,\approx\,
\frac{\delta^2 S}{\delta A^\mu_a(x)\delta A^\nu_b(y)}\bigg|_{
{\cal A}}\delta{\cal A}_b^\nu(y)\,-\,{\cal J}_\mu^{\,a}(\vec x),\ee
with an {induced current\/}:
\be\label{JTOT}
{\cal J}_\mu^{\,a}(\vec x)\equiv-\,\frac{1}{2}\,
\frac{\delta^3 S}{\delta A^\mu_a(x)\delta A^\nu_b(y)
\delta A^\lambda_c(z)}\bigg|_{
{\cal A}}\biggl(\Bigl\langle a^\nu_b(y)a^\lambda_c(z)\Bigr\rangle
+ \Bigl\langle \delta A^\nu_b(y) \delta A^\lambda_c(z) \Bigr\rangle\biggr)
\,,\ee
which involves two types of contributions:
({\it a\/}) A piece proportional to the 2-point function
$\langle a^\nu a^\lambda\rangle$ of the semi-fast gluons, and
({\it b\/}) a second piece involving the {\it induced} correlator 
$\langle \delta A^\nu \delta A^\lambda \rangle$ of the
soft fields, as given by eq.~(\ref{JJSIG}). 
Let us explain this in more detail:

({\it a\/}) The first piece is the expectation value
$\langle\delta J^\mu_a\rangle$ of the current (\ref{deltaJ}).
Once again, only the $\mu=+$ component, i.e., the {induced
color charge density} : 
\be\labe{JIND}\,
\hat\sigma_a(\vec x)\,\equiv\,\langle\delta \rho_a(x)\rangle\,,\ee
matters to LLA:
$\hat\sigma={\rm O}\bigl(\alpha_s\ln(1/b)\rho\bigr)$.
In Fig. \ref{SIGLIN}.a we show a diagram contributing to
 $\hat\sigma\/$ at lowest order in $\rho\,$; 
this is obtained by evaluating 
$g f^{abc} \langle (\partial^+ a_{b}^{i})a_{c}^{i}\rangle$
to linear order in $\rho$ (cf. eq.~(\ref{drhoYM})),
and will be computed in Sect. 5.1.
Note that the classical source $\rho$ is represented
here as a continuous line; this is to suggest
the physical origin of $\rho$ (namely, fast partons 
moving on straight line trajectories at the speed
af light), and also the fact that its coupling
to the semi-fast gluons is non-local in time (cf. Sect. 4.1).

{ 
({\it b\/}) To LLA, the second piece in eq.~(\ref{JTOT})
involves only the transverse field correlator 
$\langle \delta A^i \delta A^j\rangle$,
which is proportional to $\hat\chi$ (cf. eq.~(\ref{GSIG1})).
We denote this as:
\be\label{delcalJ}
\delta{\cal J}_\mu\,\equiv\,
-\,\frac{1}{2}\,
\frac{\delta^3 S}{\delta A^\mu \delta A^i \delta A^j}\bigg|_{
{\cal A}}\Bigl(G^{i-}\hat\chi \,G^{-j}\Bigr).\ee
This is of order $g\hat\chi \sim g\alpha_s\ln(1/b)\rho\rho\,$;
thus, in the saturation regime where $g\rho \sim 1$,
it is as large as the induced charge $\hat\sigma$.

\begin{figure}
\protect\epsfxsize=12.cm{\centerline{\epsfbox{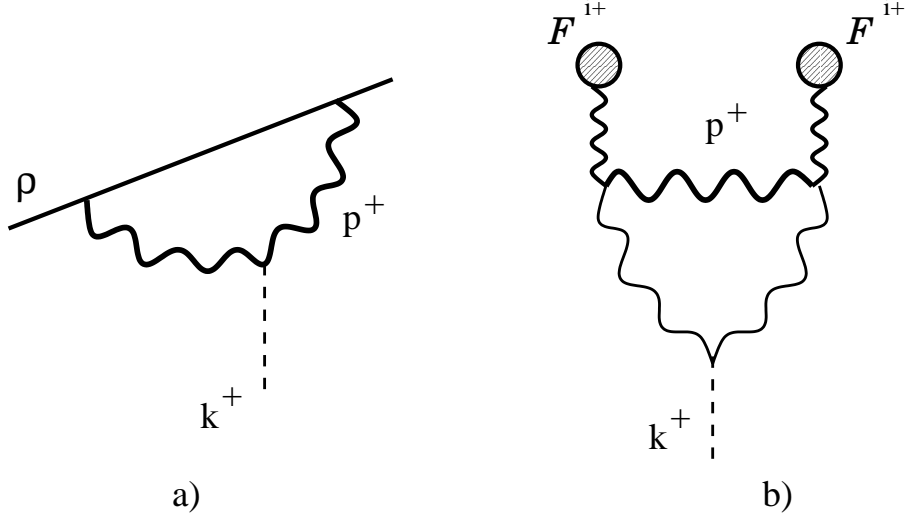}}}
         \caption{Contributions to the induced current to
low orders in $\rho$. (a) A contribution to $\sigma_a\,$;
the continuous line represents the classical source $\rho$.
(b) A contribution to $\delta{\cal J}^\mu_a\,$;
the thick internal line, with momentum $p^+$,
is the propagator $\langle a^i a^i \rangle$
of the semi-fast gluons; the other, thiner,
internal lines are the
propagators $G^{i-}$ and $G^{-i}$ of the soft fields.}
\label{SIGLIN}
\end{figure}

By using Fig. \ref{CHILIN}.a for $\hat\chi$, we obtain the contribution
 to $\delta{\cal J}_\mu$ depicted in Fig. \ref{SIGLIN}.b, 
which should be compared to Fig. \ref{SIGLIN}.a for $\hat\sigma$ :
The logarithm $\ln(1/b)$ in the latter is generated by the
integration over the internal $p^+$ momentum, which is restricted
to the strip (\ref{strip}). By contrast, in  Fig. \ref{SIGLIN}.b
the loop momentum is soft, but the propagator 
$\langle \delta A^i \delta A^j\rangle$ is itself of order 
$\alpha_s\ln(1/b)$, since proportional to $\hat\chi$.
Thus, the logarithm is now generated by the subintegral
giving $\hat\chi$.
Note that, while $\hat\sigma$ describes the {\it direct} polarization
of the semi-fast gluons, $\delta{\cal J}^\mu$ is rather an 
{\it indirect} effect of the latter, which induce a current after
first modifying the propagator of the soft fields.
}

To summarize, ${\cal J}^\mu =\delta^{\mu+}\hat\sigma
+\delta{\cal J}^\mu\,$. As we shall verify in Sect. 3.3,
where $\delta{\cal J}^\mu$ will be constructed explicitly,
the current ${\cal J}^\mu$
is covariantly conserved, ${\cal D}_\mu {\cal J}^\mu=0$,
as necessary for the mean field equation (\ref{MFE1}) to be
well defined. 
(Here, ${\cal D}_\mu \equiv \partial_\mu-ig{\cal A}_\mu$ is the 
covariant derivative constructed with the tree-level background field.)
Since ${\cal J}^\mu$ is independent of time, the LC-gauge solution to
eq.~(\ref{MFE1}) is of the form (cf. Sect. 2.3):
\be\labe{deltaCA}
\delta{\cal A}^+_a&=&
\delta{\cal A}^-_a\,=\,0\,,\nn
\delta{\cal A}^i_a(\vec x)&=&
\int d^3\vec y\,\,G_{ab}^{\,i\nu}(\vec x,\vec y,p^-=0)\,
{\cal J}_\nu^{\,b}(\vec y).\ee

At this stage, the gluon density in eq.~(\ref{2PL}) can be
compactly written as follows:
\be\labe{2PL1}
{\cal G}(\vec x, \vec y)=
{\cal A}^i_{\vec x}{\cal A}^i_{\vec y}+
(G^{\,i\nu}{\cal J}_\nu^{})_{\vec x}\,{\cal A}^i_{\vec y}+
{\cal A}^i_{\vec x}\,(G^{\,i\nu}{\cal J}_\nu^{})_{\vec y}+
(G^{\,i-}\hat\chi \,G^{\,-i})_{\vec x\,\vec y}\,.\ee
The logarithmic enhancement is implicit in
$\hat\sigma$ (which enters ${\cal J}^\mu$) and $\hat\chi$.
In fact, to LLA, the induced
charge $\hat\sigma(\vec x)$ can be replaced in eq.~(\ref{2PL1})
with a two-dimensional charge density localized at $x^-=0$ :
\be \labe{sigchi1}
\hat\sigma_a(\vec x)&\to&\delta(x^-)\,
\alpha_s\ln{1\over b}\,\sigma_a ({x}_\perp),\ee
where
\be\labe{sigperp}
\alpha_s\ln{1\over b}\,\sigma_a ({x}_\perp)\,\equiv \,
\int dx^- \,\hat\sigma_a(x^-, x_\perp)\ee
is the effective color charge density in the transverse plane.
As shown by the last equality, the logarithmic
enhancement becomes manifest only after integrating over $x^-$.

{ To understand 
this last operation, recall the relevant longitudinal scales:
$\sigma_a(\vec x)$ is the color charge induced by the semi-fast gluons,
and it spreads over longitudinal distances
$1/\Lambda^+ \simle x^- \simle 1/b\Lambda^+$ (this will be verified
by the explicit calculations in Sect. 5.1 and Paper II). Thus,
$\sigma_a$ sits further away
from the light-cone than the original source $\rho_a$ (which,
we recall, has support at $0\simle x^- \simle 1/\Lambda^+$),
but it is still very close to $x^-=0$ on the resolution scale
of the soft fields $\delta A^i$. In convolutions like 
\be 
(G^{\,i-}\hat\sigma)_{\vec x}\equiv 
\int d^3\vec y\,G^{\,i-}(\vec x,\vec y)\,\hat\sigma(\vec y),\ee
the integral over $y^-$ runs typically over large distances
$y^-\gg 1/b\Lambda^+$, where it is indeed legitimate to replace
$\hat\sigma(\vec x)$ as shown in eq.~(\ref{sigchi1}).

Since
eqs.~(\ref{GSIG1}) and (\ref{delcalJ}) involve similar
convolutions, there too one can replace
\be\labe{sigchi2}
\hat\chi_{ab}(\vec x,\vec y)&\to&\delta(x^-)\,
\alpha_s\ln{1\over b}\,
\chi_{ab}(x_\perp, y_\perp)\,\delta(y^-),\nn
\alpha_s\ln{1\over b}\,
\chi_{ab}(x_\perp, y_\perp)&\equiv&\int dx^-\int dy^-\,
\hat\chi_{ab}(\vec x,\vec y).
\ee
In fact, $\hat\chi$ turns out to be
even more localized than $\hat\sigma\,$: Indeed,
$\hat\chi$ is explicitly proportional to the 
tree-level source $\rho$ and/or electric field ${\cal F}^{+i}$, 
since it is generated by vertices which have this property
(like the first term in the r.h.s. of eq.~(\ref{drhoYM});
see also Fig. \ref{CHILIN}.a).
Thus,  $\hat \chi(\vec x, \vec y)$ has support at 
$0\simle x^-,y^- \simle 1/\Lambda^+$, like the original
source $\rho$.

The previous considerations show that the longitudinal and temporal
structure of the quantum corrections
do not matter when computing soft correlation functions:
when ``seen'' by the soft fields, the (semi)fast gluons appear 
as a static color charge localized near the light-cone.
This is consistent with the physical picture in Sect. 2.1,
and also clarifies its limitations: This picture holds only to LLA,
which is the accuracy to which we have a strong separation of scales.}
To this accuracy, only the one-point ($\hat\sigma$) and
two-point ($\hat\chi$) correlations of the
quantum charge $\delta\rho$ must be kept (cf. Sect. 4.3 below).
This is why, for matching purposes, it is enough
to consider the two-point function $\langle A^i
A^i\rangle$, as we do here. This is finally given by:
\be\label{2PL2}
{\cal G}(\vec x, \vec y)\,\approx\,
{\cal A}^i_{\vec x}{\cal A}^i_{\vec y}\,+\,\alpha_s\ln(1/b)
\Bigl\{
(G^{\,i\nu}{\cal J}_\nu)_{\vec x}\,{\cal A}^i_{\vec y}+
{\cal A}^i_{\vec x}\,(G^{\,i\nu}{\cal J}_\nu)_{\vec y}+
(G^{\,i-}\chi \,G^{\,-i})_{\vec x\,\vec y}
\Bigr\},\ee
where the logarithmic factor $\ln(1/b)$ is now explicit.
After averaging over $\rho$ with weight function 
$W_{\Lambda}[\rho]$, eq.~(\ref{2PL2})
gives the gluon density at the
scale $b\Lambda^+$, as computed in Theory I to LLA.
By construction, this must be the same as the
gluon density in Theory II at tree-level. That is, 
the following equality must hold:
\be\label{evolG}
\langle{\cal A}^i_{\vec x}{\cal A}^i_{\vec y}
\rangle_{b\Lambda}=\langle{\cal A}^i_{\vec x}{\cal A}^i_{\vec y}
\rangle_{\Lambda}+
\alpha_s\ln(1/b)\Bigl\langle
(G^{\,i\nu}{\cal J}_\nu)_{\vec x}\,{\cal A}^i_{\vec y}+
{\cal A}^i_{\vec x}\,(G^{\,i\nu}{\cal J}_\nu)_{\vec y}+
(G^{\,i-}\chi \,G^{\,-i})_{\vec x\,\vec y}
\Bigr\rangle_{\Lambda},\,\,\,\,\ee
where, e.g.,
\be\labe{clascorrL}
\langle{\cal A}^i_{\vec x}{\cal A}^i_{\vec y}
\rangle_{\Lambda}\,\equiv\,
\int {\cal D}\rho\,\,W_\Lambda[\rho]\,\,{\cal A}_a^i({\vec x})
{\cal A}_a^i({\vec y})\,,\ee
together with a similar definition for $
\langle{\cal A}^i{\cal A}^i\rangle_{b\Lambda}$
in terms of $W_{b\Lambda}$.  Eq.~(\ref{evolG}) is the evolution
equation for the gluon density, and can be used to derive 
the evolution $W_\Lambda[\rho]\,\to\,W_{b\Lambda}[\rho]$ 
of the classical weight function, as it will be explained shortly.

To avoid cumbersome notations, in what follows we shall 
not always distinguish between $\hat\sigma$ and $\sigma$, or 
between $\hat\chi$ and $\chi$; these quantities differ just by
their arguments and by a factor of $\alpha_s\ln(1/b)$ (cf. 
eqs.~(\ref{sigchi1})--(\ref{sigperp}) and (\ref{sigchi2})),
 which will be written down explicitly whenever needed.

\subsubsection{Tree-level calculation in Theory II}

In this subsection, the evolution equation for $W_\tau[\rho]$ 
will be established in two steps: First, we shall show that the
quantum corrections described previously can be generated
by adding a Gaussian ``noise term'' to the Yang Mills equation
(\ref{cleq0}). Then, we shall absorb this noise term into
a redefinition of the classical source and weight function.

Our starting point is the following Yang Mills equation:
\be
[D_{\nu}, F^{\nu \mu}]_a\, =\, \delta^{\mu +} 
(\rho_a(\vec x)+\nu_a(\vec x)),
\labe{cleqn}
\ee
which in addition to eq.~(\ref{cleq0}) contains
a random source $\nu_a(\vec x)$
(the ``noise term'') chosen such as to generate, via the
solution to eq.~(\ref{cleqn}), the same 
two-point function $\langle A^i A^i\rangle$ 
as obtained in Theory I after including the
quantum corrections (cf. eq.~(\ref{2PL2})).
Thus, $\nu_a$ plays the role of the fluctuating 
charge $\delta\rho_a$ of the semi-fast gluons.
Based on this analogy, we take $\nu_a$ to be static and have
the same non-trivial correlators as $\delta\rho_a$, namely
(with $\langle\cdots\rangle_{\nu}$ denoting the average over $\nu$) :
\be\labe{noise}\,
\langle\nu_a(\vec x)\rangle_{\nu}\,=\,
\sigma_a (\vec x),\qquad
\langle\nu_a(\vec x)\nu_b(\vec y)\rangle_{\nu}\,=\,
\chi_{ab}(\vec x,\vec y).\ee
{ From the previous subsection, we recall that the precise longitudinal
structure of $\sigma$ and  $\chi$ does not matter for the calculation
of soft correlations. As we shall shortly discover, to a large extent
this is also true for the evolution equation, which involves only
the integrated densities 
$\sigma_a(x_\perp)$ and $\chi_{ab}(x_\perp,y_\perp)$,
eqs.~(\ref{sigperp}) and (\ref{sigchi2}).
Still, for this equation to be well defined,
it will be important to recall that the induced charge 
$\sigma_a (\vec x)$ has a longitudinal support which is separated
from that of the original source $\rho$
(cf. the discussion after eq.~(\ref{sigperp})).
This property is relevant here since the solution to the classical
equation (\ref{cleqn}) involves the color source via
path-ordered exponentials in $x^-$, cf. Sect. 2.3.
To keep trace of that, and for
consistency with the first equation (\ref{noise}),
we choose $\nu_a(\vec x)$ to have support in the range
$1/\Lambda^+ \simle x^- \simle 1/b\Lambda^+$, so like $\sigma$.
}

We first verify that the classical stochastic problem
in eqs.~(\ref{cleqn})--(\ref{noise}) generates the 
same correlator $\langle A^i A^i\rangle$ as the quantum problem
in Sect. 3.2.1.
The solution $A^\mu[\rho,\nu]$ to eq.~(\ref{cleqn}) can be
read off Sect. 2.3: 
if ${\cal A}^i[\rho]$ is the solution (\ref{Ansatz})
to eq.~(\ref{cleq0}), then, clearly, 
\be\label{Anu} A^\mu[\rho,\nu](x)\,=\,\delta^{\mu i}
{\cal A}^i[\rho+\nu](\vec x).\ee
For fixed $\rho$, this is a random variable whose correlations
can be obtained by expanding in powers of $\nu$, and then
averaging over $\nu$ with the help of eq.~(\ref{noise}).
To the order of interest, we need the expansion
of (\ref{Anu}) to {\it quadratic} order 
(below, ${\cal A}^i_x\equiv {\cal A}^i_a[\rho](\vec x)$) :
\be\label{calAexp}
{\cal A}^i_x[\rho+\nu]\,\approx\,{\cal A}^i_{ x}[\rho]
\,+\,\frac{\delta {\cal A}^i_{ x}}{\delta \rho_{ y}}
\bigg|_\rho\nu_{ y}\,+\,\frac{1}{2}\,
\frac{\delta^2 {\cal A}^i_{ x}}
{\delta \rho_{ y} \delta \rho_{ z}}\bigg|_\rho\nu_{ y}
\nu_{ z}\,\equiv\,{\cal A}^i_{ x}[\rho]\,+\,
\delta{ A}^i_{ x}[\rho,\nu] \,.\ee
Indeed, after averaging over $\nu$, both the linear and the
quadratic term will contribute to order $\alpha_s\ln(1/b)$,
while the terms neglected in eq.~(\ref{calAexp})
would be of higher order.
The classical two-point correlation function is then obtained as:
\be\labe{2PN}\,
\langle{\cal A}^i_{ x}[\rho+\nu]\,{\cal A}^i_{ y}[\rho+\nu]
\rangle_\nu\,=\,{\cal A}^i_{ x}{\cal A}^i_{ y}
\,+\,\langle\delta{ A}^i_{ x}\rangle_\nu {\cal A}^i_{ y}
\,+\,{\cal A}^i_{ x}\langle\delta{ A}^i_{ y}\rangle_\nu
\,+\,\langle\delta{ A}^i_{ x}
\delta{ A}^i_{ y}\rangle_\nu\,,\ee
where, to the order of interest,
\be\label{dAnn}
\langle\delta{ A}^i_{ x}\rangle_\nu&=&
\frac{\delta {\cal A}^i_{ x}}{\delta \rho_{ y}}
\bigg|_\rho\sigma(y)
\,+\,\frac{1}{2}\,
\frac{\delta^2 {\cal A}^i_{ x}}
{\delta \rho_{ y} \delta \rho_{ z}}\bigg|_\rho\chi(y,z),\nn
\langle\delta{ A}^i_{ x}
\delta{ A}^i_{ y}\rangle_\nu&=&
\frac{\delta {\cal A}^i_{ x}}{\delta \rho_{ z}}
\bigg|_\rho\chi(z,u)\,
\frac{\delta {\cal A}^i_{ y}}{\delta \rho_{ u}}\,.\ee
These classical correlators coincide with the 
corresponding quantum corrections $\delta{\cal A}^i$,
eq.~(\ref{deltaCA}), and $\langle\delta A^i\delta A^i\rangle$,
eq.~(\ref{GSIG1}), as we demonstrate now:

For the two-point function, one can use the 
following identity, to be derived shortly,
\be\labe{delAG}
\frac{\delta {\cal A}^i_{ x}}{\delta \rho_{ y}}
\bigg|_\rho\,\equiv\,
G^{\,i-}(\vec x,\vec y,p^-=0),\ee
to conclude that the expression (\ref{dAnn}) for
$\langle\delta{ A}^i\delta{ A}^i\rangle_\nu$
is indeed the same as eq.~(\ref{GSIG1}) for 
$\langle\delta A^i\delta A^i\rangle$.
Eq.~(\ref{delAG}) can be proven by noticing that,
to linear order in $\nu$, $\delta{ A}^i\approx
G^{\,i-}\nu$, which must be the same as the linear-order
term in eq.~(\ref{calAexp}). 

Eq.~(\ref{delAG}) also shows that 
the first term in the r.h.s. of eq.~(\ref{dAnn}) for
$\langle\delta{ A}^i\rangle_\nu$ is the same as $G^{\,i-}\sigma$,
which is one of the contributions to 
$\delta{\cal A}^i$ in eq.~(\ref{deltaCA}). The other contribution,
involving $\delta{\cal J}^\mu_a$, can be similarly identified
with the term proportional to $\chi$ in eq.~(\ref{dAnn}).
More rapidly, the equality between $\langle\delta{ A}^i\rangle_\nu$
and $\delta{\cal A}^i$ can be established by showing that
they both satisfy the same equation of motion.
In the case of $\langle\delta{ A}^i\rangle_\nu$, this is obtained
by averaging eq.~(\ref{cleqn}) over $\nu$, and can be rewritten as:
\be\labe{cleqAV}
\left\langle \frac{\delta S_{YM}}{\delta A^\mu_a}\bigg|_{
{\cal A}+\delta A}\right\rangle_\nu\,=\,\delta_{\mu -}
(\rho_a+\sigma_a),
\ee
where $S_{YM}$ is the Yang-Mills action.
For $\delta{\cal A}^i$, the corresponding equation is rather
(\ref{MFE1}), which is equivalent to:
\be\labe{QAV}
\left\langle \frac{\delta S}{\delta A^\mu_a}\bigg|_{
{\cal A}+\delta A}\right\rangle\,\approx\,\delta_{\mu -}\sigma_a .
\ee
Since\footnote{By ``$\delta A^-=0$'', we mean here that
$\langle\delta A^-\rangle=0$, and the two-point functions
involving $\delta A^-$ --- like $\langle\delta A^-\delta A^i\rangle$
--- vanish as well (cf. the discussion after eq.~(\ref{GSIG1})).}
 ${\cal A}^-=\delta A^-=0$, this is formally similar
to eq.~(\ref{cleqAV}). The final identification between
these two equations then follows after recalling that the corresponding
two-point functions $\langle\delta A^i\delta A^i\rangle$
(which enter both eq.~(\ref{cleqAV}) and eq.~(\ref{QAV}))
are also the same.

To summarize, the classical two-point function 
(\ref{2PN}) coincide with the corresponding quantum 
function ${\cal G}(\vec x, \vec y)$, eq.~(\ref{2PL2}).
This allows us to write:
\be\labe{AAnu}\,
{\cal G}(\vec x, \vec y)\,=\,
\langle{\cal A}^i_x[\rho+\nu]\,{\cal A}^i_{ y}[\rho+\nu]
\rangle_\nu\,\equiv\,\int {\cal D}\nu\,{\cal W}[\nu\,;\rho]\,\,
{\cal A}^i_{ x}[\rho+\nu]\,{\cal A}^i_{ y}[\rho+\nu]\,,\ee
where the second equality is just a rewritting of the average
over $\nu$ as a functional integral with the
Gaussian weight function:
\be\labe{Wnu}\,
 {\cal W}[\nu\,;\rho]\,\equiv\,{\rm e}^{-{1\over 2}{\rm Tr}\ln\chi}
\,\,{\rm exp}\left\{-{1\over 2}\,(\nu-\sigma)_{x}
\chi_{x,y}^{-1}(\nu-\sigma)_{y}\right\}.\ee
(This depends upon $\rho$ via $\chi$ and $\sigma$.)
After also averaging over $\rho$, 
with weight function $W_{\Lambda}[\rho]$, eq.~(\ref{AAnu}) must
be the same as the gluon density 
generated in Theory II at tree-level (cf. eq.~(\ref{evolG})). 
This requires:
\be\label{recurW0}
\int {\cal D}\rho\,\,W_{b\Lambda}[\rho]\,\,{\cal A}_x^i[\rho]
{\cal A}_y^i[\rho]\,=\,\int {\cal D}\rho\,\,W_\Lambda[\rho]
\int {\cal D}\nu\,{\cal W}[\nu\,;\rho]\,\,
{\cal A}^i_{ x}[\rho+\nu]\,{\cal A}^i_{ y}[\rho+\nu]\,,\ee
which is satisfied provided
\be\labe{recurW}
W_{b\Lambda}[\rho]\,=\,\int {\cal D}\nu\,W_\Lambda[\rho-\nu]\,
{\cal W}[\nu\,;\rho-\nu]\,.\ee
This is a functional recurrency formula for $W[\rho]$, 
whose r.h.s. can be expanded as follows:
\be
W_\Lambda[\rho-\nu]\,
{\cal W}[\nu\,;\rho-\nu]\,\approx\,\left(1\,-\,
\nu_x\,\frac{\delta}{\delta \rho_x}\,+\,\frac{1}{2}\,\nu_x\,
\nu_y\,\frac{\delta^2}{\delta \rho_x \delta \rho_y}\right)
W_\Lambda[\rho]\,{\cal W}[\nu\,;\rho]\,,\ee
and then integrated over $\nu$ to give, to the order of interest,
\be\label{EVOLW}
W_{b\Lambda}[\rho]\,-\,W_\Lambda[\rho]\,=\,
-\,\frac{\delta}{\delta \rho_x}\,[W_\Lambda\sigma_{x}] 
\,+\,\frac{1}{2}\,\frac{\delta^2}{\delta \rho_x \delta \rho_y}
 [W_\Lambda \chi_{xy}].\ee
A priori, the convolutions in the r.h.s. of eq.~(\ref{EVOLW})
involve three-dimensional integrals, like
\be\label{rhoW1}
\frac{\delta}{\delta \rho_x}\,[W_\Lambda\sigma_{x}] \,\equiv\,
\int d^3{\vec x} \,\frac{\delta}{\delta \rho_a(\vec x)}\,
\Bigl[W_\Lambda\sigma_a(\vec x)\Bigr].\ee
Recall, however, that $\sigma$ is distributed over distances 
$1/\Lambda^+ \simle x^- \simle 1/b\Lambda^+$, and that the
logarithmic enhancement is seen only after integration
over $x^-$, cf. eq.~(\ref{sigperp}).
In the limit $b\to 1$, in which we are eventually interested,
eq.~(\ref{rhoW1}) can be replaced with
\be\label{rhoW2}
\frac{\delta}{\delta \rho_x}\,[W_\Lambda\sigma_{x}] \,=\,
\alpha_s\ln{1\over b}\,
\int d^2{x_\perp} \,\frac{\delta}{\delta \rho_a(x^-_\Lambda,x_\perp)}\,
\Bigl[W_\Lambda\sigma_a(x_\perp)\Bigr],\ee
where the functional derivative is to be evaluated at
$x^-_\Lambda\equiv 1/\Lambda^+$.

It is convenient to introduce the rapidity variable\footnote{We prefer
to denote this as $\tau$, rather than the more conventional $y$,
to avoid possible confusion with the coordinate variable $y$,
which also enters the subsequent equations.}
 $\tau\equiv\ln(P^+/\Lambda^+)
=\ln(1/x)$, where $P^+$ is the total longitudinal momentum of the
hadron, and $x$ is Bjorken's $x$ (cf. Sect. 2.1). 
Then, $\ln(P^+/b\Lambda^+)=\tau+\Delta \tau$,
with $\Delta \tau \equiv\ln(1/b)$. After also
renoting $W_\Lambda\equiv W_\tau$,
$W_{b\Lambda}\equiv W_{\tau+\Delta\tau}$, and
$x^-_\Lambda=1/\Lambda^+\equiv x^-_\tau$,
eq.~(\ref{EVOLW}) is rewritten as
\be\label{DISCEVOLW}
W_{\tau+\Delta\tau}[\rho]-W_\tau[\rho]\,=\,\alpha_s\Delta\tau
\left\{ {1 \over 2} {\delta^2 \over {\delta
\rho_\tau(x) \delta \rho_\tau(y)}} [W_\tau\chi_{xy}] - 
{\delta \over {\delta \rho_\tau(x)}}
[W_\tau\sigma_{x}] \right\}\,,
\ee
where $\rho_\tau(x_\perp)\equiv \rho(x^-_\tau,x_\perp)$,
and the convolutions are now to be understood
as two dimensional integrals, cf. eq.~(\ref{rhoW2}). For instance:
\be\label{rhoW3}
\frac{\delta}{\delta \rho_\tau(x)}\,[W_\tau\sigma_{x}] \,=\,
\int d^2{x_\perp} \,\frac{\delta}{\delta \rho_a(x^-_\tau,x_\perp)}\,
\Bigl[W_\tau\sigma_a(x_\perp)\Bigr].\ee
According to eqs.~(\ref{DISCEVOLW})--(\ref{rhoW3}), the
evolution from $W_\tau[\rho]$ to $W_{\tau+\Delta\tau}[\rho]$
is due to changes in $\rho$ within the rapidity interval
$(\tau, \tau+\Delta\tau)$, where the
quantum corrections are located.

Note that the variable $\tau$ in the above equations acts simultaneously
as a {\it momentum} rapidity [$\tau=\ln(P^+/\Lambda^+)$], and
as a {\it space-time} rapidity [$\tau=\ln(x^-_\tau/x^-_0)$,
with $x^-_0$ some arbitrary scale of reference; e.g.,
$x^-_0=1/P^+$]. This is so since the support of the quantum
corrections is correlated to the longitudinal momenta of the modes
which have been integrated out. 
These equations suggest a space-time picture of the
quantum evolution where the color source is built up by adding
contributions in successive layers of (space-time) rapidity,
from $x^-=0$ up.

By taking the limit
$\Delta \tau \equiv \ln(1/b)\to 0$,
we finally obtain a functional differential equation for
the evolution of the weight function with $\tau\equiv\ln(1/x)\/$:
\be\labe{RGE}
{\del W_\tau[\rho] \over {\del \tau}}\,=\,\alpha_s
\left\{ {1 \over 2} {\delta^2 \over {\delta
\rho_\tau(x) \delta \rho_\tau(y)}} [W_\tau\chi_{xy}] - 
{\delta \over {\delta \rho_\tau(x)}}
[W_\tau\sigma_{x}] \right\}\,.
\ee
This renormalization group equation (RGE) has been originally
derived in Ref. \cite{JKLW99a}, although its longitudinal structure
--- i.e., the fact that the functional differentiations are to be
taken with respect to the color source in the highest bin
of rapidity $(\tau, \tau+d\tau)$ ---
has not been recognized there.
(In Ref. \cite{JKLW99a} the differentiations are 
rather taken  w.r.t. the integrated charge density 
$\rho(x_\perp)\equiv \int dx^- \rho(\vec x)$.)
To understand the importance of this structure, recall
from Sect. 2.3 that the classical solution ${\cal A}^i[\rho]$
--- and therefore also the quantum corrections $\sigma$ and $\chi$,
which are functionals of ${\cal A}^i$ \cite{ILM00II} --- depends upon $\rho$
via Wilson lines in $x^-$ (cf. eq.~(\ref{UTAF})). Thus, in evaluating
eq.~(\ref{RGE}), we shall be led to differentiate \cite{ILM00II}
\be\labe{UTAF1}
V^{\dagger}(y_{\perp})\,\equiv\,
 {\rm P} \exp
 \left \{
ig \int_{-\infty}^{x^-_\tau} dz^- \int d^2z_\perp
\langle y_\perp|\,\frac{-1}{\grad^2_\perp}\,|z_\perp\rangle\,
\tilde\rho_a(z^-,z_\perp) T^a \right \},\ee
with respect to $\tilde\rho_\tau(x)$. Note the upper limit
$x^-_\tau$ in the integral above: this occurs since
the original source $\tilde\rho$ has support up to $x^-\simeq
x^-_\tau$. Because of the path-ordering in $x^-$, the functional
derivative of eq.~(\ref{UTAF1}) is as simple as a normal derivative:
\be\label{DIFFU}
{\delta \over \delta\tilde \rho_a(x^-_\tau,x_\perp)}
\,V^\dagger (y_{\perp})\,=\,
ig\,\langle x_\perp|\,\frac{-1}{\grad^2_\perp}\,|y_\perp\rangle\,T^a\,
V^\dagger (y_{\perp}).\ee
By contrast, its derivative 
w.r.t. the integrated charge
$\tilde\rho_a(x_\perp)$ is not well defined.

Note also that the color source which appears
in eqs.~(\ref{UTAF1})--(\ref{DIFFU})
is the {\it covariant-gauge}
source $\tilde \rho_a$, and not the LC-gauge source $\rho_a$.
It is only with respect to $\tilde \rho_a$ that we know the
classical solution explicitly. This brings us to the necessity
to rewrite the RGE (\ref{RGE}) as an evolution equation for
$W_\tau[\tilde\rho]$, which will shall do in the next subsection.

Let us conclude this subsection with a few remarks on
eq.~(\ref{RGE}). This can be recognized as a functional
Fokker-Planck equation \cite{Keizer}, 
with  $\tau$ playing the role of time.
It describes diffusion in the functional space spanned by $\rho$,
with ($\rho$-dependent) ``drift velocity''
$\alpha_s\sigma$ and ``diffusion constant'' $\alpha_s\chi$.
(In this language, the recurrency formula (\ref{recurW})
can be viewed as a functional Chapman-Kolmogorov equation \cite{Keizer}.)
Alternatively, eq.~(\ref{RGE}) is like a functional Schr\"odinger equation 
in imaginary ``time'' $\tau$. 
It can be transformed into an hierarchy of ordinary differential 
equations for the correlation functions $\langle \rho\rho
\cdots\rho\rangle_{\tau}$ generated by $W_\tau[\rho]\,$.
For instance, by multiplying it with $\rho(x)\rho(y)$
and functionally integrating over $\rho$, one obtains an evolution
equation for the two-point function:
\be\labe{RGE2p}
{d\over {d\tau}}\,
\Big\langle\rho_a(\vec x)\rho_b(\vec y)\Big\rangle_\tau&=& \alpha_s\,
\Big\langle \delta(x^--x^-_\tau)\sigma_a(x_\perp)\rho_b(\vec y)
\,+\,\delta(y^--x^-_\tau)\rho_a(\vec x)\sigma_b(y_\perp)\nn
&{}&\,\,\,\,\,\,+\,\delta(x^--x^-_\tau)\delta(y^--x^-_\tau)
\chi_{ab}(x_\perp,y_\perp)\Bigr\rangle_\tau\,,\ee
where $\langle \cdots \rangle_\tau$ denotes the 
average over $\rho$ with weight function $W_\tau[\rho]\,$.
For strong fields and sources, that is, in the saturation
regime, this equation involves also the higher $n$-point correlators
($n\ge 3$), since $\sigma$ and $\chi$ are non-linear in $\rho$ \cite{ILM00II}.
But in the weak source limit, where the non-linear effects can be 
neglected,  eq.~(\ref{RGE2p}) closes at the level of the
two-point functions, and becomes equivalent to the BFKL equation,
as demonstrated in Ref. \cite{JKLW97} (see also Sect. 5 below,
and Paper II, where the BFKL equation will be recovered as 
the weak field limit of the general evolution equations).

\subsection{Evolution equation with a covariant-gauge source}

For the RGE (\ref{RGE}) to be of practical use, its
coefficients $\sigma$ and $\chi$ must be
known explicitly in terms of the tree-level source $\rho$.
This requires, in particular, an explicit expression for
the classical solution ${\cal A}^i[\rho]$ 
(since $\sigma$ and $\chi$ are functionals of ${\cal A}^i$; 
cf. Sects. 4, 5 and Paper II). 
From Sect. 2.3, we know that such an explicit expression
can be obtained if the {\it light-cone} gauge solution 
${\cal A}^i$ is expressed in terms of the 
{\it covariant} gauge (COV-gauge) charge density $\tilde\rho$.
This observation, together with the gauge invariance of the 
weight function ($W_\Lambda[\rho]= W_\Lambda[\tilde\rho]$),
makes it more convenient to use $\tilde\rho$
as the classical source variable to be averaged over.
In this subsection, we shall rewrite the RGE in this new
variable. This is non-trivial since, as it will be explained
shortly, the gauge rotation from $\rho$ to $\tilde\rho$
is itself subjected to quantum evolution.

Recall indeed that the definition of the COV-gauge
depends upon the classical color source in the problem:
For a given static source $\rho_a(\vec x)$ in the LC-gauge,
the COV-gauge is the gauge where the classical field generated
by $\rho$ has just a plus component, $\tilde A^\mu=
\delta^{\mu +}\alpha$, with $\alpha$ linearly related to
$\tilde\rho$ (the COV-gauge version of $\rho$). Thus, if
the classical source is modified by quantum corrections 
(say, from $\rho$ to $\rho+\nu$ in the LC gauge), then what we 
mean by ``the covariant gauge'' changes as well: This is now the
gauge where the {\it new} classical field, as generated by
the total charge $\rho+\nu$, has just a plus component, say 
$\bar A^\mu= \delta^{\mu +}\bar\alpha$, with
$\bar\alpha$ linearly related to the COV-gauge
source $\bar\rho+\bar\nu$. [Note that we denote with a tilde 
(a bar) quantities in the COV-gauge
associated with $\rho$ (respectively, $\rho+\nu$).]
Clearly, the rotation 
between the old and the new COV-gauges depends on $\nu$,
and thus upon the quantum corrections.
In what follows, we shall construct this rotation explicitly.

Recall first the classical solution corresponding to $\tilde\rho$,
as constructed in Sect. 2.3. This is 
summarized in the following equations:
\be\labe{Atilde}\,
{\cal A}^i[\tilde\rho]\,
(\vec x) &=&{i \over g}\, U(\vec x) \,\partial^i  U^\dagger(\vec x),\nn
U^{\dagger}(x^-,x_{\perp})&=&
 {\rm P} \exp
 \left \{
ig \int_{-\infty}^{x^-} dz^-\,{\alpha}(z^-,x_{\perp})
 \right \},\nn
- \nabla^2_\perp \alpha({\vec x})&=&{\tilde \rho}(\vec x)\,,
\ee
where $U^\dagger(\vec x)$ is the gauge rotation
from the LC-gauge to the original COV-gauge (cf. eq.~(\ref{gtr})).

When the classical source changes from $\rho$ to $\rho+\nu$,
the corresponding solution is obtained by replacing 
$\tilde\rho \rightarrow \bar\rho+\bar\nu$
in eqs.~(\ref{Atilde}) :
\be\labe{Abar}\,
{\cal A}^i[\bar\rho+\bar\nu]\,
(\vec x) &=&{i \over g}\, \bar U(\vec x) 
\,\partial^i \bar U^\dagger(\vec x),\nn
 \bar U^{\dagger}(x^-,x_{\perp})&=&
 {\rm P} \exp
 \left \{
ig \int_{-\infty}^{x^-} dz^-\,\bar{\alpha}(z^-,x_{\perp})
 \right \},\nn
-\grad^2_\perp \bar{\alpha}&=&\bar\rho+\bar\nu \,.\ee
We write $\bar U^{\dagger}\equiv \delta U^{\dagger}\,U^{\dagger}$,
with $\delta U^{\dagger}$ the gauge rotation from the original
COV-gauge towards the new one. It turns out that this has a relatively
simple expression in terms of $\bar\nu$.

To see this, we set $\bar\alpha=\alpha+ \delta\alpha$, with $\alpha$ 
defined by the last equation (\ref{Atilde}) and $\delta\alpha$
proportional to $\nu$. These two fields have no overlap in the
longitudinal direction: $\alpha$, like $\rho$, has 
support at $0\simle x^- \simle 1/\Lambda^+$, while $\nu$, and therefore 
$\delta\alpha$, is rather localized at
$1/\Lambda^+ \simle x^- \simle 1/b\Lambda^+$.
Then, the path-ordered exponential in eq.~(\ref{Abar}) simply factorizes:
\be\labe{expU}\,
 \bar U^{\dagger}(x^-,x_{\perp})\,\equiv\,
\delta U^{\dagger}(\vec x) \, U^{\dagger}(\vec x)\,\approx\,
{\rm exp}\Bigl\{{i g\theta(x^--x^-_\tau)\delta {\alpha}(x_{\perp})}\Bigr\}\,
 U^{\dagger}(\vec x)\,,\ee
where  $x^-_\tau=1/\Lambda^+$, 
$\delta {\alpha}(x_{\perp})\equiv \int dx^- \delta {\alpha}(\vec x)$,
and the approximate equality holds as $b\to 1$.
In particular, within the support of $\rho$,
$\bar U^{\dagger}$ reduces to $U^{\dagger}$, and therefore 
\be\label{tilderhobar}
\bar\rho\,=\,\bar U^{\dagger}\rho \bar U \,=\,U^{\dagger}\rho U
\,=\,\tilde\rho.\ee
Thus, remarkably, the original source $\tilde\rho$ is {\it not}
affected by the supplementary gauge rotation 
$\delta U^{\dagger}$ induced by the quantum corrections.
By using this property, and subtracting the equation satisfied by
$\alpha$ from the equation for $\bar {\alpha}$, we deduce 
\be \label{dalpha}
-\grad^2_\perp\delta{\alpha}(\vec x)\,=\,\bar \nu(\vec x)\,,\ee
which shows that $\delta {\alpha}$ is linear in $\bar \nu$.

We are now in a position to reformulate the RGE by using the
COV-gauge source as a variable. The analog of eq.~(\ref{recurW0})
reads:
\be
\int {\cal D}\tilde\rho\,\,W_{b\Lambda}[\tilde\rho]\,\,{\cal A}_x^i[\tilde\rho]
{\cal A}_y^i[\tilde\rho]\,=\,\int {\cal D}\tilde\rho\,\,W_\Lambda[\tilde\rho]
\int {\cal D}\bar\nu\,{\cal W}[\bar\nu\,;\tilde\rho]\,\,
{\cal A}^i_{ x}[\bar\rho+\bar\nu]\,{\cal A}^i_{ y}[\bar\rho+\bar\nu]\,.\ee
Note the assignments of tilde's and bar's in the r.h.s.: This is
what we need since, first, $\tilde\rho$ is the
COV-gauge source at tree-level, and, second,
the classical solution corresponding to $\rho+\nu$ is the
{\it same} functional of $\bar\rho+\bar\nu$ as the tree-level
field in terms of $\tilde\rho$ (cf. eqs.~(\ref{Atilde})
and (\ref{Abar})). By also using $\tilde\rho=\bar\rho$
(cf. eq.~(\ref{tilderhobar})), the r.h.s.
can be rewritten as:
\be
\int {\cal D}\bar\rho\,\,W_\Lambda[\bar\rho]
\int {\cal D}\bar\nu\,{\cal W}[\bar\nu\,;\bar\rho]\,\,
{\cal A}^i_{ x}[\bar\rho+\bar\nu]\,{\cal A}^i_{ y}[\bar\rho+\bar\nu]\,,\ee
which leads to the following recurrency formula
(compare to eq.~(\ref{recurW})):
\be \labe{recurW1}
W_{b\Lambda}[\tilde\rho]\,=\,\int {\cal D}\bar\nu\,
W_\Lambda[\tilde\rho-\bar\nu]\,
{\cal W}[\bar\nu\,;\tilde\rho-\bar\nu]\,.\ee
Thus, in this framework,
it is natural to define the noise correlations directly for $\bar\nu$. 
We write (compare to eq.~(\ref{noise})):
\be\labe{noiseCOV}\,
\langle\bar\nu^a(\vec x)\rangle_{\nu}\,=\,
\bar\sigma^a_{cl} (\vec x),\qquad
\langle\bar\nu^a(\vec x)\bar\nu^b(\vec y)\rangle_{\nu}\,=\,
\bar \chi^{ab}_{cl}(\vec x,\vec y),\ee
which together with eq.~(\ref{recurW1}) leads to a RGE
for $W_\tau[\tilde\rho]$ with the same formal structure as 
eq.~(\ref{RGE}), but with $\rho \to \tilde\rho$,
$\chi\to \bar\chi_{cl}$, and $\sigma_{cl}\to \bar\sigma_{cl}$.
It just remains to specify the functions
$\bar\sigma_{cl}$ and $\bar\chi_{cl}$ in eq.~(\ref{noiseCOV}).

These can be obtained from the corresponding correlations in
the LC gauge, eq.~(\ref{noise}), by an appropriate gauge rotation.
We have:
\be\label{tildenubar}
\bar \nu\,=\,\bar U^{\dagger}\nu \bar U \,=\,
\delta U^{\dagger}\Bigl(U^{\dagger}\nu U\Bigr)\delta U
\,\equiv\,\delta U^{\dagger}\tilde\nu\,\delta U\,,\ee
with the correlations of $\tilde\nu$ following trivially from 
eq.~(\ref{noise}):
\be\label{noiseCOV1}\,
\langle\tilde\nu_a(\vec x)\rangle_{\nu}&=&
\tilde\sigma_a (\vec x)\,\equiv\,
 U^{\dagger}_{ab}(\vec x)\,\sigma_b(\vec x),
\nn
\langle\tilde\nu_a(\vec x)\tilde\nu_b(\vec y)\rangle_{\nu}&=&
\tilde \chi_{ab}(\vec x,\vec y)\,\equiv\,
U^{\dagger}_{ac}(\vec x)\,
\chi_{cd}(\vec x,\vec y)\,U_{d b}(\vec y).\ee
But the relation between $\bar \nu$ and $\tilde\nu$ is non-linear,
since the ``small'' gauge rotation $\delta U^{\dagger}$ is itself
dependent on $\bar \nu$. To the order of interest,
one can expand 
eq.~(\ref{tildenubar}) to quadratic order in $\tilde\nu$, to
obtain (cf. eq.~(\ref{expU})) :
\be
\bar \nu\,\approx\,\tilde\nu\,+\,ig\Bigl
[\theta(x^--x^-_\tau)\delta {\alpha}(x_{\perp}),\,\tilde\nu\Bigr]\,=\,
\tilde\nu\,+\,g\,{i\over 2}\,[\delta 
{\alpha}(x_{\perp}),\,\tilde\nu],\ee
where $\delta {\alpha}$
is to be taken to linear order in $\tilde\nu$, and
the factor $1/2$ occurs since, as $b\to 1$, $\tilde\nu(\vec x)\approx
\delta(x^--x^-_\tau)\tilde\nu(x_{\perp})$ and $\theta(x)\delta(x)
= (1/2)\delta(x)$.
By using the equations above, and after simple algebra, one
finally obtains
\be\labe{NUCOV}\,
\bar\sigma^a_{cl}\,=\,\tilde\sigma^a - \delta\bar\sigma^a_{cl},
\qquad \bar \chi^{ab}_{cl}\,=\,\tilde \chi^{ab},\\
\label{sigcl}
\delta\bar\sigma^a_{cl}(\vec x)\,\equiv\,
{g\over 2}\,f^{abc}\int d^3\vec y\,\,
\tilde\chi_{bc}(\vec x,\vec y)\,
\langle y_\perp|\,\frac{1}{\grad^2_\perp}\,|
x_\perp\rangle\,.\ee
The correction $-\delta\bar\sigma_{cl}$ is a ``counterterm''
which takes care of the spurious
classical polarization which appears in the covariant gauge (cf.
eq.~(\ref{LT}) below).
Note that $\delta\bar\sigma_{cl} \sim g\alpha_s\ln(1/b)\rho\rho$, which
in the saturation regime ($g\rho \sim 1$) is as large as $\sigma$.

To summarize, in addition to the straightforward gauge rotations
(\ref{noiseCOV1}), the RGE for $W_\tau[\tilde\rho]$ differs from
the corresponding equation (\ref{RGE}) for $W_\tau[\rho]$ also
by the correction (\ref{sigcl}) to the induced charge density.
This latter is the consequence of the quantum evolution of the 
``covariant gauge'' itself.

By exploiting the previous results, it is now straightforward
to derive explicit expressions for the classical
correlations generated by $\bar\nu$ (cf. eq.~(\ref{dAnn})),
to be then compared with the quantum correlations in Sect. 3.2.1.
This will allow us to understand the structure of these correlations
in more detail. To this aim, one needs the expansion of 
${\cal A}^i[\bar\rho+\bar\nu]$ to quadratic order in $\bar\nu$,
which reads (cf. eqs.~(\ref{Abar}) and (\ref{expU})):
\be\label{CLINMF}
{\cal A}^i[\bar\rho+\bar\nu]&=&{\cal A}^i[\tilde\rho] +
\delta A^i[\tilde\rho,\bar\nu],\nn
\delta A^i[\tilde\rho,\bar\nu] &=&
{i \over g}\,U\,\left({\rm e}^{-ig\delta {\alpha}}\,
\partial^i {\rm e}^{ig\delta {\alpha}}\right)U^{\dagger}
\,\equiv \,U\,\delta \tilde{A}^i \,U^{\dagger},\\
 \label{dtA}
 \delta \tilde { A}^i&\approx& -\,
 \partial^i\delta {\alpha}\,+\,g\,{i\over 2}\Bigl[\delta {\alpha},\,
\partial^i\delta {\alpha}\Bigr].\ee
Here, $\delta {\alpha}\equiv \theta(x^--x^-_\tau)
\delta {\alpha}(x_{\perp})$, so that the induced field
$\delta A^i$ has support only at positive and (relatively) large
$x^-$, $x^-\simge 1/\Lambda^+$ (compare with the original field
in eq.~(\ref{APM})). This property is the consequence of the fact 
that the induced source itself has support at $x^-\simge 1/\Lambda^+$,
together with the retarded boundary conditions used when solving
eq.~(\ref{cleqn}).

Note the relatively simple expression for the linear
term in eq.~(\ref{dtA}) for
 $\delta \tilde { A}^i$. This is the same as (cf. eq.~(\ref{dalpha}))
\be\label{LIN1}
\delta \tilde { A}^i(\vec x)\,\Big|_{linear}\,=\,\theta(x^--x^-_\tau)
\left(\frac{\partial^i}{\grad^2_\perp}\,\tilde\nu\right)(x_\perp)
\,=\,\int d^3\vec y\,\,
G^{i-}_0(\vec x,\vec y,p^-=0)\,\tilde\nu(\vec y),\ee
where $G^{i-}_0(p^-=0)=-p^i/(p^+p_\perp^2)$,
with $1/p^+\equiv 1/(p^++i\epsilon)\,$,
is the free LC-gauge propagator,
cf. eqs.~(\ref{aaimom}) and (\ref{LCPROP}).
A priori, the emergence of the {\it free} propagator might seem
surprising; one would rather expect the following
Green's function (with $p^-=0$ and
$G^{i-}[{\cal A}^i]$ as introduced in eq.~(\ref{JJSIG})) :
\be\label{LIN2}
\tilde G^{i-}(\vec x,\vec y)[\alpha]&\equiv&
U^{\dagger}(\vec x)\,G^{i-}(\vec x,\vec y)[{\cal A}^i]
\,U(\vec y),\ee
which is the propagator of the
LC-gauge fluctuations $\delta A^\mu$ in the presence of the
COV-gauge background field\footnote{Note that the fluctuations 
$\delta A^\mu_a$ transform {\it homogeneously} under the gauge rotations
of the background field; see, e.g., eq.~(\ref{CLINMF}).}
$\tilde A^\mu=\delta^{\mu +}\alpha\/$.
In fact, eqs.~(\ref{LIN1}) and (\ref{LIN2}) are consistent
each other: indeed, $\alpha(\vec x)$ is localized near
$x^-=0$, while the integral over $y^-$
in eq.~(\ref{LIN1}) runs from $y^- =x^-_\tau >0$ to $x^->y^-$.
That is, the whole propagation takes place far outside the support
of $\alpha$, where $\tilde G^{i-}=G^{i-}_0$ indeed (cf. Sect. 6).

The correlations $\langle\delta{ A}^i\rangle_\nu$
and $\langle\delta{ A}^i\delta{ A}^i\rangle_\nu$ are 
simply gauge rotations of the corresponding correlations
of $\delta \tilde{A}^i$ (cf. eq.~(\ref{CLINMF})).
To the order of interest, one immediately obtains:
\be\label{sigcl3}
\Bigl\langle\delta \tilde{A}^i_a(\vec x)\delta \tilde{A}^j_b(\vec y)
\Bigr\rangle_{\nu} &=&\int d^3\vec z\int d^3\vec u\,
\langle x_\perp|\,\frac{\partial^i}{\grad^2_\perp}\,|
z_\perp\rangle\,\tilde\chi_{ab}(\vec z,\vec u)\,
\langle u_\perp|\,\frac{\partial^j}{\grad^2_\perp}\,|
y_\perp\rangle\nn
&=&\Bigl(G^{i-}_0\,\tilde\chi\,G^{-j}_0\Bigr)_{ab}(\vec x,\vec y),\ee
which, as expected, is the (gauge rotation of the) quantum 
correlator (\ref{GSIG1}).

The one-point function $\langle\delta \tilde{ A}^i\rangle_{\nu}$
is more interesting. It reads (for $x^- > x^-_\tau$) :
\be\label{CLOP}
\langle\delta \tilde{ A}^i\rangle_{\nu}\,=\,\,
\frac{\partial^i}{\grad^2_\perp}\,\bar\sigma_{cl}\,+
\,g\,{i\over 2}\left\langle\Bigl[\delta {\alpha},\,
\partial^i\delta {\alpha}\Bigr]\right\rangle_{\nu}\,.\ee
where the second term in the r.h.s. is proportional to $\tilde \chi$.
After simple algebra, one finds
\be \label{LT}
{i\over 2}\left\langle\Bigl[\delta {\alpha},\,
\partial^i\delta {\alpha}\Bigr]\right\rangle_{\nu}\,=\,
\frac{\partial^i}{\grad^2_\perp}\,\delta\bar\sigma_{cl}\,+\,
\frac{1}{\grad^2_\perp}\,\delta\tilde{\cal J}^i,\ee
where $\delta\bar\sigma_{cl}$ is defined in eq.~(\ref{sigcl}), and
the vector current
\be\label{calJi}
\delta\tilde{\cal J}^i_a\,\equiv\,{1\over 2}\,f^{abc}(\partial^i\partial^j
-\delta^{ij}\grad^2_\perp)\langle\delta {\alpha}_b\,
\partial^j\delta {\alpha}_c\rangle_{\nu}\,=\,f^{abc}
\Bigl\langle\Bigl((\partial^i\partial^j
-\delta^{ij}\grad^2_\perp)\delta {\alpha}_b\Bigr)\,
\partial^j\delta {\alpha}_c\Bigr\rangle_{\nu}\ee
is transverse, $\partial^i \delta\tilde{\cal J}^i_a =0$,
as required by gauge symmetry. (In the background field LC gauge, 
this becomes ${\cal D}^i \delta{\cal J}^i=0$, with ${\cal D}^i=
\partial^i -ig{\cal A}^i$.) By the matching conditions, this is
the same as the current 
(\ref{delcalJ}) generated by soft fluctuations in Theory I.
In particular, we note that $\delta\tilde{\cal J}^+_a=0$ :
no charge density is induced by these fluctuations\footnote{This
could be also verified via a direct calculation of
$\delta{\cal J}^+_a=gf^{abc}\langle(\partial^+ \delta A^i_b)
\delta A^i_c\rangle$ (cf. eq.~(\ref{rho2})).}.

The induced mean field reads finally
(cf. eqs.~(\ref{CLOP}), (\ref{LT}) and (\ref{NUCOV}))
\be\label{CLMF}
\langle\delta \tilde{ A}^i\rangle_{\nu}\,=\,\,
\frac{\partial^i}{\grad^2_\perp}\,\tilde\sigma\,+\,
\frac{1}{\grad^2_\perp}\,\delta\tilde{\cal J}^i\,\equiv\,
G^{i\nu}_0\tilde {\cal J}_\nu,\ee
and coincides, as expected, with the (gauge rotation of the)
quantum correction $\delta{\cal A}^i$, eq.~(\ref{deltaCA}).

Note that, unlike the tree-level field ${\cal A}^i$, 
the total mean field including quantum corrections,
${\cal A}^i +\delta {\cal A}^i$,
is not a ``pure gauge'' anylonger (because the current
 ${\cal J}_\nu$ has non-trivial
transverse components ${\cal J}_i$, cf. eq.~(\ref{calJi})).
Rather, it is the {\it fluctuating} field
${\cal A}^i[\bar\rho+\bar\nu]$, i.e. the solution to the
classical equations (\ref{cleqn}) for a {\it fixed}
value of $\bar\nu$, which is a pure gauge (cf. eq.~(\ref{Abar})).
This is as expected, since, in the classical effective
theory defined by $\tilde\rho$, it is the solution
${\cal A}^i[\tilde\rho]$ at {\it fixed} $\tilde\rho$
which has this property.

\subsection{RG evolution in $p^-$}

Previously, the ``semi-fast gluons'' (i.e., the
quantum fluctuations to be integrated over in one step of the
renormalization group analysis) have
been defined as the fluctuations with longitudinal momenta inside 
the strip $b\Lambda^+\le |p^+|\le \Lambda^+$. 
But we have also mentioned that the leading logarithmic effects 
involve nearly on-shell quanta ($2p^+p^-\approx p_\perp^2$), 
for which the strip restriction on $p^+$
implies a similar restriction on $p^-$:
\be\labe{strip-}\,
\Lambda^- \,\simle\, |p^-| \,\simle\, \Lambda^-/b\,,\ee
with $\Lambda^-\equiv Q_\perp^2/2\Lambda^+$, and generic $Q_\perp$
(cf. eq.~(\ref{strips})). Thus, in practical calculations,
the strip restriction can be imposed either on $p^+$, or 
on $p^-$.

In Sect. 5, we shall perform calculations in the weak source
limit and show that, to LLA, both strip restrictions lead
indeed to the same results (namely, to the BFKL equation).
For general, non-linear, calculations however, it turns
out that it is more convenient to work with a strip 
restriction on $p^-$. The reason is that the tree-level 
fields and sources are independent of time, so any
constraint on the energy of the quantum fields will be
 automatically preserved during their propagation.
By contrast, the inhomogeneity of the background fields 
in $x^-$ makes it difficult to maintain a similar restriction on $p^+$.

To be more specific, let us note
that the key ingredient in the computation of
$\sigma$ and $\chi$ is the 2-point function 
$\langle a^\mu(x) a^\nu(y)\rangle$ of the semi-fast gluons 
in the background of the tree-level fields and sources
${\cal A}^i$ and $\rho$. For instance, eqs.~(\ref{deltaJ}) and 
(\ref{JIND}) imply:
\be\labe{JIND1}\,
\hat\sigma_a(\vec x)\,=\,-\,\frac{1}{2}\,
\frac{\delta^3 S}{\delta A^-_a(x)\delta A^\mu_b(y)
\delta A^\nu_c(z)}\bigg|_{
{\cal A}}\,\langle a^\mu_b(y)a^\nu_c(z)\rangle\,.\ee
As we shall see in Sect. 4, this 2-point 
function can be obtained from the following time-ordered 
(or Feynman) propagator:
\be\labe{delAcorr}
iG^{\mu\nu}_{ab}(x,y)[{\cal A},\,\rho]&\equiv&
\langle {\rm T}\,a_a^\mu(x) a_b^\nu(y)\rangle\nn&=&
Z_{b\Lambda\to\Lambda}^{-1}\int_{b\Lambda}^{\Lambda}
{\cal D}a\,\delta(a^+)\,\,a_a^\mu(x)a_b^\nu(y)\,
{\rm e}^{\,iS_0[{\cal A},a]}\,,\ee
where, as compared to eq.~(\ref{PARTQUANT}), the functional
integral is now restricted to semi-fast fluctuations,
and the action is evaluated in the Gaussian approximation:
\be\labe{SEXP2}
S[{\cal A}+a,\,\rho]\,\approx\,\frac{1}{2}\int d^4x
\int d^4y\,\,a^\mu_a(x)\,
\frac{\delta^2 S}{\delta A^\mu_a(x)\delta A^\nu_b(y)}\bigg|_{
{\cal A}}\,a_b^\nu(y)\,\equiv\,S_0[{\cal A},a]\,.\ee
[The first two terms in this expansion vanish since ${\cal A}^i$
is a solution of the classical equations of motion 
$\delta S/\delta A^\mu=0$, and, moreover, the action itself
vanishes on this solution: $S[{\cal A},\,\rho]=0\,$; cf. Sect. 4.1.]
Thus, this propagator is obtained by inverting the
differential operator (\ref{invG}) in the LC gauge, and
in the subspace of fields with momenta restricted to the strip.

For sufficiently weak fields and sources, the propagator
can be computed via an expansion in powers of ${\cal A}^i$ and 
$\rho$. In particular, when $\rho\to 0$, and thus ${\cal A}^i\to 0$, 
it must reduce to the free propagator 
given in eq.~(\ref{LCPROP}) below.
But in the saturation regime, where the background fields are
as  strong as $\rho\sim 1/g$ and ${\cal A}^i\sim 1/g$, this
propagator must be known {\it exactly}
(i.e., to all orders in ${\cal A}^i$ and $\rho$).
In that case, it is easier to work with a strip restriction on
$p^-\,$: then, the $p^+$ momenta are unrestricted,  which makes it
convenient to use the $x^-$-representation and thus 
exploit the specific geometry of the problem (namely, the fact 
that the source $\rho$ is localized near $x^-=0$) in order
to construct the propagator. This construction will be
presented in Sect. 6.

To conclude, from now on we shall impose the strip restriction
on $p^-$, cf. eq.~(\ref{strip-}).
This is consistent with the {\it background-field} gauge symmetry, 
since all the gauge transformations of the background fields
to be performed (like the rotations between the LC-gauge 
and the COV-gauge) are time-independent, and thus
cannot change $p^-$. But this leaves the place for a potential
ambiguity in the {\it quantum} gauge condition, as associated
with the axial pole at $p^+=0$ in the LC-gauge propagator.
That is, when inverting eq.~(\ref{invG}), one has to specify 
not only the $i\epsilon$ prescription at the physical pole
for on-shell excitations (here, the Feynman prescription),
but also the prescription for the unphysical pole in $1/p^+$.
The latter is, however, fixed by our choice of the
boundary conditions for the classical solution in Sect. 2.3,
cf. eq.~(\ref{Atilde}):
For this to be consistent with the quantum evolution,
one must use the ``retarded'' prescription $1/(p^++i\epsilon)$
in the $G^{i-}$ component of the propagator (see the discussion
after eq.~(\ref{GFIX})).
The prescriptions in the other components then
follow by hermiticity. 

In particular, the free propagator  reads as follows:
\be\label{LCPROP}
G_0^{i-}(p)&=&{p^i\over p^++i\epsilon}\,G_0(p),\quad
G_0^{-i}(p)={p^i\over p^+-i\epsilon}\,G_0(p),\nn
G_0^{ij}(p)&=&\delta^{ij}G_0(p),\qquad
G_0^{--}(p)=\,{\rm PV}\, {2p^-\over p^+}\,G_0(p)\,,\ee
where $G_0(p)=1/(2p^+p^--p_\perp^2+i\epsilon)$, and PV denotes
the principal value prescription:\be
{\rm PV}\ {1 \over p^+} \equiv {1 \over 2}
\left ( 
{1 \over p^+\ -i \varepsilon} + {1 \over p^+\ +i \varepsilon}
\right ) \,.
\labe{PV}\ee
Other prescriptions to be later referred to
are the ``advanced'' prescription previously used by
Mueller and Qiu \cite{MQ}, and by
Kovchegov and Mueller \cite{K96,KM98} --- this is obtained by changing
the sign of $i\epsilon$ for the axial poles 
in eq.~(\ref{LCPROP}) ---, and the
PV-prescription, for which $1/p^+ \equiv {\rm PV}(1/p^+)$
in all the components of the propagator.

\setcounter{equation}{0}
\section{Quantum dynamics on a complex-time contour }

In this section, we  describe the action $S[A,\,\rho]$ which
enters the quantum McLerran-Venugopalan model,
eq.~(\ref{PART}), and derive Feynman rules for the
computation of $\sigma$ and $\chi$.
Because of the temporal non-locality of the classical equations
of motion (\ref{cleq1}), this action, and the
associated quantum dynamics, must be formulated along a
contour in the complex-time plane.
It turns out, however, that in the approximations of interest
the contour structure is not essential, and one can restrict
oneself to the dynamics in real time, as done a priori
in Refs. \cite{JKLW97,JKLW99a}.

\subsection{The contour action}

By definition, the action $S[A,\,\rho]$ must be such as
to reproduce the classical equations of motion
(\ref{cleq1}) at the tree-level. Thus, 
$S[A,\,\rho]$ has the following structure:
\be\labe{ACTION0}
S[A,\rho]\,=\,- \int d^4x \,{1 \over 4} F_{\mu\nu}^a F^{\mu\nu}_a
\,+\,S_W[A^-,\rho]\,,
\ee
with $S_W[A^-,\rho]$ constrained by (cf. eqs.~(\ref{JTIME}) and
(\ref{WLINE1}))
\be\labe{SWJ}
\frac{\delta S_W}{\delta A^-_a(x)}=-\,J^+_a(x)\,\equiv
\,-\,
\frac{1}{N_c}\,{\rm Tr}\left\{\rho(\vec x)
W^\dagger(x)\,T^a\,W(x)\right\}.\ee
However, with the retarded current $J^+(x)$ in
eqs.~(\ref{JTIME})--(\ref{WLINE1}), 
the constraint (\ref{SWJ}) cannot be satisfied,
for any action. For instance, 
eq.~(\ref{SWJ}) implies the following symmetry property,
or {\it Maxwell relation} :
\be
\frac{\delta J^+_a(x)}{\delta A^-_b(y)}
\,=\,\frac{-\delta^2 S_W}{\delta A^-_a(x) \delta A^-_b(y)}
\,=\,\frac{\delta J^+_b(y)}{\delta A^-_a(x)},\ee
which is however inconsistent with the current (\ref{JTIME}), 
which rather yields
\be
\frac{\delta J^+_a(x)}{\delta A^-_b(y)}\bigg |_{A^-=0}&=&
\frac{ig}{N_c}\,\theta(x^+-y^+)\,\delta^{(3)}(\vec x-\vec y)\,
{\rm Tr}\,\Bigl\{\rho(\vec x)[T^a,T^b]\Bigr\}\nonumber\\
&=&-gf^{abc}\rho_c(\vec x)\,\theta(x^+-y^+)\,
\delta^{(3)}(\vec x-\vec y)
\,\equiv\,\Pi_{ab}^{R}(x,y).\ee
That is, the two-point polarization function $\Pi^R$,
and all the higher vertices generated by
$J^+(x)$, are necessarily retarded with respect to $x^+$,
while the corresponding amplitudes generated by $S_W$ should be
rather symmetrical (e.g., $\Pi_{ab}(x,y)= \Pi_{ba}(y,x)$).

Thus,  the non-local equation
(\ref{cleq1}) cannot be generated from an action.
This reflects the fact that this is an {\it effective}
equation, obtained after some degrees of freedom
(the fast partons) have been integrated out,
with retarded boundary conditions, to generate the non-local
current in the r.h.s. There is a standard procedure to circumvent
this difficulty and construct a quantum version of the non-local
theory (see, e.g., Ref. \cite{BI00} and Refs. therein):
This involves a generalization of eq.~(\ref{cleq1}) to complex
time variables which take values along the Schwinger-Keldysh 
contour depicted in Fig.~\ref{CONT}. 

\begin{figure}
\protect \epsfxsize=16.cm{\centerline{\epsfbox{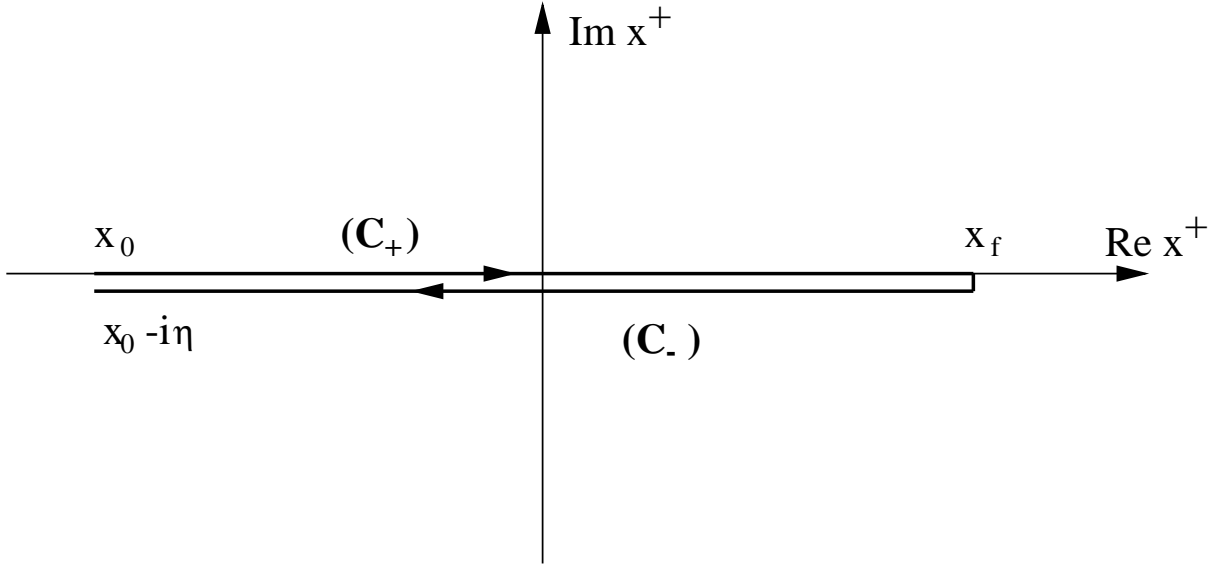}}}
         \caption{Complex-time contour for the quantum MV model:
 $C=C_+\cup C_-$.}
\labe{CONT}
\end{figure}

We call $z$ the (complex) time variable along the contour, 
and reserve the notation $x^+$ for real times.
The contour $C$ may be seen as the juxtaposition of two pieces:
$C=C_+\cup C_-$. On $C_+$, $z=x^+$ takes all the
real values between $x^+_0$ and $x^+_f$ 
(eventually, we let
$x^+_0 \to -\infty$ and  $x^+_f\to \infty$).
On  $C_-$, $z=x^+-i\eta$, with $\eta\to 0_+$
and $x^+$ running backward from $x^+_f$ to $x^+_0$.
We define a contour $\theta$-function $\theta_C\,$:
$\theta_C(z_1,z_2)=1$ if $z_1$ is further than $z_2$ 
along the contour (we write then $z_1 \succ z_2$),
while $\theta_C(z_1,z_2)=0$ if the opposite situation holds
($z_1 \prec z_2$).
We can formalize this by introducing a real parameter $u$ which
is continously increasing along the contour; then, the contour $C$
is specified by a function $z(u)$, and $\theta_C(z_1,z_2)=
\theta(u_1-u_2)$. The contour $\delta$-function is defined by:
\be
\delta_C(z_1,z_2)\equiv \left(\frac {\del z}{\del u}\right)^{-1}
\delta(u_1-u_2).\ee
The quantum extension of the MV model is a quantum
field theory living on this contour. Specifically, we allow the
quantum fluctuations of the fields $A^\mu$ to take arbitrary 
values on both sides of the contour,
and define a contour ``evolution operator'' by (for $z_2 \succ z_1$)
\be\labe{WLC0} 
W_{z_2,z_1}(\vec x)\,\equiv\,{\rm T}_C\, \exp\left\{\,
ig\int_{z_1}^{z_2} dz A^-(z, \vec x) \right\},
\ee
where $\int_{z_1}^{z_2}$ is the integral running along $C$ from $z_1$
to $z_2$, and the operator ${\rm T}_C$
orders the color matrices $A^-(z)$ from right to left in 
increasing sequence of their $u$ arguments. (Note
that the ordering along the contour coincides 
with the chronological ordering on $C_+$, and with
antichronological ordering on $C_-$.)
In particular, for $z_1=x^+_0$ and $z_2=x^+_0-i\eta$ 
(the end points of the contour),
we write:
\be\labe{WLC} 
W_C(\vec x)\,\equiv\,{\rm T}_C\, \exp\left\{\,
ig\int_C dz A^-(z, \vec x) \right\},
\ee
with $\int_C$ running all the way along the contour.

We are now in a position to present the quantum action
 for the non-local effective theory (below, 
$\int_C d^4x \equiv \int_C dz\int d^3 \vec x$) :
\be\label{ACTION}
S[A,\rho]\,=\,- \int_C d^4x \,{1 \over 4} F_{\mu\nu}^a F^{\mu\nu}_a
\,+\,{i \over {gN_c}} \int d^3 \vec x\, {\rm Tr}\,\Bigl\{ \rho(\vec x)
\,W_C[A^-](\vec x)\Bigr\}\,\equiv\,S_{YM}\,+\,S_W.\,\,
\ee
This is real, as it should, since the field $A^-$ in the Wilson line
(\ref{WLC}) is in the adjoint representation: $A^-\equiv A^-_aT^a$.
In the saddle point approximation, eq.~(\ref{ACTION})
generates eq.~(\ref{cleq1}), as we verify now. 
Note first that (with $x^+_0 \to -\infty$) :
\be\labe{SWJ1}
\frac{\delta S_W}{\delta A^-_a(z,\vec x)}\,=\,-\,
\frac{1}{N_c}\,{\rm Tr}\Bigl\{\rho(\vec x)
W_{-\infty-i\eta,z}(\vec x)\,T^a\,
W_{z,-\infty}(\vec x)\Bigr\}.\ee
The tree-level field $A^\mu(z)$ (the solution to
$\delta S/\delta A^\mu(z)=0$) takes identical values
on both sides of the contour:
$A^\mu_a(x^+) = A^\mu_a(x^+-i\eta)$ for any real $x^+$.
For such a field, eq.~(\ref{SWJ1})
reduces to the current $J^+(x)$ in eq.~(\ref{SWJ}),
for $z$ irrespectively on $C_+$ or on $C_-$. To verify this,
consider, e.g., $z=x^+\in C_+$; then,
$W_{x^+,-\infty}=W(x^+)$ (cf. eq.~(\ref{WLINE1})), and
\be
W_{-\infty-i\eta,x^+}\,=\,W_{-\infty-i\eta,x^+-i\eta}\,
W_{x^+-i\eta,\infty-i\eta}\,W_{\infty,x^+},\ee
where we have used the group property
of the contour operators (\ref{WLC0}). 
Since the tree-level field $A^-$ is the same on both sides of the
contour, the forward evolution in time from
$x^+$ to $\infty$ is compensated by the backward 
evolution in time from $\infty-i\eta$ to $x^+-i\eta\,$:
$W_{x^+-i\eta,\infty-i\eta}\,W_{\infty,x^+}=1$.
Similarly (with ${\tilde {\rm T}}$ denoting anti-chronological
ordering),
\be
W_{-\infty-i\eta,x^+-i\eta}\,=\,W^\dagger(x^+)\,
\equiv\,{\tilde {\rm T}}\, \exp\left\{\,
-ig\int_{x^+_0}^{x^+} dz^+ A^-(z^+, \vec x) \right\}.
\ee
Thus, at tree-level, and for $z=x^+\in C_+$,
$W_{-\infty-i\eta,z}\,T^a\, W_{z,-\infty}
\to W^\dagger(x^+)\,T^a\,W(x^+)$, in agreement
with eq.~(\ref{SWJ}). A similar conclusion holds for
$z=x^+-i\eta\in C_-$. In both cases, the saddle point equation
$\delta S/\delta A^\mu(z)=0$ turns out to be equivalent to
eq.~(\ref{cleq1}).

\begin{figure}
\protect\epsfxsize=8.cm{\centerline{\epsfbox{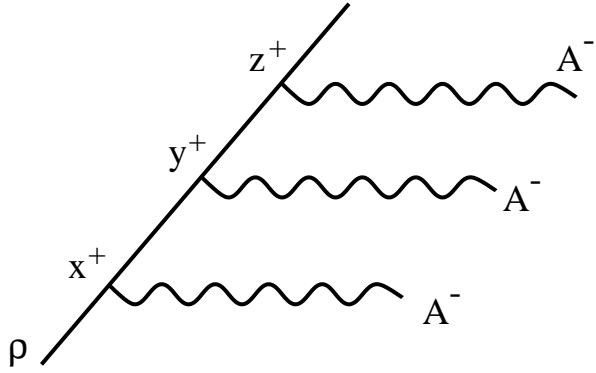}}}
         \caption{A typical $n$-point vertex (here, $n=3$)
generated by the action $S_W$. This is non-local in time, but local
in the spatial coordinates. The continuous line represents the
source $\rho$, while the wavy lines are $A^-$ gluon fields.}
\label{SWvertex}
\end{figure}

The non-local piece of the action $S_W$ 
describes the propagation of the color source $\rho$
($=$ the fast degrees of freedom of the hadron) at the
speed of light, in the background of the soft fields
$A^\mu$ (the classical field radiated by $\rho$ plus the
soft quantum fluctuations), and in the eikonal approximation.
This gives rise to vertices of the type 
$\rho(A^-)^n$, $n\ge  1$, which are
non-local in time (see Fig. \ref{SWvertex}).


$S_W$ is invariant under the {\it periodic} gauge
transformations $\Omega(x)\in {\rm SU}(N)$ which satisfy
\be \labe{periodic}
\Omega(-\infty-i\eta,\vec x)\,=\,\Omega(-\infty,\vec x) \ee
for any $\vec x$. Indeed, under such a transformation,
\be\labe{gtr1}
A^\mu(x)&\longrightarrow& 
\Omega(x)\Bigl(A^\mu +
(i/g)\partial^\mu\Bigr)\Omega^{\dagger}(x)\,,\nonumber\\
\rho(\vec x)&\longrightarrow& 
\Omega(-\infty,\vec x) \,\rho(\vec x)\,
\Omega^{\dagger}(-\infty,\vec x)
\,,\nonumber\\
W_C(\vec x)&\longrightarrow& \Omega(-\infty-i\eta,\vec x)
\,W_C(\vec x) \,\Omega^{\dagger}(-\infty,\vec x),\ee
and $S_W$ is invariant because of the property (\ref{periodic}).
Here, we are mostly interested in time-independent gauge
transformations, which preserve the static character of the 
background fields, and for which the periodicity condition 
(\ref{periodic}) is trivially satisfied.

The transformation law for $\rho(\vec x)$ in eq.~(\ref{gtr1})
reflects
its interpretation as the initial value of the color charge
$J^+$ at $x^+_0=-\infty$ (cf. Sect. 2.3)). 
In fact, $S_W$ can be also written as 
follows\footnote{To see this, use the identity
$i\del_zW_{z,-\infty}=gA^-(z)W_{z,-\infty}$ to 
perform the integral over $z$ in eq.~(\ref{SW2}).} :
\be\labe{SW2}
S_W\,=\,-\,{1\over {N_c}} \int_C d^4x\, {\rm Tr}\,\Bigl\{A^-(z,\vec x)
W_{z,-\infty}(\vec x) \rho(\vec x)\Bigr\},\ee
which emphasizes the fact that this is a gauge-invariant
generalization of the linear vertex $\int d^4x \,\rho_a A^-_a$.

\subsection{Contour Green's functions}

The generating functional for contour Green's functions
is defined as in eq.~(\ref{PARTQUANT}) where the time 
variables now run along the contour, and the external source
is extended to a function $j^\mu_a(z,\vec x)$ on $C$.
For instance, the (connected) contour two-point function 
is obtained as (with  color indices omitted):
\be\labe{QcorrC}
iG_C^{\mu\nu}(z_1,z_2)&\equiv&\langle {\rm T}_C\,
A^\mu(z_1)A^\nu(z_2)\rangle_c\,=\,-\,\frac{\delta^2 \ln Z_{\Lambda}
[\rho,\,j]}
{\delta j_\mu(z_1)\delta j_\nu(z_2)}\bigg|_{j=0}\nonumber\\
&=&\theta_C(z_1,z_2) G^{>\,\mu\nu}(z_1,z_2)\,+\,
\theta_C(z_2,z_1) G^{<\,\mu\nu}(z_1,z_2),\ee
where, to lighten the notations, we have indicated only
the time variables. In the second line above, we
have introduced the two-point correlation, or {\it Wightman},
functions:
\be\labe{G><}
G^{>\,\mu\nu}_{ab}
(z_1,z_2)\equiv\langle A^\mu_a(z_1)A^\nu_b(z_2)\rangle_c,\qquad
G^{<\,\mu\nu}_{ab}
(z_1,z_2)\equiv\langle A_b^\nu(z_2)A_a^\mu(z_1)\rangle_c,\ee
which, unlike the contour propagator $G_C^{\mu\nu}(z_1,z_2)$,
are continuous at $z_1=z_2$. 
These are the functions which enter the quantum corrections
$\sigma$ and $\chi$; for instance,
the contribution of eq.~(\ref{drhoYM}) to $\sigma$ reads
(see also Sect. 4.3 below):
\be\label{sigmaYM}
\sigma_a (x)\big|_{YM}
\,=\,g f^{abc}\partial^+_y\langle a_{b}^{i}(y)a_{c}^{i}(x)\rangle
\big|_{y=x}
\,=\,\,g f^{abc}\partial^+_yG^{<\,ii}_{cb}(x,y)\big|_{y=x}\,
.\ee
For real time variables, the
functions $G^>$ and $G^<$  can be used to construct the 
retarded $(G_R)$ and
advanced $(G_A)$ propagators, which will also be needed:
\be\labe{A}
G_R(x,y)&\equiv &
i\theta(x^+-y^+)\Bigl[G^>(x,y)\,-\,G^<(x,y)\Bigr]\,,\nonumber\\
G_A(x,y)&\equiv & -
i\theta(y^+-x^+)\Bigl[G^>(x,y)\,-\,G^<(x,y)\Bigr]\,.\ee

As a simple example, let us construct the contour propagator
for a free scalar field (the generalization to a free gluon field
in the LC gauge is straightforward). For both time arguments
$x^+$ and $y^+$ on $C_+$, this coincides with the usual Feynman
propagator, which reads (with $\vec p=(p^+,{\bf p}_\perp)$) :
\be\labe{GF0}
iG_0(x^+-y^+,\vec p)&=&\int {dp^- \over 2 \pi}\ 
e^{-ip^-(x^+-y^+)} \,\frac{i}{2p^+p^- - p_\perp^2 +i\epsilon}
\\&=&{1\over 2 p^+}\
\left \{\theta(p^+) \theta(x^+-y^+) - \theta(-p^+) \theta(y^+-x^+)
\right \}
{\rm e}^{-i {p_{\perp}^2 \over 2p^+}(x^+-y^+)},\nonumber\ee
from which one can identify the free  Wightman functions:
\be
G^>_0(x^+,\vec p)\,=\,\frac{\theta(p^+)}{2 p^+}\,
{\rm e}^{-i {p_{\perp}^2 \over 2p^+}x^+},\qquad
G^<_0(x^+,\vec p)\,=\,-\,\frac{\theta(-p^+)}{2 p^+}\,
{\rm e}^{-i {p_{\perp}^2 \over 2p^+}x^+},\ee
or, in momentum space (with $p^2=2p^+p^- - p_\perp^2$),
\be\labe{G><0}
G^>_0(p)\,=\,2\pi\theta(p^+)\,\delta(p^2),\qquad
G^<_0(p)\,=\,2\pi\theta(-p^+)\,\delta(p^2),\ee
The free contour propagator $G^{\,0}_{C}$ is finally obtained
by inserting $G^>_0$ and $G^<_0$ in eq.~(\ref{QcorrC}).

Returning to the gauge theory with action (\ref{ACTION}), the
corresponding contour propagator $G_C$ satisfies the 
Dyson-Schwinger equation with time arguments on $C$:
\be \labe{DS0}
\int_C {\rm d}^4z\,\left\{
\frac{\delta^2 S}
{\delta A^\mu(x)\delta A^\nu(z)}\bigg|_{{\cal A}}\,+\,
\delta \Sigma_{\mu\nu}^C(x,z)\right\}G_C^{\nu\lambda}(z,y)\,=\,
\delta^\lambda_\mu\delta_C(x,y)
\,.\ee
Here, $\delta \Sigma_{\mu\nu}^C(x,y)$ is the (contour) self-energy,
and describes quantum corrections, and
\be
\frac{\delta^2 S}
{\delta A^\mu(x)\delta A^\nu(y)}\bigg|_{{\cal A}}
\,=\,{\cal D}_{\mu\nu}(x)\delta_C(x,y)
\,+\,\delta_{\mu -}\delta_{\nu -}\Pi^C(x,y),\ee
where we have denoted 
\be\labe{HATSIG}
{\cal D}_{\mu\nu}(x)\equiv\,g_{\mu\nu}{\cal D}^2
-{\cal D}_\mu {\cal D}_\nu- 2ig
{\cal F}_{\mu\nu}\,,\qquad
\Pi^C_{ab}(x,y)\equiv\,
\frac{\delta^2 S_W}{\delta A^-_a(x) \delta A^-_b(y)}
\bigg |_{{\cal A}}\,.\ee
In these equations, ${\cal D}^\mu=\del^\mu-ig
{\cal A}^\mu$, and ${\cal A}^\mu$ is the background field, to be
eventually identified with the classical field in
Sect. 2.3 (hence the notation). In particular, ${\cal A}^-=0$,
and a second differentiation in eq.~(\ref{SWJ1}) 
yields (with $x^+$, $y^+\in C$):
\be\labe{PIC}
\Pi_{ab}^C(x,y)
&=&-\,\frac{g}{2}\,f^{abc}\rho_c(\vec x)
\,\delta^{(3)}(\vec x-\vec y)\Bigl\{\theta_C(x^+,y^+)
-\theta_C(y^+,x^+)\Bigr\}\nn&\equiv&\theta_C(x^+,y^+)\Pi^>_{ab}(x,y)
\,+\,\theta_C(y^+,x^+)\Pi^<_{ab}(x,y),\ee
with time-independent $\Pi^>$ and $\Pi^<$ :
\be\labe{P><}
 \Pi^>_{ab}(x,y)\,=\,-(g/2)\,f^{abc}\rho_c(\vec x)
\,\delta^{(3)}(\vec x-\vec y)\,=\,-\Pi^<_{ab}(x,y)\,.\ee
The contour self-energy admits a similar decomposition:
\be\labe{Sig<>}
\delta \Sigma^C(z_1,z_2)\,=\,\theta_C(z_1,z_2) \delta \Sigma^>(z_1,z_2)
+\theta_C(z_2,z_1)\delta \Sigma^<(z_1,z_2).\ee
The Dyson-Schwinger equation is then conveniently rewritten as
\be \labe{D1}
{\cal D}_{\mu\nu}(x)
G_C^{\nu\lambda}(x,y)\,
+\int_C {\rm d}^4z\,\Sigma_{\mu\nu}^C(x,z) 
\,G_C^{\nu\lambda}(z,y)\,=\,\delta^\lambda_\mu\delta_C(x,y)\,,\ee
with the total self-energy $\Sigma_{\mu\nu}^C\equiv
\delta_{\mu -}\delta_{\nu -}\Pi^C+\delta \Sigma_{\mu\nu}^C$.

By choosing $x^+\in C_+$ and $y^+\in C_-$ in eq.~(\ref{D1}),
and using the decompositions (\ref{QcorrC})
and (\ref{Sig<>}),
we obtain, after simple manipulations, an
equation for $G^<(x,y)$ in {\it real} time :
\be\labe{KBYM1}
{\cal D}_\mu^{\,\,\,\rho}(x)
G^<_{\rho\nu}(x,y)\,=\,
\int {\rm d}^4z\,\bigl(
\Sigma_{R}\,G^<+\Sigma^<G_{A}\bigr)_{\mu\nu}(x,y).\ee
[The retarded ($\Sigma_R$) and advanced ($\Sigma_A$) self-energies
are defined  in terms of the Wightman self-energies
 $\Sigma^>$ and $\Sigma^<$ as in eq.~(\ref{A}).]
One similarly derives an equation for $G^>(x,y)$, as
well as the following equation for $G_R(x,y)$:
\be\labe{eqGR}
{\cal D}^{\mu}_{\,\,\rho}(x)
G^{\rho\nu}_R(x,y)
\,-\,\int {\rm d}^4z\,\bigl(
\Sigma_R\,G_R\bigr)^{\mu\nu}(x,y)\,=\,g^{\mu\nu}\delta^{(4)}(x-y)\,.\ee
The two equations above imply a relation
between $G^<$ and $\Sigma^<$ :
\be\labe{WRA}
G^<_{\mu\nu}(x,y)\,=\,\int {\rm d}^4z\,{\rm d}^4u\,\Bigl(
G_R(x,z)\,\Sigma^<(z,u)\,G_A(u,y)\Bigr)_{\mu\nu}
.\ee
A similar relation holds between $G^>$ and $\Sigma^>$.

Let us now apply this general formalism to the calculation
of the quantum corrections discussed in Sect. 3.2.1.
We then need the following contour Green's functions:

({\it a}) The two-point function
$G^{<\,\mu\nu}_{ab}(x,y)\equiv \langle \delta A^\nu_b(y)
\delta A^\mu_a(x)\rangle$ of the soft fields
induced by their coupling to the semi-fast gluons.
This is given by eq.~(\ref{WRA}) with $\Sigma^<$ replaced by
\be\labe{DELSIG}
\delta \Sigma_{\mu\nu}^<(x,y)\,=\,
\langle \delta J_\nu(y)\delta J_\mu(x)\rangle\,,\ee
where $\delta J_\mu^a$ is the quantum color current in eq.~(\ref{deltaJ}).
To lowest order in $\alpha_s\,$, the soft field propagators $G_R$
and $G_A$ in eq.~(\ref{WRA}) can be computed in the mean field
approximation; that is, one can neglect the quantum self-energy
$\delta \Sigma^{\mu\nu}_R$ when solving  eq.~(\ref{eqGR}). 
With these approximations,  eq.~(\ref{WRA}) reduces 
to eq.~(\ref{JJSIG}), which has been used in Sect. 3.2.1.

({\it b}) The background field propagator 
$iG_C^{\mu\nu}(x,y)\equiv \langle {\rm T}_C\,a
^\mu(x) a^\nu(y)\rangle$ of the semi-fast gluons
in the Gaussian, or mean field, approximation
(cf. eq.~(\ref{delAcorr})). 
This is given by eq.~(\ref{D1}) where the quantum self-energy
$\delta \Sigma^C_{\mu\nu}$ is neglected; that is,
\be \labe{D2}
{\cal D}_{\mu\nu}(x)
G_C^{\nu\lambda}(x,y)\,+\,\delta_{\mu -}\int_C {\rm d}^4z\,
\Pi^C(x,z)\,G_C^{-\lambda}(z,y)
\,=\,\delta^\lambda_\mu\delta_C(x,y)\,.\ee
Since the tree-level self-energy (\ref{PIC}) is non-local,
the integration over $z^+$ in this equation 
runs along the whole contour $C$. Thus, even for real time 
arguments  $x^+$ and $y^+$, eq.~(\ref{D2}) is still
sensitive to the complex side $C_-$ of the contour.
This is to be contrasted to
a {\it local} field theory, where the contour structure becomes
irrelevant in the mean field approximation \cite{BI00}.

However, a closer inspection reveals that, even for 
eq.~(\ref{D2}), the contour structure is {\it not} essential,
in the sense that one can ignore $C_-$ 
when computing the {Feynman} propagator
(i.e., the propagator $G_C^{\mu\nu}(x,y)$ with both time
arguments on $C_+$). Indeed, for $x^+$, $y^+ \in C_+$,
the contribution of $C_-$ to eq.~(\ref{D2}) reads, schematically,
\be
\int_{C_-} {\rm d}^4z\,\Pi^C(x,z)\,G_C(z,y)\,=\,-
\int{\rm d}^4z\,\Pi^<(x,z)\,G^>(z,y)&\propto&\,\nn\propto
\int {\rm d}z^+ \,G^>(z^+-y^+, \vec x, \vec y)\,=\,
G^>(p^-=0,\vec x, \vec y)&=&0,\ee
where we have used eqs.~(\ref{PIC})--(\ref{P><}),
the time-homogeneity of the problem (i.e., the fact that
$\Pi^<$ is independent of time, while
$G^>$ depends only upon the relative time $z^+-y^+$),
and the fact that $G^>(p^-=0)=0$ (since, by definition,
this has support only for $p^-$ in the strip (\ref{strip-})).

To conclude, the background field Feynman propagator 
$iG^{\mu\nu}(x,y)\equiv \langle {\rm T}\,a ^\mu(x) a^\nu(y)\rangle$ 
can be obtained by solving the following equation
\be \labe{D3}
{\cal D}_{\mu\nu}(x)
G^{\nu\lambda}(x,y)\,+\,\delta_{\mu -}\int{\rm d}^4z\,
\Pi(x,z)\,G^{-\lambda}(z,y)
\,=\,\delta^\lambda_\mu\delta^{(4)}(x-y)\,,\ee
where all the time variables are real, and $\Pi(x,z)$
is the restriction of $\Pi^C(x,z)$,  eq.~(\ref{PIC}),
to time arguments on $C_+$. Then, the corresponding
Wightman functions
(which enter $\sigma$ and $\chi\,$; see, e.g., eq.~(\ref{sigmaYM}))
can be extracted from the Feynman propagator by considering
appropriate time orderings, as 
we did for the free propagator in eqs.~(\ref{GF0})--(\ref{G><0}).

Note that any explicit use of the contour has been avoided in this way.
This has been possible because, first, of the static
character of the background, and, second, of the simple nature
of the present, mean field, approximations.
For more general problems, where higher-order quantum effects
should be computed in the presence of inhomogeneities in
time, a complete use of the contour would be generally unavoidable.

\subsection{Feynman rules for $\sigma$ and $\chi$}

As explained in the previous subsection, the use of the
contour techniques can be avoided for the calculation of
$\sigma$ and $\chi$.
It is then sufficient to consider the restriction of the quantum
theory to the real time axis, as defined by the following
action (compare to eq.~(\ref{ACTION})):
$S=S_{YM}+S_W$, with
\be\labe{ACTION1}
S_W\,=\,{i \over g{N_c}} \int d^3 \vec x\, {\rm Tr}\,\Bigl\{ \rho(\vec x)
\,W_{\infty, -\infty}[A^-](\vec x)\Bigr\}\,\ee
and the {\it real-time} Wilson line:
\be\labe{WLINE}
     W_{\infty,-\infty}[A^-](\vec x)\, =\,{\rm T}\, \exp\left[\,
ig\int dx^+ A^-(x) \right].
\ee
This is the original version of the quantum MV model
proposed in Ref. \cite{JKLW99a}.
In particular, the tree-level self-energy generated by this 
action reads:
\be\labe{R--}
\Pi^{ab}(x,y) &\equiv&
\frac{\delta^2 S_W}{\delta A^-_a(x) \delta A^-_b(y)}
\bigg |_{A^-=0} \,=\,-
\frac{g}{2}\,f^{abc}\rho_c(\vec x)
\,\epsilon(x^+-y^+)\,\delta^{(3)}(\vec x-\vec y)\nn
&=&g\rho^{ab} (\vec x)\,\delta^{(3)}(\vec x-\vec y)\,
\langle x^+ |{\rm PV}\,{1 \over i\partial^-} |y^+ \rangle\,,\ee
where $\rho^{ab} \equiv -if^{abc} \rho^c$, and 
$\epsilon(x)\equiv \theta(x)- \theta(-x)$. 
Eq.~(\ref{R--}) coincides, as it should, with the restriction of the
contour self-energy $\Pi^C$, eq.~(\ref{PIC}),
to the real time axis.
Note the emergence of the principal value prescription in $1/p^-$ 
(this is defined as in eq.~(\ref{PV})).

Since we are mainly interested in the strong
field regime, where ${\cal A}^i\sim 1/g$ and $\rho\sim 1/g$, 
it is convenient for power counting to perform a rescaling of the 
background fields and the color sources:
\be\labe{rescale}
{\cal A}_a^\mu \longrightarrow \frac{1}{g}\,{\cal A}_a^\mu,\qquad
\rho_a \longrightarrow \frac{1}{g}\,\rho_a, \qquad
\delta\rho_a \longrightarrow \frac{1}{g}\,\delta\rho_a\,.\ee
In the saturation regime, the new tree-level quantities
${\cal A}^\mu$ and $\rho$ are of order 1, while the new  
quantum corrections $\chi$ and $\sigma$ --- which are defined
in terms of the rescaled charge fluctuations $\delta\rho$
by the same equations as before (i.e., 
(\ref{CHIDEF}) and (\ref{JIND})) --- are of order
$\alpha_s$.
After this rescaling, the
coupling constant $g$ completely drops out from 
the classical analysis in Sect. 2. On the other hand, we prefer
not to rescale the quantum fluctuations $a^\mu\,$; thus,
their propagator  $G^{\mu\nu}(x,y)[{\cal A},\rho]$ remains of
order one, and all the factors of $g$ can be explicitly read
off the vertices (see below). 

The quantum charge fluctuations
$\delta \rho_a(x)$ are defined by eq.~(\ref{deltaJ})
with $\mu=-\,$.
After the rescaling (\ref{rescale}), this yields:
\begin{equation}\labe{delta12}
\delta \rho_a (x) =\delta \rho_a^{(1)} (x)+\delta \rho_a^{(2)} (x),
\end{equation}
where $\delta \rho^{(1)}$ is linear in $a^\mu$, and
$\delta \rho^{(2)}$ is quadratic:
\be\labe{rho10}
 \delta \rho_a^{(1)} (x) & = & 
-2i g{\cal F}^{+i}_{ac} (\vec x) a^{ic} (x) + \nonumber \\
& & +g\rho^{ac} (\vec x)
\int dy^+ \langle x^+ |{\rm PV}\,{1 \over i\partial^-} |y^+ \rangle
 a^{c-}(y^+,\vec x), \\
\delta \rho_a^{(2)}(x)& = & 
 g^2 f^{abc} [\partial^+ a^{b}_{i}(x)
  ]a^{c}_{i}(x) 
 \nonumber\\ &{-}& {{g^2}\over{N_c}} \,\rho^{b}({\vec x})
  \int dy^+ a^{-c}(y^+,{\vec x}) \int dz^+ a^{-d}(z^+,{\vec x})
\nonumber\\ &{}& \nonumber\,\,\,\, \times\,
  \biggl\{\theta (z^+ -y^+)
  \theta (y^+ -x^+) {\rm Tr} \,(T^a T^c T^d T^b)
\\ &{}&\nonumber 
  \qquad+\ \theta (x^+ -z^+) \theta (z^+ -y^+) {\rm Tr} \,(T^a T^b T^c T^d)
\\ &{}&\qquad
+\ \theta (z^+ -x^+) \theta (x^+ -y^+) {\rm Tr}\,(T^a T^d T^b T^c) \biggr\},
\labe{rho2}
\ee
where all the factors of $g$ are now explicit.
In the right hand sides of these equations, the terms involving
$a^i$ come from the three-gluon vertex in
$S_{YM}$, while the terms involving $a^-$
come from the two- and three-point vertices in $S_W$.

In eqs.~(\ref{rho10})--(\ref{rho2}) it is understood that only
the soft modes (with $k^+\simle b\Lambda^+$) are kept in the
products of fields. This has
been used to simplify the first contribution to $\delta \rho^{(1)}$ 
by writing:
\be
(\del^+ {\cal A}^i_{ac})\, a^{ic}\,-\,{\cal A}^i_{ac} \,\del^+a^{ic}
\,\approx\,2 {\cal F}^{+i}_{ac}a^{ic}\,.\ee
Thus, the above expressions can be used only for calculations
to LLA.

The quantum corrections $\chi$ and $\sigma$ are
given by eq.~(\ref{CHIDEF}) (with $x^+=y^+$) and (\ref{JIND}),
where the average over the semi-fast fluctuations $a^\mu$
is defined as in eqs.~(\ref{delAcorr})--(\ref{SEXP2}).
This generates Wightman functions of the fields
$a^\mu$ (see, e.g., eq.~(\ref{sigmaYM})), which are related 
to the corresponding time-ordered, or Feynman, propagator
as explained in Sect. 4.2. In particular, the equal time limit
of a two-point function is obtained as:
\be
\langle a_b^\nu(x^+) a_a^\mu
(x^+) \rangle\, \equiv \,
G^{<\,\mu\nu}_{\,\,\,ab}(x^+=y^+)\,=\,
iG^{\mu\nu}_{ab}(y^+=x^++\epsilon).\ee
In momentum space, this instructs us to perform the
integration over $p^-$ by closing the contour 
in the upper half of the complex $p^-$ plane:
\be \labe{prescripW} G^< (x^+,x^+)\,=\,i
\int {dp^-\over 2\pi} \,{\rm e}^{ip^-\epsilon}\,G(p^-)\,.\ee
In practice, this prescription will play a role only for the
vacuum piece $G_0$ (i.e., the limit of $G$ as $\rho\to 0$,
which to the order of interest coincides with the free propagator
$G_0$; cf. Sect. 3.4). Indeed, 
within the present approximations,
the induced piece $\delta G\equiv G-G_0$ is continuous at $x^+=y^+$,
so its equal-time limit is unambiguous.

\begin{figure}
\protect\epsfxsize=14.cm{\centerline{\epsfbox{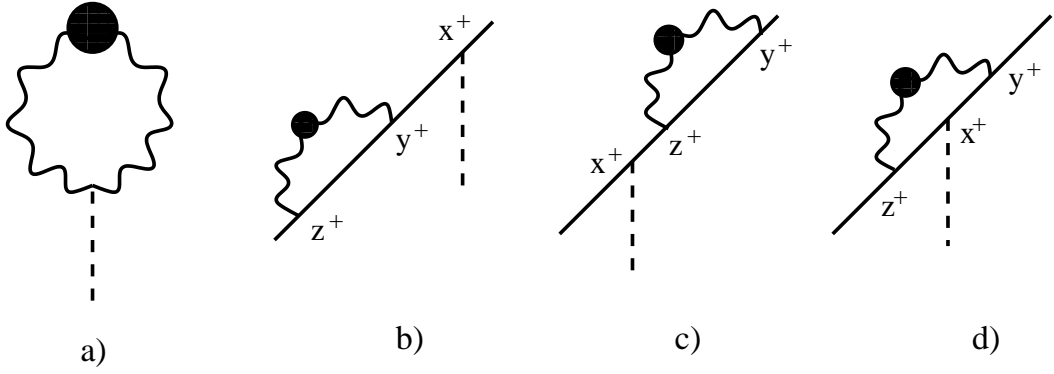}}}
         \caption{Feynman diagrams for $\hat\sigma_1$ (a) and
 $\hat\sigma_2$ (b,c,d). The wavy line with a blob denotes the
background field propagator of the semi-fast gluons;
the continuous line represents the source $\rho$; the precise
vertices can be read off eq.~(\ref{rho2}).}
\label{SIGFIG}
\end{figure}

We are now prepared to express $\sigma$ and $\chi$
in terms of the Feynman propagator $G^{\mu\nu}_{ab}(x,y)$ of the
semi-fast gluons. For $\sigma\equiv\langle\delta \rho\rangle
=\langle\delta \rho^{(2)}\rangle\/$, one obtains
(with the ``approximately equal'' sign denoting
 an equality which holds to LLA) :
\be\labe{sigma12}
\alpha_s\ln{1\over b}\,\sigma_a ({\bf x}_\perp)&\approx&
\int dx^- \, {\rm Tr } \,(T^a \hat\sigma(\vec x)),\nonumber\\
\hat\sigma(\vec x)&\equiv&-g^2 \partial^+_y G^{ii}\Big |_{x=y}\,
+\,ig^2\rho({\vec x})\Bigl\langle x\bigg|
 {1 \over i \partial^-} \,G^{--}  {1 \over i\partial^-}\bigg|x 
\Bigr\rangle\nn
&\equiv&\hat\sigma_1(\vec x)\,+\,\hat\sigma_2(\vec x).
\ee
In writing $\hat\sigma_2$ as above, we have used compact but
formal notations for the second, non-local, contribution
to $\delta \rho_a^{(2)}$, eq.~(\ref{rho2}).
A pictorial representation of the two pieces of $\hat\sigma$
is given in Fig. \ref{SIGFIG}.

\begin{figure}
\protect\epsfxsize=12.cm{\centerline{\epsfbox{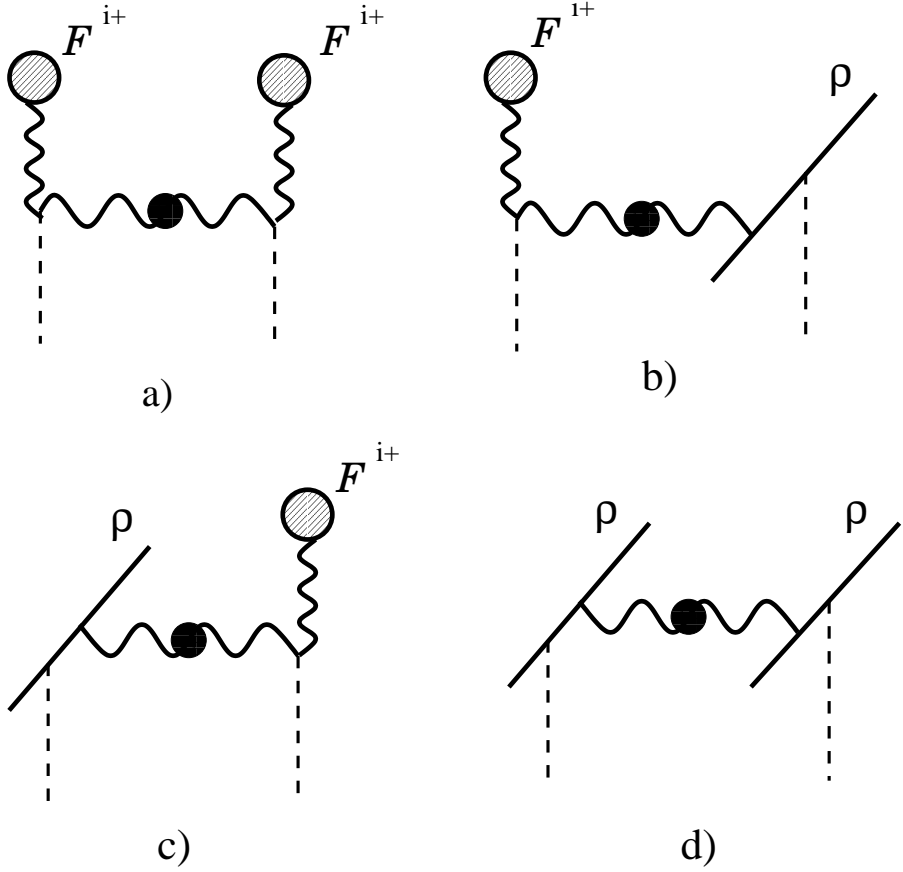}}}
         \caption{Feynman diagrams for the four contributions
to $\chi$ given in eq.~(\ref{chi2}).}
\label{CHIFIG}
\end{figure}

Concerning $\chi$, we note that to order $\alpha_s$
this involves $\delta \rho^{(1)}$, but not $\delta \rho^{(2)}$.
[This is clear by counting powers of $g$ according
to eqs.~(\ref{rho10}) and (\ref{rho2}).] Therefore
(with $y^+=x^++\epsilon$):
\be\labe{chi1}
\alpha_s\ln{1\over b}\,
\chi_{ab}({\bf x}_\perp,{\bf y}_\perp)\,\approx\,
\int dx^- \int dy^-\,\hat\chi_{ab}(\vec x,\vec y),
\qquad\qquad\qquad\nonumber
\\
\hat\chi_{ab}(\vec x,\vec y)\,\equiv\,g^2
\left\langle \left(-2i{\cal F}^{+i}
a^i + \,\rho {1 \over i\partial^-} a^- \right)_x^a
\left(
2i a^i {\cal F}^{+i} +a^- {1 \over 
i\partial^-} \rho \right)_y^b \right\rangle
\ee
where we have used also the symmetry property
\be
\langle x|{\rm PV}\, {1 \over i\partial^-} |y \rangle \, = \, -
\langle y|{\rm PV}\, {1 \over i\partial^-} |x \rangle.
\ee
This further yields, in matrix notations (the
PV prescription in $1/p^-$ is implicit) :
\be\labe{chi2}
\frac{1}{g^2}\,\hat \chi(\vec x, \vec y) &=&i\,
2{\cal F}^{+i}_x\, \langle x|G^{ij}|y\rangle \,2{\cal F}^{+j}_y \, +\, 
2{\cal F}^{+i}_x\,\langle x|G^{i-}\,{1 \over i\partial^-}|y\rangle \, \rho_y 
\nonumber\\&{}&\,\,\,\,-\,
\rho_x \, \langle x|{1 \over i\partial^-}\, G^{-i}|y\rangle \, 
2{\cal F}^{+i}_y\,+\,i
\rho_x \langle x|{1 \over i\partial^-} G^{--} {1 \over i\partial^-}
|y\rangle\,\rho_y.\ee
Diagramatically, all the above contributions to $\chi$ 
are represented by {\it tree-like} Feynman graphs (no loops),
with vertices proportional to $\rho$ or ${\cal F}^{+i}$ (see Fig. \ref{CHIFIG}).
Thus, like the tree-level source $\rho$ itself,
$\hat \chi(\vec x, \vec y)$ is localized near the LC, at
$0\simle x^-,y^- \simle 1/\Lambda^+$.

It is now easy to verify, by using eqs.~(\ref{delta12}) and (\ref{rho2})
and straightforward power counting, that all the $n$-point correlators
of $\delta\rho$ beyond $n=2$ are of higher order in $\alpha_s$.
Indeed, each factor of $a^\mu$ in these
equations is accompanied by a power of $g$. Then,
for instance, the three-point function $\langle \delta\rho\,
\delta\rho\,\delta\rho\rangle$ --- which involves, at least,
four fields $a^\mu$ --- is of order $\alpha_s^2$.
This explains why, to the present accuracy, we can restrict
ourselves to the one- and two-point functions in
eqs.~(\ref{sigma12}) and (\ref{chi2}).
The remaining part of this paper, and also Paper II,
will be devoted to their explicit evaluation.

\setcounter{equation}{0}
\section{The weak source approximation and BFKL}

As a simple illustration of the formalism
developed in the previous sections, let us explicitly
compute here some of the contributions to $\sigma$ and $\chi$
in the weak source ($\rho\ll 1$), or linear response, approximation.
We already know from Ref. \cite{JKLW97} that, in this approximation,
eq.~(\ref{RGE2p}) reduces to the BFKL equation. Our
intention here is not to repeat the derivation in Ref. \cite{JKLW97},
but rather to illustrate in this simple setting some 
of the general arguments which have
appeared in the previous discussion.
These include the longitudinal structure of the induced source, 
the equivalence between the strip restrictions on $p^+$ and on $p^-$, 
and the sensitivity to the $i\epsilon$ prescription in the LC-gauge 
propagator.

To this aim, we shall compute just one contribution to
$\sigma$ and one contribution to $\chi$,
namely (cf. eqs.~(\ref{sigma12}) and (\ref{chi2})):
\be\label{sigma10}
\hat\sigma_1^a(\vec x)\,\equiv\,-g^2 \,\partial^+_y{\rm Tr}
\Bigl(T^a G^{ii}(x,y)\Bigr)\Big |_{x=y},\qquad
\hat\chi_1(\vec x, \vec y) \,\equiv\,
4ig^2{\cal F}^{+i}_x\, \langle x|G^{ij}|y\rangle \,{\cal F}^{+j}_y\,.\ee
These quantities will be evaluated here to lowest order
in $\rho$ (that is, to linear order in the case of
$\sigma$, and to quadratic order for $\chi$), 
and to LLA. In this approximation, 
there is no distinction between the LC-gauge source $\rho$ 
and the COV-gauge source $\tilde\rho$, 
so we shall use the notation $\rho$ to refer to any of them.
Also, to simplify the notations, we shall not distinguish
below between $\hat\sigma$ and $\sigma$, neither between 
$\hat\chi$ and $\chi$.

\subsection{The weak source limit of $\sigma$}

 Consider $\sigma_1$ first: because of the color trace
in eq.~(\ref{sigma10}), the vacuum piece  $G^{ii}_0$ of the propagator
drops out from this equation, which can thus be rewritten as
\be\label{sigma101}
\sigma_1^a(\vec x)=-g^2 \,\partial^+_y{\rm Tr}
\Bigl(T^a \delta G^{ii}(x,y)\Bigr)\Big |_{x=y}
={g^2 \over 2}
(\partial^+_x-\partial^+_y){\rm Tr}
\Bigl(T^a \delta G^{ii}(x,y)\Bigr)\Big |_{x=y},\ee
where the second equality follows from
the symmetry 
property $\delta G^{ij}_{ab}(x,y)=\delta G^{ji}_{ba}(y,x)$.

A contribution linear in $\rho$ to $\delta G^{ii}$
can be generated either by a direct insertion of $\rho$,
which couples to the fields $a^-$ via the
vertices in $S_W$ (cf. eqs.~(\ref{ACTION1})--(\ref{WLINE})), 
or by an insertion of the mean field ${\cal A}^i$, which 
couples to $a^i$ via the usual vertices in $S_{YM}$.
However, the insertion of ${\cal A}^i$ does not contribute 
to $\sigma_1$ to LLA: indeed, this produces the standard one-loop 
polarization tensor of QCD, which has no logarithmic enhancement. 
On the other hand, such a logarithm {\it is} generated 
by the insertion of $\rho$, as we show now.

The corresponding contribution to $\sigma_1$ is displayed
in Fig.  \ref{SIGLIN}.a, and reads 
\be\label{S11}
\sigma_1(\vec x)&=&-g^2 \int d^4z \int d^4y\,
G^{i-}_0(x-z)\Pi(z,u)\,\partial^+_y G^{-i}_0(u-y)\Big |_{x=y},\ee
where $\Pi(z,u)$ is given by eq.~(\ref{R--}),
and $G^{i-}_0$ and $G^{-i}_0$ are
the free propagators in eq.~(\ref{LCPROP}).
Below, we shall compute this quantity in two different ways:\\
\hspace{.8cm}{\bf a\/)} First, we shall directly compute the 
two-dimensional charge density 
$\sigma_a (x_\perp)\equiv \int dx^- \sigma_a(\vec x)$,
for which the logaritmic enhancement is manifest
(cf. eq.~(\ref{sigma12})). We shall verify that the result
is independent upon the choice of the strip restriction, and
also upon the $i\epsilon$ prescription in the LC-gauge 
propagator.\\
\hspace{.8cm}{\bf b\/)} By using a strip restriction on $p^-$,
we shall recalculate $\sigma_1(\vec x)$ in the $x^-$
representation, and thus exhibit its longitudinal structure.

{\bf a\/)} We first rewrite eq.~(\ref{S11}) in momentum space,
as follows (with $\vec q=\vec p-\vec k$):
\be\label{S110}
\sigma_1(\vec x)&=&-ig^2\int {dp^-\over 2\pi} \,
\int \frac{d^3 \vec p}{(2\pi)^3}\,\frac{d^3 \vec k}{(2\pi)^3}
\,{\rm e}^{-i\vec q\cdot \vec x}\,
G^{i-}_0(p^-,\vec p)\left[{\rm PV}\,{1\over p^-}\,
\rho(\vec q)\right]k^+ G^{-i}_0(p^-,\vec k).
\nonumber
\ee
Here, $k^+$ could be as well replaced by $p^+$, or
$(k^++p^+)/2$ : this can be seen by using the symmetry of the
integrand under the simultaneous exchange 
$\vec k \leftrightarrow -\vec p$ and
$p^- \leftrightarrow -p^-$, or, alternatively, by using directly
the second, more symmetrical, representation
for $\sigma_1$ in eq.~(\ref{sigma101}). By also using 
eq.~(\ref{LCPROP}), we can write:
\be
\frac{k^++p^+}{2}\,G^{i-}_0(p)\,G^{-i}_0(k)
&\equiv&\frac{k^++p^+}{2}\,{p^i\over p^++i\epsilon}\,G_0(p)\,
{k^i\over k^+-i\epsilon}\,G_0(k)\nn &=&
\frac{p_{\perp}\cdot k_{\perp}}{2}
\left({1\over p^++i\epsilon}+ {1\over k^+-i\epsilon}\right)G_0(p)\,G_0(k).
\ee
The integration over $x^-$ then sets $k^+=p^+$, so
we are left with the following, manifestly symmetrical,
integrations over $p^+$ and $p^-$ :
\begin{equation}\label{p+-}
\int {dp^- \over 2\pi}\,
{\rm PV}\, {1 \over p^-}\,\int {dp^+ \over 2 \pi}
\,{\rm PV}\,{1 \over p^+}\,\,
{1 \over 2p^+p^- - p^2_{\perp}+i\epsilon}
\, {1 \over 2p^+p^- - k^2_{\perp}+i\epsilon}\,.
\end{equation}
Note the effective PV-prescription in $1/ p^+$: this
is a consequence of the symmetry properties
of the integrand in eq.~(\ref{S110}), and is largerly
insensitive to the original $i\epsilon$
prescription in the LC-gauge propagator (\ref{LCPROP}). This
same effective prescription would have been generated by 
starting, e.g., with the advanced prescription of
Refs. \cite{MQ,KM98}, or with the
principal value prescription. Thus, the present calculation 
of $\sigma_1$ is independent of the axial prescription.

The integral in eq.~(\ref{p+-}) must be supplemented with
a strip restriction either on $p^+$ (cf. eq.~(\ref{strip})), 
or on $p^-$ (cf. eq.~(\ref{strip-})), with the two restrictions
related by eq.~(\ref{strips}). Given the symmetry of the
integrand under $p^+ \leftrightarrow p^-$, and the fact that
the integral is only logarithmically sensitive to its cutoff,
it is clear that both restrictions will lead to the same result.
For instance, with a strip restriction on $p^+$,
the unrestricted integral over $p^-$ can
be performed by contour techniques, to yield:
\be\label{p-}
\int {dp^- \over 2\pi}\,
{\rm PV}\, {1 \over p^-}\,
{1 \over 2p^+p^- - p^2_{\perp}+i\epsilon}
\, {1 \over 2p^+p^- - k^2_{\perp}+i\epsilon}\,
=\,{i\epsilon (p^+) \over 2 p^2_{\perp} k^2_{\perp}}\,.\ee
Then, the restricted integration over $p^+$ generates the
expected logarithmic enhancement:
\be\label{LOGX}
\int_{strip} {dp^+ \over 2 \pi} \,{\epsilon(p^+)\over p^+}\,\equiv\,
\left(\int_{-\Lambda^+}^{-b\Lambda^+}+\int_{b\Lambda^+}^{\Lambda^+}
\right){dp^+ \over 2 \pi}\,{\epsilon(p^+)\over p^+}
\,=\,{1 \over \pi}\,\ln (1/b)\,.\ee
One thus obtains:
\begin{equation}\label{sig1res}
\sigma_1^a (x_{\perp}) \, = \,
{g^2 N_c\over 2 \pi} \ln {1 \over b} 
\int \frac{d^2 p_\perp}{(2\pi)^2}
\int \frac{d^2 k_\perp}{(2\pi)^2}\,{\rm e}^{i(p_\perp-k_\perp)
\cdot x_\perp}\,
{p_{\perp} \cdot k_{\perp} \over p_{\perp}^2 k_{\perp}^2}
\,\rho^a(q_\perp),
\end{equation}   
in agreement with Ref. \cite{JKLW97}. In this equation,
\be\label{rhoq}
\rho_a(q_\perp)\equiv \rho_a(q^+=0, q_\perp)\,=\,
\int dx^- \int d^2x_\perp
\,{\rm e}^{iq_\perp\cdot x_\perp}\,\rho_a(x^-,x_\perp)\,.\ee
After simple manipulations, eq.~(\ref{sig1res}) can be finally rewritten as:
\be\label{sigma1q}
\sigma_1^a (q_\perp)\,=\,-2 N_c\,\alpha_s\ln {1 \over b} \,\rho_a(q_\perp)
\int \frac{d^2 p_\perp}{(2\pi)^2}\,\left[
{q^2_{\perp} \over 2p_{\perp}^2 
(p_{\perp}- q_{\perp})^2}\,-\,{1 \over p_{\perp}^2}\right].\ee

{\bf b\/)} The longitudinal structure of eq.~(\ref{S11}) 
is contained in (cf. eq.~(\ref{R--}))
\be
\int dz^-\,G^{i-}_0(x^--z^-)\rho(z^-)\,
\partial^+_x G^{-i}_0(z^--x^-)\,\approx\, \rho\,G^{i-}_0(x^-)
\partial^+_x G^{-i}_0(-x^-)
,\ee
where we have used
$\rho(z^-,z_\perp) \approx \delta(z^-)\rho(z_\perp)$, 
which is appropriate since
$\rho(z^-)$ is localized at $0\simle z^- \simle 1/\Lambda^+$,
while $\sigma(x^-)$
is rather needed at large $x^-\simge 1/b\Lambda^+$.

 With the strip restriction (\ref{strip-}) on $p^-$,
the integrations over $p^+$ and $k^+$ are unrestricted,
so we can use the standard expressions for the propagators
$G^{i-}_0$ and $G^{-i}_0$ in the $x^-$ 
representation\footnote{By contrast, it would be more 
difficult to study the $x^-$ structure by using a strip 
restriction on $p^+\/$; cf. the discussion in Sect. 3.4.} :
\be\label{intp+}
G^{i-}_0(x^-,p^-,{ p}_\perp)&=&
\int {dp^+ \over 2 \pi}\  e^{-ip^+x^-}\,
{1\over p^++i\epsilon}\
\frac{p^i}{2p^+p^- - p_\perp^2 +i\epsilon}\\&=&
{ip^i \over p_{\perp}^2}
\left \{
\theta(x^-) -
\left [
\theta(x^-)\theta(p^-) -\theta(-x^-)\theta(-p^-) \right]
e^{-i {p_{\perp}^2 \over 2p^-}x^-}\right \},\nn
\partial^+_x G^{-i}_0(-x^-,p^-,{ k}_\perp)&=&
{k^i \over 2p^-}\,\left \{
\theta(-x^-)\theta(p^-) -\theta(x^-)\theta(-p^-)\right \}
e^{i {k_{\perp}^2 \over 2p^-}x^-}.\ee
This yields then:
\be\label{s1long}
G^{i-}_0(x^-)\partial^+_x G^{-i}_0(-x^-)\,=\,-i\theta(x^-)\,
{\theta(-p^-)\over 2p^-}\,
{p_{\perp} \cdot k_{\perp} \over p_{\perp}^2}\,
e^{i {k_{\perp}^2 \over 2p^-}x^-},\ee
which implies that $\sigma_1(x^-)$ has support 
at $1/\Lambda^+ \simle x^- \simle 1/b\Lambda^+\/$. Indeed, the
typical range of $x^-$ is fixed by the exponential
$e^{ip^+_{eff}x^-}$, with $p^+_{eff}\equiv k_{\perp}^2/2p^-\/$;
since $p^-$ is restricted to
the strip (\ref{strip-}), $p^+_{eff}$ is constrained by 
$b\Lambda^+ \simle |p^+_{eff}|\simle \Lambda^+ $, hencefrom
the support of $\sigma_1(x^-)$ alluded to before.

In fact, in this case it is easy to pursue the calculation
and obtain $\sigma_1(x^-)$ in explicit form. After inserting
eq.~(\ref{s1long}) into eq.~(\ref{S11}), one has to compute
the restricted integral over $p^-$. This yields:
\be\label{LOGX1}
\int_{strip} {dp^- \over 2 \pi}\,
{\theta(-p^-)\over 2(p^-)^2}\,e^{i {k_{\perp}^2 \over 2p^-}x^-}\,=\,
\frac{1}{k_{\perp}^2}\,\frac{e^{-ib\Lambda^+ x^-}-
e^{-i\Lambda^+ x^-}}{2\pi i x}\,,\ee
where we have replaced $ k_{\perp}^2/2\Lambda^-\approx \Lambda^+$
to LLA (cf. eq.~(\ref{strips})).
The final expression for $\sigma_1(x^-)$ reads then
(compare to eq.~(\ref{sig1res})) :
\begin{equation}\label{sig1x}
\sigma_1^a (\vec x) \, = \,
{g^2 N_c\over 2 \pi} \,F(x^-)
\int \frac{d^2 p_\perp}{(2\pi)^2}
\int \frac{d^2 k_\perp}{(2\pi)^2}\,{\rm e}^{i(p_\perp-k_\perp)
\cdot x_\perp}\,
{p_{\perp} \cdot k_{\perp} \over p_{\perp}^2 k_{\perp}^2}
\,\rho^a(q_\perp),
\end{equation}   
with the ``form factor''
\be\label{FormF}
F(x^-)\,\equiv\,\theta(x^-)\,\frac{e^{-ib\Lambda^+ x^-}-
e^{-i\Lambda^+ x^-}}{x}\,,\ee
which is indeed localized at 
$1/\Lambda^+ \simle x^- \simle 1/b\Lambda^+\/$, and yields
the expected logarithmic enhancement after the 
integration over $x^-\/$:
\be 
\int dx^- F(x^-) \,=\,\ln {1 \over b}\,.\ee

Of course, the specific longitudinal structure
in eqs.~(\ref{sig1x})--(\ref{FormF}) is directly
related to our choice of the retarded prescription 
in the LC-gauge propagator (\ref{LCPROP}). If the advanced 
prescription were chosen instead, the corresponding 
$\sigma_1(x^-)$ would be localized at negative
$x^-$, while with a PV-prescription it would have
support at both $x^->0$ and $x^-<0$ (with
$1/\Lambda^+ \simle |x^-| \simle 1/b\Lambda^+$ in all
cases). Nevertheless, the integrated 
charge density $\sigma_1({x}_\perp)\equiv
\int dx^- \sigma_1(x^-, x_\perp)$ --- which, we recall,
is the relevant quantity to LLA ---,  would come up the same
with {\it any} prescription, as already demonstrated by calculation (a).

To linear oder in $\rho$, the second contribution
to the induced charge (\ref{sigma12}) reads
\be\label{sigma20}
\sigma_2(\vec x)\,=\,
-\,ig^2\rho({\vec x})\,\langle x|
 {1 \over i \partial^-} \,G^{--}_0  {1 \over i\partial^-}|x \rangle,
\ee
and involves the prescription-independent component
$G^{--}_0$ of the free LC-gauge propagator.
To evaluate eq.~(\ref{sigma20}), one has to recall the
precise temporal and color structure of eq.~(\ref{rho2}).
After some simple algebra, one obtains\cite{JKLW97}
\be\label{sigma2q}
\sigma_2^a (q_\perp)\,=\,-2 N_c\,\alpha_s\ln {1 \over b} \,\rho_a(q_\perp)
\int \frac{d^2 p_\perp}{(2\pi)^2}\,{1 \over p_{\perp}^2}\,,\ee
which precisely cancels the tadpole piece $1/p_{\perp}^2$ 
in the integrand of eq.~(\ref{sigma1q}). Thus, finally,
\be\label{sigmaq}
\sigma^{(0)}_a (q_\perp)\,\equiv\,\sigma^a_1+\sigma^a_2
\,=\,-N_c\,\alpha_s\ln {1 \over b} \,\rho_a(q_\perp)
\int \frac{d^2 p_\perp}{(2\pi)^2}\,{q^2_{\perp} \over p_{\perp}^2 
(p_{\perp}- q_{\perp})^2}\,,\ee
where the upper script on $\sigma^{(0)}$ is to recall that this
is the result to linear oder in $\rho$. This result
holds independently of the axial prescription.


 \subsection{The weak source limit of $\chi$}

To lowest, quadratic, order in $\rho$, we have (see
Fig.  \ref{CHILIN}.a) :
\be\label{CHI11}
 \chi_1^{ab}(\vec x, \vec y) &=&
4ig^2{\cal F}^{+i}_{ac}(\vec x)
\, \langle x|G^{ij}_0|y\rangle \,{\cal F}^{+j}_{cb}(\vec y),\ee
where $y^+=x^++\epsilon$, and ${\cal F}^{+i}$ is
related to $\rho$ by the linear-response,
or Abelian, approximation (cf. Sect. 2.2): ${\cal F}^{+i}\approx 
-(\partial^i/\grad_\perp^2)\rho\,$. 
Eq.~(\ref{CHI11}) involves:
\be\label{CMEL}
\langle x|G^{ij}_0|y\rangle\, =\,\delta^{ij}
\int {dp^-\over 2\pi} \,{\rm e}^{ip^-\epsilon}
\int {dp^+ \over 2 \pi}\int \frac{d^2 p_\perp}{(2\pi)^2}\,
\frac{{\rm e}^{-ip^+(x^--y^-)}\,\,{\rm e}^{ip_\perp\cdot (x_\perp-y_\perp)}}
{2p^+p^- - p_\perp^2 +i\epsilon}\,,\ee
which we shall first evaluate with a strip restriction on $p^+$.
Then, e.g., $|p^+ x^-|\ll 1$ (since typically $|p^+|\ll \Lambda^+$,
and $x^- \simle 1/\Lambda^+$), 
so we can neglect $x^-$ and $y^-$ in eq.~(\ref{CMEL}). 
By using the convergency factor 
${\rm e}^{ip^-\epsilon}$,
the integral over $p^-$ can be computed by closing the contour
in the upper half plane. This yields:
\be\label{chi1res}
 \chi_1(\vec x, \vec y)&=&-4g^2{\cal F}^{+i}(\vec x)
\left[\int_{strip} {dp^+ \over 2 \pi}
\,{\theta(-p^+) \over 2p^+}\int \frac{d^2 p_\perp}{(2\pi)^2}\,
{\rm e}^{ip_\perp\cdot (x_\perp-y_\perp)}\right]{\cal F}^{+i}(\vec y)
\nonumber\\ &=&\frac{g^2}{\pi}\,\ln(1/b)\,
\delta^{(2)}(x_\perp-y_\perp)\,{\cal F}^{+i}(\vec x)\,
{\cal F}^{+i}(\vec y).
\ee

It is a little bit trickier, but at the same time instructive,
to show that the same result
can be obtained by using a strip restriction on $p^-$.
Then, the convergency factor plays no role, but the 
(restricted) integration over $p^-$ can still 
be performed in the first place, by using:
\be\label{indstrip}
\int_{strip}
 {dp^-\over 2\pi} \,\frac{1}{2p^+p^- - p_\perp^2 +i\epsilon}
\,\approx \,
\,\frac{-i\epsilon(p^+)}{4p^+}\,\Theta_{strip}(p^+),\ee
which holds to LLA. Here, $\Theta_{strip}(p^+)$ is a
step function which implements the
strip restriction (\ref{strip}): $\Theta_{strip}(p^+)=1$ 
for  $b\Lambda^+ \ll |p^+| \ll \Lambda^+$, and
$\Theta_{strip}(p^+)=0$ otherwise. Thus, to LLA,
the restricted integration over $p^-$ generates an effective
strip restriction on $p^+$, with the two restrictions related by
eq.~(\ref{strips}). This is so since, to this accuracy, the integral
in eq.~(\ref{indstrip}) is saturated by on-shell modes.
It is now easy to verify that the ensuing integral over $p^+$
leads to eq.~(\ref{chi1res}) again.

The other components of $\chi$ in eq.~(\ref{chi2}) can be similarly
computed, with the following final result (in matrix notations
where, e.g., $\rho_x\equiv \rho_{ab}(\vec x)$) \cite{JKLW97}:
\be\label{CHIB0}
\chi^{(0)}(\vec x,\vec y)&=&4\alpha_s \ln (1/b)
\int {d^2p_\perp \over (2 \pi)^2}\,
\frac{{\rm e}^{ip_{\perp}\cdot(x_{\perp}-y_{\perp})}}
{p_{\perp}^2}\nonumber\\
&{}&\,\,\,\,\times\left\{\rho_x\rho_y+ i\rho_x({\cal F}^{+i}p^i)_y
-i({\cal F}^{+i}p^i)_x\rho_y+ p^2_\perp {\cal F}^{+i}_x 
{\cal F}^{+i}_y\right\}.\ee
The result (\ref{CHIB0}) is independent of the axial prescription
in the LC-gauge propagator, as can be easily verified by
repeating the calculations with various prescriptions.

\subsection{The BFKL equation}

In the present weak field approximation,
the only non-trivial evolution equation is that for the
2-point function $\langle \rho\rho\rangle_{\tau}$,
which reads (cf. eq.~(\ref{RGE2p}), which we also integrate
over $x^-$ and $y^-$):
\be\labe{RGE2pLIN}
{d\over {d\tau}}\,
\Big\langle\rho_a(x_\perp)\rho_b(y_\perp)\Big\rangle_\tau\,=\,\alpha_s
\Big\langle\sigma_a^{(0)}(x_\perp)\rho_b(y_\perp)
\,+\,\rho_a(x_\perp)\sigma_b^{(0)}(y_\perp)\,+\,
\chi_{ab}^{(0)}(x_\perp,y_\perp)\Bigr\rangle_\tau\,.\ee
It is interesting to consider the diagonal element
of this equation in momentum space. Indeed, this is the equation
satisfied by 
\be\label{varphi}
\Big\langle\rho_a(k_\perp)\rho_a(-k_\perp)\Big\rangle_\tau
\,\approx\,k^2_{\perp}
\Bigl\langle\,|{\cal F}^{i+}_a(k_\perp)|^2\Bigr\rangle_\tau
\,\equiv\,\varphi(x,k^2_{\perp})\,,\ee
which, according to eq.~(\ref{GCL}),
is proportional to the density of gluons with transverse momentum 
$k_\perp$ and rapidity 
$\tau\equiv \ln(1/x)$ (i.e., longitudinal momentum $k^+=xP^+$).
This quantity is often referred to as
 the {\it unintegrated gluon distribution}. Note that the
simple relation (\ref{varphi}) between $\varphi(x,k^2_{\perp})$ and 
$\langle \rho\rho\rangle_{\tau}$
holds only in the weak source limit,
where the electric field and the color source are linearly related: 
${\cal F}^{+i}\approx -i(k^i/k^2_\perp)\rho$.

By combining eqs.~(\ref{RGE2pLIN})--(\ref{varphi}) with the
previous expressions for $\sigma^{(0)}$, eq.~(\ref{sigmaq}),
and $\chi^{(0)}$, eq.~(\ref{CHIB0}), one finally obtains the
following evolution equation for $\varphi(x,k^2_{\perp})$ \cite{JKLW97}:
\begin{eqnarray}
x {\partial \varphi(x,k^2_{\perp}) \over \partial x} & = & \,\,\,
{\alpha_s N_c \over \pi^2}\,
\int d^2 p_{\perp}
 {k^2_{\perp} \over p^2_{\perp} (k_{\perp}-p_{\perp})^2}\,
 \varphi(x,p^2_{\perp}) \nonumber \\
& & -\,
{\alpha_s N_c \over2 \pi^2}\,
\int d^2 p_{\perp}
 {k^2_{\perp} \over p^2_{\perp} (k_{\perp}-p_{\perp})^2}\,
 \varphi(x,k^2_{\perp})\,,
\label{BFKL}
\end{eqnarray}
which coincides, as anticipated, with the BFKL equation \cite{BFKL}.
The first term in the r.h.s., which here is generated by 
$\chi^{(0)}$, is the {\it real} BFKL kernel, 
while the second term, coming from $\sigma^{(0)}$,
is the corresponding {\it virtual} kernel.

Thus, outside the saturation regime, the general evolution
equation (\ref{RGE}) reduces to the BFKL equation, as necessary
on physical grounds. This is a very non-trivial
check of the {\it quantum} structure of the effective theory 
under consideration 
(i.e., eq.~(\ref{PART}) with the action (\ref{ACTION})).
Indeed, unlike the classical MV model, which is sensitive
only to the eikonal vertex $\int d^4x \rho A^-$ (the
one-point vertex in the action $S_W$), the quantum corrections
$\sigma$ and $\chi$ are generated by the 2-point and 3-point
vertices in $S_W$. These are also the vertices which enter
the calculation of $\sigma$ and $\chi$ in the general
non-linear case, to be described in Paper II.

As we have explicitly verified here, the BFKL equation is
obtained independently of the gauge-fixing prescription in the 
LC-gauge propagator. This is as expected since, in this approximation,
the 2-point function $\langle \rho\rho\rangle_{\tau}$
of the color charge coincides with the physical gluon density, cf.
eq.~(\ref{varphi}). In the presence of non-linear effects,
this relation may not hold anylonger. Then, the charge correlators
are not physical observables by themselves, and need not be 
gauge-fixing independent. And, indeed, we shall find in Paper II
that the non-linear effects in $\sigma$ and
$\chi$ are sensitive to the axial prescription.

To anticipate, the simplest results will be obtained by using the retarded,
or the advanced, prescriptions discussed in relation with
eq.~(\ref{LCPROP}). With
the {retarded} prescription, both the classical mean field
${\cal A}^i_a$, eq.~(\ref{APM}), and the induced source
$\sigma$, eq.~(\ref{sig1x}),
sit at $x^- > 0$, or $z < t\,$; that is, the soft gluons fields
are behind their source, the hadron (which is located at $z=t$). 
It is intuitively plausible
that within this picture there are no initial state gluon interactions.
This is the counterpart of the conclusion of Mueller and
Kovchegov \cite{KM98,AM2} who used an {advanced} prescription 
$1/(p^+-i\epsilon)$, and showed that all the {\em final}
state gluon interactions disappeared. 
These two prescriptions are simple in that one
can either put gluon interactions in the final or initial state. 
Other gauge conditions such as
Leibbrandt-Mandelstam or principal value do not have this simple feature,
and lead to other complications for computations as well \cite{ILM00II}.

\setcounter{equation}{0}
\section{The background field gluon propagator}

In this section, we construct the background field propagator 
$iG^{\mu\nu}(x,y)\equiv \langle {\rm T}\,a ^\mu(x) a^\nu(y)\rangle$ 
of the semi-fast gluons, which is the key ingredient for the
the calculation of $\sigma$ and $\chi$ to be presented in Paper II.
(See also Refs. \cite{MV94a,AJMV95,HW98} for previous related work.)
The propagator will be obtained by solving  eq.~(\ref{D3}) in the LC gauge
$a^+_a=0$, and in the subspace of fields with energies $p^-$
restricted to the strip (\ref{strip-}).
The latter restriction causes no difficulty, since the problem
being homogeneous in time (recall that the background field is 
static), we can construct the propagator for any fixed  $p^-$.
Concerning the gauge condition, we still need to fix the
residual gauge freedom in the LC gauge; this will be
done by giving a prescription for the ``axial'' pole at 
$p^+=0$, in agreement with eq.~(\ref{LCPROP}).

For more clarity, we shall first briefly review the
corresponding construction for the {\it scalar} propagator, 
whose definitive form has been given 
by Hebecker and Weigert \cite{HW98}. Then, in Sects. 6.2 and 6.3,
we shall construct the gluon propagator in two steps:
First, from the scalar propagator,
we shall derive the gluon propagator in the temporal gauge
$a^-=0$. Then, the propagator in the LC gauge will be obtained
via a gauge rotation. The reasons for this particular
strategy should become clear in a moment.

\subsection{The background field scalar propagator}

The background field scalar propagator $G(x,y)[{\cal A}]$
is defined as the solution to
\be \labe{GS0}
-{\cal D}^2_x\,G(x,y)\,=\,\delta^{(4)}(x-y)\,,\ee
where\footnote{Recall that the gauge 
coupling constant has been absorbed into
the normalization of the classical field, cf.
eq.~(\ref{rescale}).}
 ${\cal D}^2\equiv {\cal D}^+{\cal D}^-
+ {\cal D}^-{\cal D}^+  - {\cal D}_\perp^2=
2\partial^+\partial^-- {\cal D}_\perp^2\,$, 
 ${\cal D}^\mu\equiv\del^\mu-i
{\cal A}^\mu$, and ${\cal A}^\mu=\delta^{\mu i}{\cal A}^i$ 
is the LC-gauge color field generated by $\rho$
and constructed in Sect. 2.3. To solve this equation, it is convenient
to perform a rotation to the covariant gauge where the 
background field has just a plus component:
${\cal A}^\mu\to\tilde A^\mu=\delta^{\mu +}\alpha$
(cf. eq.~(\ref{gtr})). We thus write: 
\be \labe{GS1}
G(x,y)\,=\,U(\vec x)\,\tilde G(x,y)\,U^{\dagger}(\vec y)\,,\ee
where $\tilde G(x,y)$ obeys the same eq.~(\ref{GS0}), but with
${\cal D}^2=
2\partial^- {\cal D}^+  - \grad_\perp^2$ and 
${\cal D}^+=\partial^+ -i\alpha(\vec x)$.

In order to solve this equation non-perturbatively in $\alpha$, 
it is important to recall the separation of scales in the problem:
the semi-fast quantum fluctuations whose propagator we are about
to compute have relatively small longitudinal momenta
(typically, in the strip (\ref{strip})), and therefore
a resolution scale $\Delta x^-$ which is much larger
than the longitudinal extent $\sim 1/\Lambda^+$ of the source $\rho$.
When ``seen'' by these fluctuations, the source $\rho$ and 
the associated field $\alpha$ are effectively delta functions at $x^-=0$.
We are thus led to solve the following, singular, equation:
\be \labe{GS2}
\left\{-2\partial^-\Bigl(\partial^+ -i\delta(x^-)\alpha(x_\perp)
\Bigr)+ \grad_\perp^2\right\}\tilde 
G(x,y)\,=\,\delta^{(4)}(x-y)\,,\ee
whose precise meaning is as follows:
for $x^-$ either positive or negative,
$\tilde G(x,y)$ obeys the free equation
\begin{equation}
\label{GFREE}
        -\partial^2_x\, \tilde G(x,y) \,=\, \delta^{(4)} (x-y)\quad
{\rm for}\quad x^-\ne 0,
\end{equation}
but it is discontinuous at $x^-=0$ (and similarly at $y^-=0$).
To evaluate its discontinuity, it is easier to argue in the LC gauge:
the transverse background field ${\cal A}^i$ is effectively
a step function (cf. eq.~(\ref{APM})), so it
is discontinuous but finite at $x^- =0$.  Then, eq.~(\ref{GS0})
implies that $G(x,y)$ must be continuous at $x^- = 0$ :
\be
G(x,y)\mid_{x^- \rightarrow 0^+}
 \,=\,G(x,y)\mid_{x^- \rightarrow 0^-},\ee
and similarly at $y^-=0$. After the gauge rotation (\ref{GS1}),
this provides the sought for boundary conditions for $\tilde G(x,y)$ :
\be\label{BC}
\tilde G(x,y) \mid_{x^- = 0^+}& =& V^\dagger(x_\perp)\tilde
 G(x,y) \mid_{x^- = 0^-},\nn
\tilde G(x,y) \mid_{y^- = 0^+}& =&\tilde G(x,y) 
 \mid_{y^- = 0^-}V(y_\perp) \,,\ee
with $V^{\dagger}(x_\perp)$ defined in  eqs.~(\ref{UTAF})--(\ref{v}).

To summarize, to construct the scalar Green's function, 
one must solve the free equation (\ref{GFREE}) subject to the
boundary conditions (\ref{BC}). The solution reads \cite{HW98} :
\begin{eqnarray}\label{SPROP}
      \tilde  G(x,y) &  =  & 
G_0(x,y) -2\int d^4z \, G_0(x,z)\, \delta (z^-)
\,\partial^-_zG_0(z,y) \nonumber \\
 & & \times \left\{ \theta (x^-) \theta (-y^-) (V^\dagger(z_\perp) -
1)
-\theta (-x^-) \theta(y^-) (V (z_\perp) -1) \right\},
\end{eqnarray}
where $G_0$ is the free scalar propagator:
\begin{equation}
         G_0(x-y)\, =\, \int {{d^4p} \over {(2\pi )^4}} \,\,
  e^{-ip\cdot(x-y)} \, {1 \over {p^2 + i\epsilon}}\,.
\end{equation}
To verify that this is the correct solution indeed, note that:
{\it a\/}) In the absence of the background field, $V=
V^\dagger =1$, and $\tilde G$ reduces to $G_0$, 
as it should. {\it b\/}) When acting on eq.~(\ref{SPROP})
with $\partial^2$, one 
gets zero everywhere except at $x^- =0$
when operating on the left, or $y^- = 0$
when operating on the right.  Therefore 
$\tilde G$ solve the equations of motion
everywhere except where the source sits. {\it c\/})
By using
\be\label{G0X}
G_0(x)=- i\int {{dp^-} \over {(2\pi)}}\,   {{\rm e}^{-ip^-x^+}
\over {2p^-}}\, \left\{ \theta (x^-) \theta (p^-)
- \theta (-x^-) \theta (-p^-) \right\}
\int {d^2p_\perp \over (2 \pi)^2}\,
{\rm e}^{ip_{\perp}\cdot x_{\perp} -i\frac{p_{\perp}^2}{2p^-}x^-},\,\,
\ee
one deduces that
\begin{equation}
        \lim_{x^- \rightarrow 0} \,  
G_0(x-z)\mid_{z^- = 0} \, =- i
\delta^{(2)}(x_\perp - y_\perp) \int {{dp^-} \over {(2\pi)}} \,  {
{\rm e}^{-ip^-(x^+-z^+)}\over {2p^-}} \left\{ \theta (x^-) \theta (p^-)
- \theta (-x^-) \theta (-p^-) \right\}.
\end{equation}
Inserting this into eq.~(\ref{SPROP}) shows that $\tilde G$
satisfies the correct  boundary conditions across the source.

\begin{figure}
\protect\epsfxsize=14.cm{\centerline{\epsfbox{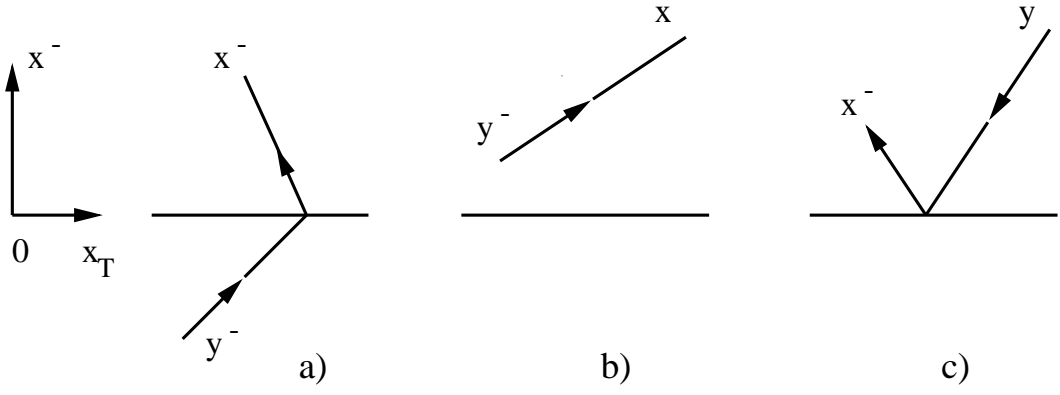}}}
         \caption{The propagation of the scalar particle in the 
delta potential at $x^-=0$: a) a crossing trajectory ($y^-<0$ and
 $x^-> 0$; b) a non-crossing trajectory (both $x^-$ and $y^-$ are positive);
c) a reflexion on the potential; this last process is actually forbidden
 by energy conservation.}
\label{REFLEX}
\end{figure} 
After a Fourier transform to the energy representation,
eq.~(\ref{SPROP}) can be rewritten as:
\be\label{GP-}
\tilde G(\vec x,\vec y, p^-)&=&G_0(\vec x-\vec y, p^-)
\Bigl\{\theta (x^-)\theta (y^-) + \theta (-x^-)\theta (-y^-)\Bigr\}\nn
&+& 2ip^-\int d^3\vec z\,\, G_0(\vec x-\vec z, p^-)
\, \delta (z^-)\,G_0(\vec z-\vec y, p^-) \nn
&{}&\times\,\left\{ \theta (x^-) \theta (-y^-)V^\dagger(z_\perp) -
\theta (-x^-) \theta(y^-) V(z_\perp)\right\}.\ee
This result is quite intuitive. There are two types
of terms (see Figs. \ref{REFLEX}.a and b): 
A {\it crossing} term, where the endpoints $x^-$ and
$y^-$ are on opposite sides with respect to the plane $x^-=0$
(where the potential $\alpha$ is located), and a {\it noncrossing} 
term, where both $x^-$ and $y^-$ are on the same side of the potential.
In the latter case, the scalar particles propagates freely from
$y$ to $x\,$. There is no reflection on the potential, i.e.,
no process like the one depicted  in Fig. \ref{REFLEX}.c ; this
is forbidden by energy conservation, since the potential is static.
Indeed, for on-shell excitations,  $p^+$ and $p^-$ 
have the same sign; thus, a particle with positive energy
propagates forward in $x^-$, as obvious on eq.~(\ref{G0X}),
and this cannot be changed by its
scattering off the static potential, which preserves $p^-$
(see Fig. \ref{REFLEX}.b). In particular, this shows that one can
effectively replace
\be\theta (x^-)\theta (-y^-)\longrightarrow \theta (p^-),\qquad
\theta (-x^-) \theta(y^-)\longrightarrow \theta (-p^-),\ee
without changing the crossing term in eq.~(\ref{GP-}).

After also performing the rotation (\ref{GS1}) to the LC gauge for the
background field, we finally derive the following expression
for the background field scalar propagator:
\be\label{GSCALAR}
G(\vec x,\vec y, p^-)&=&G_0(\vec x-\vec y, p^-)
\Bigl\{\theta (x^-)\theta (y^-)V(x_\perp)V^\dagger (y_\perp)
 + \theta (-x^-)\theta (-y^-)\Bigr\}\nn
&+& 2ip^-\int d^3\vec z\,\, G_0(\vec x-\vec z, p^-)
\, \delta (z^-)\,G_0(\vec z-\vec y, p^-) \nn
&{}&\times\,\left\{ \theta (p^-)V(x_\perp)V^\dagger(z_\perp) -
\theta (-p^-)V(z_\perp)V^\dagger (y_\perp)\right\}.\ee
It can be easily verified that this function is
continuous at both $x^- = 0$ and $y^- = 0$.

\subsection{The gluon propagator in the temporal gauge}
                      
To construct the propagator in the light cone gauge $a^+ = 0$, it
is convenient to perform an excursion through the {\it temporal}
gauge $\acute a^-=0$, that is, to compute first the propagator
in a mixed gauge where the {\it background} field is in the 
LC-gauge\footnote{Unless otherwise specified, the background
field will always be in the LC-gauge in what follows.}
${\cal A}^+=0$, but the {\it quantum}
field is in the gauge $\acute a^-=0$, and then rotate the
quantum field to the LC gauge $a^+ = 0$.
This last rotation will be constructed in Sect. 6.3 below.
Here, we shall rather focus on the construction of 
the propagator in the temporal gauge.

Let us first explain the usefulness of such a detour.
The reason is technical: by inspection of the coefficients
in eq.~(\ref{D3}) for $G^{\mu\nu}$, that is (cf. eqs.~(\ref{HATSIG})
and (\ref{R--})),
\be
G^{-1}_{\mu\nu}(x,y)\equiv\Bigl(g_{\mu\nu}{\cal D}^2
-{\cal D}_\mu {\cal D}_\nu- 2i{\cal F}_{\mu\nu}\Bigr)_x
\delta^{(4)} (x-y)\,-\,\delta_{\mu -}\delta_{\nu -}
\frac{i}{2}\,\rho(\vec x)
\,\epsilon(x^+-y^+)\,\delta^{(3)}(\vec x-\vec y),\nn
\ee
one finds that some of the coefficients which count for the LC gauge
$a^+ = 0$,
namely $G^{-1}_{--}$, $G^{-1}_{-i}$ and $G^{-1}_{i-}$, become singular
in the limit in which we treat $\rho$ as a delta
function, $\rho(\vec x) \to \delta(x^-)\rho(x_\perp) $.
(These coefficients involve either $\rho$, or ${\cal F}^{+i}$,
which is itself singular in this limit, cf. eq.~(\ref{FDELTA}).)
On the other hand, there is no such a singularity in any of the
coefficients relevant for the gauge $\acute  a^-=0$. Thus,
in the latter gauge, the vector propagator $\acute G^{\mu\nu}$
(with $\mu,\nu=+,1$ or 2) is continuous at $x^- = 0$ and $y^- = 0$, 
and can be immediately obtained from the scalar propagator 
discussed in the previous subsection. This construction
is described now (see also Refs. \cite{MV94a,AJMV95}
for more details):

To simplify notations, in the remaining part of this subsection
we shall denote the quantum fields in the temporal gauge simply
as $a^\mu$ (thus, $a^-=0$).
The easiest way to obtain $\acute G^{\mu\nu}$ is to use its definition
in terms of a path integral (the analog of eq.~(\ref{delAcorr})):
\be\labe{acuteG0}
i\acute G^{\mu\nu}_{ab}(x,y)&=&
Z^{-1}\int {\cal D} a\,\delta( a^-)\,\, a_a^\mu(x)
 a_b^\nu(y)\,
{\rm e}^{\,iS_0[{\cal A},  a]}\,,\nn
S_0[{\cal A},  a]&=&\int d^4x\left\{
{1\over 2}\,  a^i\bigl(-{\cal D}^2\bigr)  a^i
\,+\,{1\over 2}\,\bigl(\partial^- a^+- {\cal D}^i
a^i\bigr)^2\right\},\ee
where the second line follows by evaluating the Gaussian action
(\ref{SEXP2}) in the temporal gauge $ a^-=0$.
Note that this new action does not depend upon $\rho$
explicitly, but only implicitly, via the background field
${\cal A}^i$. This action can be brought to a quadratic 
form via the redefiniton
\be
 a^+\,\equiv\, \bar a^+ \,+\,{1\over \partial^-}\,{\cal D}^i a^i\,.\ee
There is no ambiguity associated with the pole
at $p^-=0$ in this transformation since, by definition,
the LC energy is restricted to the strip (\ref{strip-}). 
We thus obtain:
\be
S[{\cal A}, \bar a^+, a^i]&=&\int d^4x\left\{
{1\over 2}\,  a^i\bigl(-{\cal D}^2\bigr)  a^i
\,+\,{1\over 2}\,\bigl(\partial^-\bar a^+\bigr)^2\right\},\ee
which immediately implies that the propagator $\acute G^{ij}$
of the transverse fields is the same as the scalar propagator:
\be\label{acuteGij}
       \acute G^{ij}(x,y)& =& \delta^{ij} G(x,y)
\ee
with $G$ given by eq.~(\ref{GSCALAR}). Also,
\be \label{contact}
\langle {\rm T}\,\bar a^+(x) \bar a^+(y)\rangle \,=\,
\int {{d^4p} \over {(2\pi )^4}} \,\,
  e^{-ip\cdot(x-y)} \, {1 \over {(p^-)^2}}\,,\ee
while, e.g., $\langle {\rm T}\,\bar a^+(x) a^i(y)\rangle =0$.
Thus, the mixed propagators read
\begin{equation}
       \acute G^{+i} (x,y)\,=\, {1 \over {\partial_x^-}} {\cal D}^i_x G(x,y),
\qquad \acute G^{i+} (x,y)\,=\,G(x,y)\, {1 \over {\partial_y^-}}\,
{\cal D}^{\dagger \,i}_y\,,
\end{equation}
where ${\cal D}^{\dagger}\equiv \buildchar{\partial}{\leftarrow}
{} + ig A$, with the derivative $\buildchar{\partial}{\leftarrow}
{}$ acting on the functions on its left.
The $++$ propagator includes also the  contact term (\ref{contact}),
and reads:
\begin{equation}
   \acute      G^{++} (x,y)\,=\,
 {1 \over {\partial_x^-}} \,{\cal D}_x^i G(x,y) {\cal D}^{\dagger \,i}_y\,
{1 \over {\partial_y^-}}\,+\, {1 \over {\partial_x^-}}\, { 1 \over {\partial_y^-}} 
\delta^{(4)}(x-y).
\end{equation}
These propagators 
are conveniently rewritten in the energy representation:
\be\label{acuteG}
\acute G^{+i}(\vec x,\vec y, p^-)\,=\,{i \over p^-}\, {\cal D}_x^i 
G(\vec x,\vec y, p^-),\qquad
\acute G^{i+} (\vec x,\vec y, p^-)\,=\,-\,{i \over p^-}\,
G(\vec x,\vec y, p^-)\,{\cal D}^{\dagger \,j}_y\,,
\nonumber\\
\acute G^{++}(\vec x,\vec y, p^-)\,=\,{1 \over (p^-)^2}\,\left\{{\cal D}_x^i
G(\vec x,\vec y, p^-){\cal D}^{\dagger \,i}_y\,+\,\delta^{(3)}(\vec x-\vec y)
\right\}\quad.
\ee
We stress again that, in  our $p^-$
renormalization group scheme, we remove explicitly the support of the 
quantum fields $a^\mu$ at $p^- = 0$, so that the operator $1/p^-$
is not singular.


\subsection{The gluon propagator in the LC gauge}

The vector propagator $G^{\mu\nu}$ is defined by eq.~(\ref{delAcorr})
where both the background field and the {quantum} field  are
in the light cone gauge: ${\cal A}^+=a^+ =0$. It is however straightforward
to perform a change of variables in the form of a gauge rotation, so as
to reexpress $G^{\mu\nu}$ as an integral over quantum fields in the
{temporal} gauge $\acute a^-=0$, and thus relate it to the propagator
$\acute G^{\mu\nu}$ in the previous subsection.

Specifically, the gauge rotation $a^\mu \to \acute a^\mu$ reads:
\be\label{aacute}
{\cal A}^\mu+ \acute a^\mu\,=\,\Delta(x)\Bigl({\cal A}^\mu+a^\mu
+i\partial^\mu\Bigr)\Delta^\dagger(x),\ee
with (the initial time $x^+_0$ is arbitrary):
\be\labe{DLINE0}
\Delta^\dagger(x)
\,\equiv\,{\rm T}\, \exp\left\{\,
ig\int_{x^+_0}^{x^+} dz^+ a^-(z^+, \vec x) \right\}.\ee
Since ${\cal A}^-=0$, this transformation insures that
$\acute a^-=0$, as it should. For consistency with 
eqs.~(\ref{delAcorr})--(\ref{SEXP2}),
where the small fluctuations are treated in the Gaussian approximation,
the gauge rotation (\ref{DLINE0}) must be expanded to linear order in
$a^-$ :
\be\labe{DLINE}
\Delta^\dagger(x)\,\approx\,1+i\Lambda(x),\qquad \partial^-\Lambda(x)
\,=\,a^-(x).\ee
This implies, in particular, that the field $\Lambda(x)$ carries
the same energy $p^-$ as the quantum field $a^-$, i.e., a $p^-$ 
restricted to the strip (\ref{strip-}). As a consequence,
\be
\Lambda(x^+)\,=\,\int_{strip} {dp^-\over 2\pi} \,{\rm e}^{ip^-x^+}
\Lambda(p^-)\ee
vanishes asymptotically, $\Lambda(x^+)\to 0$ as $|x^+|\to\infty$, so
the infinitesimal gauge transformation (\ref{DLINE})
satisfies the periodicity condition $\Delta(x^+\to\infty)=
\Delta(x^+\to-\infty)=1$, as required for this to be a
symmetry of the action (\ref{ACTION1}).

To linear order in the small fluctuations, eq.~(\ref{aacute}) reduces
to 
\be\label{aacute1}
\acute a^\mu\,=\,a^\mu\,-\,[{\cal D}^\mu,\,\Lambda]\,.\ee
This is a symmetry transformation of the full action (\ref{ACTION1}),
and also of the quadratic action (\ref{SEXP2}). [To verify the latter
property explicitly, use the gauge invariance of the full action,
together with the fact that ${\cal A}^\mu$ is a solution to
the classical equations of motion $\delta S/\delta A=0$, to deduce
that $(\delta^2 S/\delta A^\mu\delta A^\nu)_{{\cal A}}{\cal D}^\nu
\Lambda=0$ for arbitrary $\Lambda\,$.] 

Eq.~(\ref{aacute1}) with $\Lambda$
defined by eq.~(\ref{DLINE}) gives us the change of variables to be
performed in eq.~(\ref{delAcorr}). To implement this, we first rewrite
eq.~(\ref{delAcorr}) as
\be 
iG^{\mu\nu}(x,y)\,=\,\frac{
\int{\cal D}a\int{\cal D}\Lambda
\,\delta(\partial^-\Lambda-a^-)\,
\delta(a^+)\,\,a^\mu(x)a^\nu(y)\,
{\rm e}^{\,iS_0[{\cal A},a]}}
{\int{\cal D}a\int{\cal D}\Lambda
\,\delta(\partial^-\Lambda-a^-)\,
\delta(a^+)\,{\rm e}^{\,iS_0[{\cal A},a]}},\ee
where the functional integral over $\Lambda$ is just 
to relabel $a^-$ as $\partial^-\Lambda$. (This transformation is
non-singular since $p^-$ never vanishes.) Then, we change variables
according to eq.~(\ref{aacute1}):
\be\label{aacute2}
a^\mu\,\equiv\,\acute a^\mu\,+\,[{\cal D}^\mu,\,\Lambda]\,,\ee
and perform the integral over $\Lambda$ with the help of
the delta function $\delta(a^+)\equiv 
\delta(\acute a^+ + \partial^+\Lambda)$. This finally yields:
\be\labe{LCG}
iG^{\mu\nu}(x,y)\,=\,\frac{\int{\cal D}\acute a
\,\delta(\acute a^-)\,\bigl(\acute a^\mu+{\cal D}^\mu\Lambda\bigr)_x
\bigl(\acute a^\nu+{\cal D}^\nu\Lambda\bigr)_y
\,{\rm e}^{\,iS_0[{\cal A},\,\acute a]}}
{\int{\cal D}\acute a\,\delta(\acute a^-)\,
\,{\rm e}^{\,iS_0[{\cal A},\,\acute a]}}
\,,\ee
where it is now understood that
\be\label{LRET}
\Lambda(x)\,\equiv\,-\,{1 \over {\partial^+}} \,\acute a^+
\,\equiv\,-\int_{-\infty}^{x^-} dz^-\,\acute a^+(z^-,x^+,x_\perp)\,.\ee
Note the lower limit $x^-_0=-\infty$ in the above integration,
which is equivalent to choosing the retarded prescription in
$1/{\partial^+}$ :
\be\label{delret}
\langle x^-|{1 \over {i\partial^+_R}}|y^-\rangle\,\equiv\,
\int {dp^+\over 2\pi} \,\frac{{\rm e}^{ip^+(x^--y^-)}}
{p^++i\epsilon}\,=\,-i\theta(x^--y^-).\ee
This represents our {\it gauge fixing prescription}: the LC gauge
which we get in this way is defined by (cf. eqs.~(\ref{DLINE})
and (\ref{LRET})):
\be \label{GFIX}
a^+\,=\,0,\qquad  \lim_{x^-\to -\infty}\,a^-\,=\,0.\ee
(This is unique since any $x^-$--independent gauge transformation
would violate the above asymptotic condition on $a^-$.)
Equivalently, the $i\epsilon$ prescription 
for the axial pole in the LC gauge propagator (\ref{LCG})
is the {\it retarded} prescription shown in eq.~(\ref{LCPROP}).

This particular gauge condition is chosen for consistency with 
the classical solution in Sect. 2.3 : There, we have assumed the tree-level
source $\rho$ to have support at $x^->0$, and we have chosen the 
boundary conditions in such a way that the field ${\cal A}^i$
is located at positive $x^-$ as well (cf. eq.~(\ref{APM})).
These assumptions are consistent
with the quantum evolution {\it provided} we use
the retarded prescription in the LC gauge propagator of the semi-fast
gluons. Indeed, with this choice, the induced charge $\sigma$ comes out
with support at positive, and relatively large, $x^-$ (see, e.g.,
eqs.~(\ref{sig1x})--(\ref{FormF})), which then implies a similar 
property\footnote{If, say, an {\it advanced} $i\epsilon$  prescription 
was used instead, the corresponding $\sigma$ would be located at
negative $x^-$, so that a field $\delta{\cal A}^i$ would be
generated at $x^-<0$ by quantum evolution, in contradiction with
the original boundary conditions at tree-level.}
for the induced field $\delta{\cal A}^i$ (cf. the discussion
after eq.~(\ref{dtA})).

To summarize, the LC gauge propagator is obtained as:
\be
iG^{\mu\nu}(x,y)\,=\,\left\langle {\rm T}\Bigl(
\acute a^\mu-{\cal D}^\mu{1 \over {\partial^+}} \,\acute a^+ \Bigr)_x
\Bigl(\acute a^\nu-{\cal D}^\nu{1 \over {\partial^+}} \,\acute a^+
\Bigr)_y\right\rangle,\ee
or, more explicitly,
\be\label{Gij}
G^{ij} & = &
\acute G^{ij} - {\cal D}^i {1 \over \partial^+}\, \acute G^{+j}
+ \acute G^{i+} {1 \over \partial^+} {\cal D}^{\dagger j}-
{\cal D}^i {1 \over \partial^+}\,{\acute G}^{++} 
{1 \over \partial^+} {\cal D}^{\dagger j}\,,\\\label{G-i}
G^{-i} &= &- \,{\partial^- \over \partial^+}\, \acute G^{+i} -
{\partial^- \over \partial^+} {\acute G}^{++}
 {1 \over \partial^+} {\cal D}^{\dagger i}\,,\\
\label{Gi-}
G^{i-} & =&   \,\acute G^{i+}\, {\partial^- \over \partial^+} - \,
{\cal D}^i \,{1 \over \partial^+}\, {\acute G}^{++} 
{\partial^- \over \partial^+ }\,,\\
\label{G--}
G^{--} & = &- \,{\partial^- \over \partial^+}
{\acute G}^{++} {\partial^- \over \partial^+}\,,
\ee
where we use the convention that the 
derivatives written on the right act on the functions on their left;
e.g., $\partial^- F \,\partial^-\equiv \partial^-_x\partial^-_y F(x,y)$.
Also, the operators $1/\partial^+$ written
on the left (right) of $\acute G^{\mu\nu}$
are regularized with a retarded (advanced) prescription.
For instance,
\be 
{1 \over \partial^+}\,{\acute G}^{++} 
{1 \over \partial^+}\,\equiv \,{1 \over \partial^+_R}\,{\acute G}^{++} 
{1 \over \partial^+_A}\,,\ee
where $1/\partial^+_R$ has been defined in eq.~(\ref{delret}), while
\be\label{deladv}
\langle x^-|{1 \over {i\partial^+_A}}|y^-\rangle\,\equiv\,
\int {dp^+\over 2\pi} \,\frac{{\rm e}^{ip^+(x^--y^-)}}
{p^+-i\epsilon}\,=\,i\theta(y^--x^-).\ee
The following symmetry property has been used in writing the
 equations above:
\be \langle x^-|{1 \over {i\partial^+_A}}|y^-\rangle\,=\,-
\langle y^-|{1 \over {i\partial^+_R}}|x^-\rangle\,.\ee
Together with the previous expressions for the scalar propagator
$G$, eq.~(\ref{GSCALAR}),
and for the temporal-gauge propagator $\acute G^{\mu\nu}$,
eq.~(\ref{acuteG}), this concludes our construction of the
LC-gauge gluon propagator in the background field.
\setcounter{equation}{0}
\section{Conclusions}

In this paper, we have given a careful
derivation of the JKLW renormalization group equation 
describing quantum evolution in the high-density regime at small $x$, 
and prepared the framework for the explicit calculation of the 
coefficients in this equation, to be presented in Paper II.

In this context,
we have introduced the physical interpretation of the saturation
regime at small $x$ as a Color Glass Condensate.
This interpretation is supported by the effective classical theory
generated by the JKLW equation (``the McLerran-Venugopalan model''),
where (a) the average over quantum fluctuations is replaced by
a classical average over disorder (as typical for a glass)
and (b) saturation is seen explicitly, and involves
color fields as strong as $A^i\sim 1/g$ (a value typical 
for a condensate).

The derivation presented here differs from the original one
in Refs. \cite{JKLW97,JKLW99a} in that we have considered directly
the field correlators, rather than the correlators of the fluctuations 
in the color charge. We thus have explicitly verified that,
to the order of interest, the quantum correlations at small $x$ are 
correctly reproduced by the classical MV model supplemented with 
the JKLW evolution equation.

In our analysis, we have addressed and clarified several subtle 
points which are essential to make the effective theory and the JKLW
equation unambiguous.

To properly formulate the functional RGE,
and also the solution to the classical source problem, we found it 
necessary to have an extended structure in $x^-$.
This allows us to properly define the functional differentiations
with respect to $\rho$, 
as well as various expressions arising from quantum fluctuations 
which appear in the RGE.

The spatial distribution of the source strength $\rho (x^-,x_\perp)$ in
$x^-$ has been found to be gauge dependent. 
With the retarded gauge prescription which we have chosen here,
we avoid initial state interactions of the gluon field, and the source 
has support at positive $x^-$ only.
In the limit where the source strength is weak, the general RGE
reduces to the BFKL equation independently of the gauge-fixing
prescription, and the longitudinal structure is not important.

To make the classical solution tractable, we have expressed
our results in terms of a source appropriate for the covariant gauge
background field problem.
When we do quantum corrections in this gauge, we found that
it was necessary to include a classical polarization term
in the renormalization group equation.

To cope with the temporal non-locality of the action,
the quantum version of the McLerran-Venugopalan model
has been formulated along a Schwinger-Keldysh contour in complex time. 
The structure of this contour has been shown however not to be important
in the approximations of interest.

Since the background is independent of time but inhomogeneous in $x^-$,
we found it convenient to integrate the quantum fluctuations in layers 
of $p^-$, rather than of $p^+$. In this way, the longitudinal momenta
remain unrestricted, which has allowed us to exploit the geometry of
the problem and construct the background field propagator.
With our retarded axial prescription, we have obtained compact and
relatively simple expressions for this propagator.

In the sequel to this paper \cite{ILM00II}, the  expressions derived
here will be used to explicitly evaluate the coefficients $\sigma$ and
$\chi$ in the RGE,
to all orders in the strong background field which represents
the Color Glass Condensate. We shall also show that, in the large 
$N_c$ limit, our results become equivalent to the non-linear evolution
equations previously proposed
by Balitsky \cite{B} and Kovchegov \cite{K}.

\bigskip
\bigskip
\bigskip
{\large{\bf Acknowledgements}}

The authors gratefully acknowledge conversations with 
Ian Balitsky, Jean-Paul Blaizot, Elena Ferreiro,
Jamal Jalilian-Marian, Al Mueller, 
Eugene Levin, Yuri Kovchegov, Alex Kovner, Raju Venugopalan, 
and Heribert Weigert.

This manuscript has been authorized under Contract No. DE-AC02-98H10886 
with the U. S. Department of Energy.

\appendix
\setcounter{equation}{0}

\end{document}